\documentclass[12pt,a4paper]{article}
\usepackage{authblk}
\title{Studies of Barkhausen Pulses in Ferroelectrics
}
\author[1]{C. D. Tan}
\author[2]{J. Gardner}
\author[2]{F. D. Morrison}
\author[3]{E. K. H. Salje}
\author[1,2,$\dagger$]{J. F. Scott}
\affil[1]{School of Physics and Astronomy, University of St Andrews, St Andrews, Fife KY16 9SS, UK}
\affil[2]{School of Chemistry, University of St Andrews, St Andrews, Fife KY16 9ST, UK}
\affil[3]{Dept. Earth Sciences, Cambridge Univ., Cambridge CB2 3EQ, UK}
\affil[$\dagger$]{jfs4@st-andrews.ac.uk}
\date{}

\usepackage[margin=2.1cm]{geometry}
\usepackage{enumitem}
\usepackage[tbtags]{mathtools}
\usepackage{footmisc}
\usepackage[document]{ragged2e}
\usepackage[british]{babel}
\usepackage{graphicx}
\usepackage{subcaption}
\usepackage{physics}
\usepackage{multicol}
\usepackage{fancyhdr}
\usepackage{cite}
\usepackage[toc,page]{appendix}
\usepackage{textcomp}
\usepackage{hyperref} 
\usepackage{cleveref}
\usepackage[font={footnotesize}]{caption}
\usepackage[font={footnotesize}]{subcaption}
\usepackage{wrapfig}
\usepackage[table]{xcolor}
\usepackage{chngcntr}
\usepackage{authblk}
\usepackage{fixltx2e}

\counterwithin{figure}{section}
\graphicspath{{Figures/}}
\numberwithin{equation}{section}
\begin{document}
	\pagenumbering{Roman}

\setcounter{secnumdepth}{0}
\maketitle
\justify
\begin{center}
	\section{Abstract}
\end{center}
Systems that produce crackling noises such as Barkhausen pulses are statistically similar and can be compared with one another. In this project, the Barkhausen noise of three ferroelectric lead zirconate titanate (PZT) samples were demonstrated to be compatible with avalanche statistics. The peaks of the slew-rate (time derivative of current $dI/dt$) squared, defined as jerks, were statistically analysed and shown to obey power-laws. The critical exponents obtained for three PZT samples (B, F and S) were 1.73, 1.64 and 1.61, respectively, with a standard deviation of 0.04. This power-law behaviour is in excellent agreement with recent theoretical predictions of 1.65 in avalanche theory. If these critical exponents do resemble energy exponents, they were above the energy exponent 1.33 derived from mean-field theory. Based on the power-law distribution of the jerks, we demonstrate that domain switching display self-organised criticality and that Barkhausen jumps measured as electrical noise follows avalanche theory.
\tableofcontents

\vfill

\vfill
\setcounter{secnumdepth}{1}
\setcounter{secnumdepth}{2}
\setcounter{secnumdepth}{3}
\newpage
\pagenumbering{arabic}
\section{Introduction}\label{sec:introduction}
Critical phenomena in solids are among the most fascinating topics in physics and chemistry.  At certain values of temperature or pressure or applied field, various thermodynamic parameters such as specific heat or magnetization display unusual dynamics, diverging rapidly or exhibiting discontinuities, often characterized by "critical exponents" that differ significantly from the Landau-Lifshitz 1937 mean-field values.  Generally it is very difficult to measure these exponents, because the researcher must approach the critical point very slowly and with extreme accurancy.  In both ferromagnetics and ferroelectrics the presence of domains complicates dynamics near the transition critical point.\\\par

However, there is a special class of critical phenomena termed "self-organized criticality" for which this is less difficult.  As first discussed by Bak \cite{Bak1987,Bak1988}, by Kadanoff\cite{Kadanoff1989} and then by Cote and Meisel\cite{Cote1991,Meisel1992}, these include such everyday things as sand dunes, which collapse when they reach a certain height.  This kind of avalanche has its own critical exponents, and the avalanche power spectrum has recently been predicted to have a critical exponent \cite{SaljePredict} of 1.65, much greater than the mean-field prediction of 1.33.  In the present study we show that Barkhausen electrical noise in  the most common ferroelectric material (lead zirconate-titanate, PZT) satisfies avalanche theory quite precisely.\\\par

Some careful attention to statistics is required simply to prove unambiguously that the dependence is power-law (and not, for example, two power laws or a power law and an exponential decay)\cite{Salje2017}.

\subsection{Ferroelectrics and Domains}
\begin{figure}[h]
	\begin{subfigure}[t]{0.27\linewidth}
		\includegraphics[width=.8\linewidth]{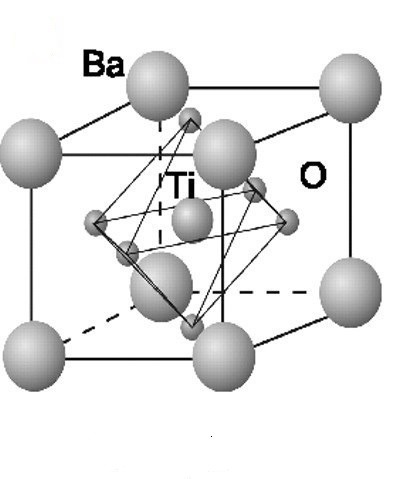}
		\caption{Paraelectric Cubic Phase}
		\label{subfig:intro_para}
	\end{subfigure}\hfill
	\begin{subfigure}[t]{0.27\linewidth}
		\includegraphics[width=.8\linewidth]{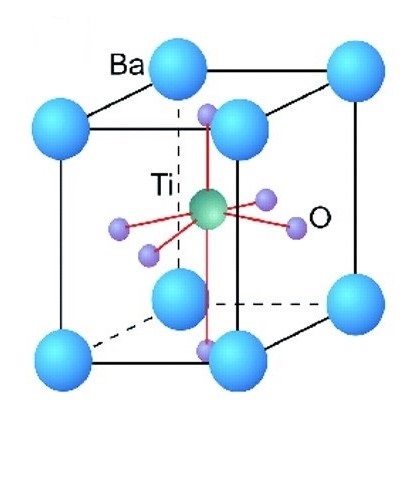}
		\caption{Ferroelectric Phase}
		\label{subfig:intro_ferro}
	\end{subfigure}\hfill
	\begin{subfigure}[t]{0.37\linewidth}
		\includegraphics[width=0.9\linewidth]{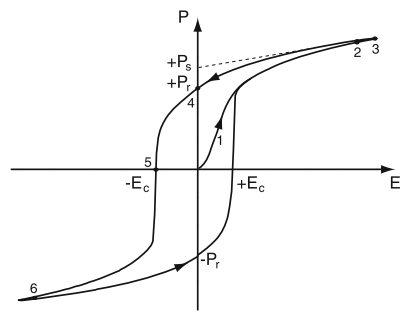}
		\caption{Ferroelectric Hysteresis}
		\label{subfig:intro_hyster}
	\end{subfigure}
	\caption{Below its Curie temperature, barium titanate, BTO transitions from a paraelectric cubic phase (a) to a ferroelectric tetragonal phase (b) by displacing the centre titanium cation with respect to other ions, resulting in a spontaneous polarisation\cite{Scott1989,Ahn2004,Auciello1998,Rabe2007,Schmidt1967}. The cation can be switched by applying an electric field larger than its coercive field $E_c$; the polarisation thus takes the form of a hysteresis loop (c)\cite{Guyonnet2014,Scott1989}. (a), (b): adapted from Ahn, Rabe and Triscone (2004) \cite{Ahn2004} and reprinted with permission from AAAS. (c): reprinted with permission from Springer Customer Service Centre GmbH: Springer Nature, Springer eBook, \href{https://doi.org/10.1007/978-3-319-05750-7}{Domain Walls in Ferroelectric Materials}, J. Guyonnet, 2014 \cite{Guyonnet2014}, Copyright 2018.}
\end{figure}

Ferroelectrics are insulating materials that display spontaneous polarisation which is switchable by an applied electric field\cite{Scott1989,Auciello1998,Rabe2007}. They were first discovered in Rochelle salt by Valasek in 1920 \cite{Valasek1921,Guyonnet2014} and later an oxide ferroelectric material was found: the perovskite-structured (\Cref{subfig:intro_para}) barium titanate, BTO\footnote{Chemical composition: BaTiO$_3$.} during World War II\cite{Scott1989}. Below their respective $T_C$, perovskite ferroelectrics such as BTO and lead zirconate titanate\footnote{Chemical composition: Pb(Zr$_x$Ti$_{1-x}$)O$_3$.} (PZT) undergo a phase transition from a paraelectric cubic phase to a tetragonal ferroelectric phase by displacing the centre titanium Ti cation with respect to other ions (\Cref{subfig:intro_ferro})\cite{Scott1989,Ahn2004,Auciello1998,Rabe2007,Schmidt1967}.\\\par
The displacement of the cation gives rise to a spontaneous polarisation, $P_s$ which can be switched when an electric field $E$ greater than the system's coercive field $E_c$ is applied\cite{Lines2001,Scott1989,Auciello1998,Rabe2007}. Thus, the polarisation of ferroelectrics takes the form of a hysteresis loop (\Cref{subfig:intro_hyster}) \cite{Lines2001,Scott1989,Auciello1998,Rabe2007}. 
Ferroelectrics have garnered a lot of attention for memory-technology applications\cite{Scott1989,Auciello1998,Kim2003,Nagel2002,Setter2006} thanks to this ability to switch their polarisation with ease.
The Ferroelectric Random-Access Memory (FeRAM), for example, capitalises on up and down polarisation switching to store information in a binary form\cite{Scott1989,Auciello1998}.\\\par

\begin{figure}[h]
	\centering
	\begin{subfigure}[t]{0.35\linewidth}
		\includegraphics[width=0.9\linewidth]{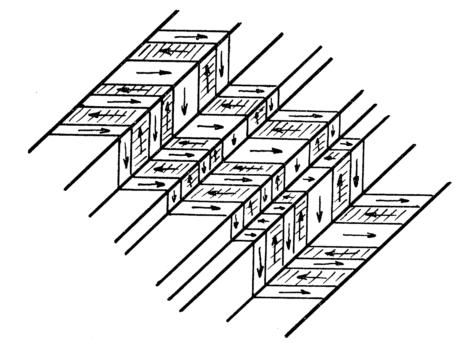}
		\caption{Complex domain structure}
		\label{subfig:intro_domains_mix}
	\end{subfigure}\hfill
	\begin{subfigure}[t]{0.35\linewidth}
		\includegraphics[width=1\linewidth]{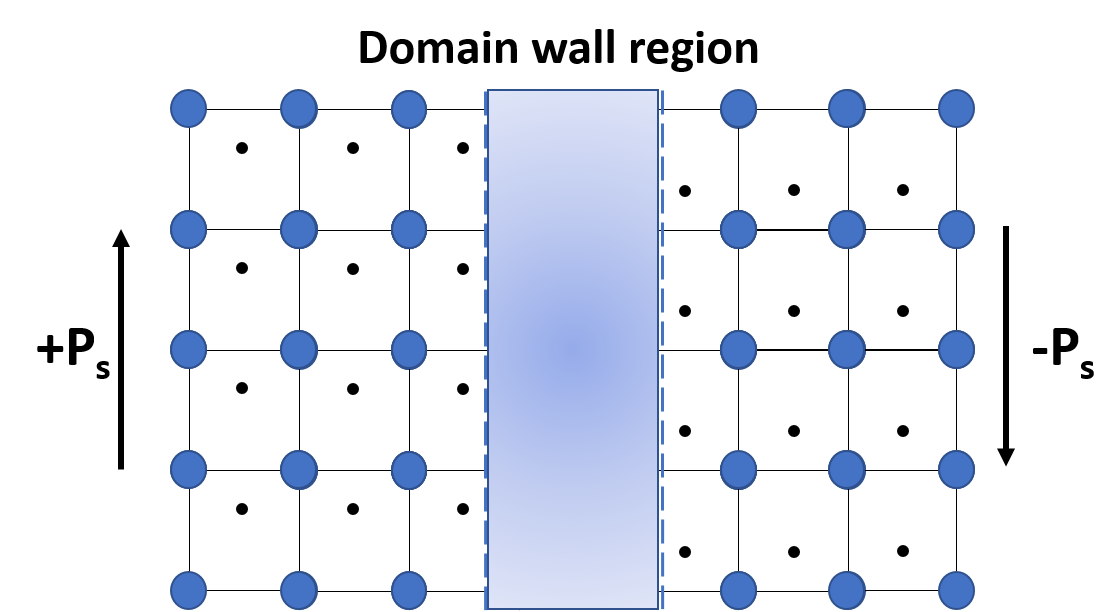}
		\caption{$180^\circ$ Domains}
		\label{subfig:intro_domains_180}
	\end{subfigure}\hfill
	\begin{subfigure}[t]{0.25\linewidth}
		\includegraphics[width=1\linewidth]{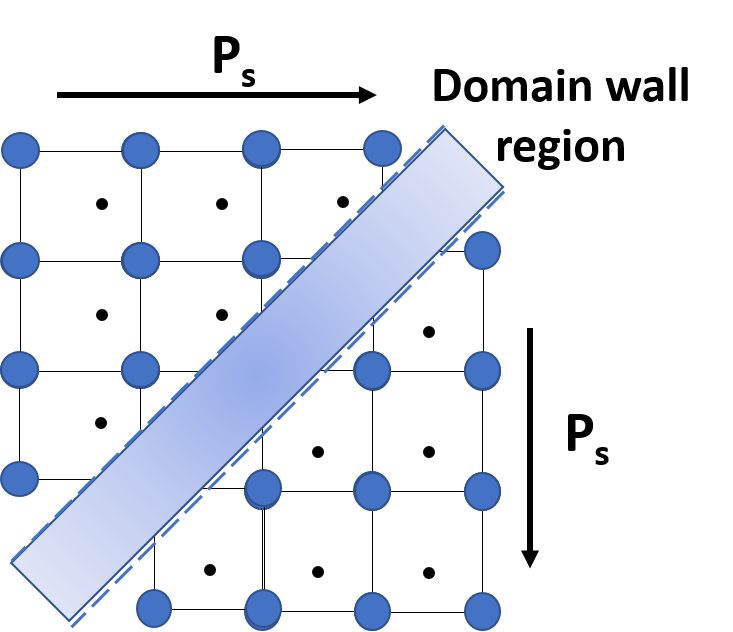}
		\caption{$90^\circ$ Domains}
		\label{subfig:intro_domains_90}
	\end{subfigure}
\caption{Regions of different polarisation (a) that is formed by a combination of $180^\circ$ (b) and $90^\circ$ \textit{domains} (c)\cite{Merz1954}. The displacements of two regions of cations (black dots) in $180^\circ$ domains (b) lead to oppositely aligned polarisation vectors; the $90^\circ$ domains (c) have the polarisation vectors aligned perpendicular to each other\cite{Damjanovic1998}. The boundaries between domains are called \textit{domain walls}\cite{Guyonnet2014,Damjanovic1998}. (a) is a reprinted figure from W. J. Merz, 
(1954) \cite{Merz1954}, Copyright (2018) by the \href{https://dx.doi.org/10.1103/PhysRev.95.690}{American Physical Society}. (b) and (c) adapted from Damjanovic (1998) \cite{Damjanovic1998}.}
\end{figure}

As a system is transitioning towards its ferroelectric phase, regions of different polarisation, called \textit{domains} (\Cref{subfig:intro_domains_mix}), are formed to compensate for the depolarising electric field caused by the volume bound charge density\footnote{From Gauss's Law, a polarised object will generate a volume bound charge density, $\rho_b=\grad\cdot\textbf{P}$\cite{Griffiths2014}.}, and to alleviate mechanical stress due to the tetragonal elongation of the unit cell\cite{Lines2001,Damjanovic1998}. 
 
The electrostatic energy is minimised by forming oppositely polarised $180^\circ$ domains (\Cref{subfig:intro_domains_180}) and perpendicularly polarised $90^\circ$ domains (\Cref{subfig:intro_domains_90}), but only $90^\circ$ domains can reduce mechanical stress\cite{Damjanovic1998}. In ferroelectric crystals\footnote{Domains in ceramics are distributed into grains. These domains can be realigned but the grains cannot. Thus, monodomains can only be achieved in poled crystals and not ceramics\cite{Damjanovic1998}.}, these domains will reorient themselves along an applied field if the external field is sufficiently large\cite{Scott1989,Damjanovic1998}.
\subsection{Barkhausen Noise}\label{subsec:intro_Barkhausen}
The jump-like change in polarisation (or magnetisation) is due to these domains; in fact, it is Barkhausen's experiment in early years that proved that domain structures exist\cite{Rudyak1971}! These jumps occur at the steepest region of the polarisation (magnetisation) hysteresis curve\cite{Rudyak1971,Tebble1955}. Going back to ferromagnets: as a magnetic field is applied, the domains initially aligned along the field will grow and can be thought of as akin to walls of the domains, conveniently called \textit{domain walls}\cite{Guyonnet2014,Damjanovic1998}, extending through space. While the domain walls move, they may come across regions of non-magnetic materials in crystals from defects and become pinned\cite{Rudyak1971,Tebble1955}. As the magnitude of the field increases, the domain walls will eventually gain enough energy to overcome the inclusion and snap forward, causing an increase in magnetisation\cite{Rudyak1971,Tebble1955}. Therefore, the crackling noise in ferromagnetic arises from the jump-like motion of the domain wall\cite{Rudyak1971}.\\\par
Barkhausen noise from a ferroelectric predominantly arises from rapid domain nucleation as an electric field is applied to it\cite{Rudyak1971,Chynoweth1958}. In a fully poled crystal, there are regions in the structure that are stressed due to crystal inhomogeneities\cite{Rudyak1971,Chynoweth1958}. As an external switching field is applied, these stressed sites will easily realign themselves along the more favourable polarisation direction compared to other regions. These rapid nucleations of new opposite domains are responsible for the jerky change in polarisation.
\subsection{Avalanche Theory}\label{subsec:intro_avaltheo}
From unwrapping your favourite confectionery\cite{Kramer1996} to pouring milk over a bowl of Rice Krispies\textsuperscript{\textregistered}\cite{Sethna2001,Salje2014}, Barkhausen noise and many other crepitations are statistically similar: they resemble avalanches and the size distribution of these crackling noises obeys a power-law\cite{Sethna2001,Salje2014,Dahmen2009}. As mentioned previously, these avalanche systems are usually categorised by one or more critical exponents\cite{Sethna2001,Salje2014,Dahmen2009,Dahmen1996,Dahmen2017}.\\\par
Due to universality, these exponents allow avalanche systems to be compared with one another, leading to a lot of interesting physics being unveiled\cite{Sethna2001,Salje2014,Dahmen2009}. For example, Dahmen and Ben-Zion (2009) in \cite{Dahmen2009} showed that Barkhausen noise and earthquake events share many statistical similarities. Meanwhile, Bar\'{o} et al. (2013) in \cite{Baro2013} demonstrated that earthquakes and compressions of porous materials are statistically alike too. Performing these experiments under the safety of the laboratory allows researchers to learn more about seismology and improve earthquake risk assessments\cite{Salje2014}!

\subsection{Project Work}\label{subsec:project-work}

Ferroelectric Barkhausen noises have commonly been addressed as observations\cite{Colla2002,Zhang2017} in experiments but never statistically analysed. We have successfully demonstrated that Barkhausen noises in ferroelectrics take the form of avalanches. Three PZT samples (labelled B, F and S) were categorised using a conventional hysteresis apparatus and the jerky changes in the polarisation switching current responses were shown to obey power-laws (critical exponents listed in \Cref{tbl:intro_sumresult} below). If treated as proxies for energy exponents, our results were in excellent agreement with theoretical predictions ($1.65$\cite{SaljePredict}) according to avalanche theory rather than the energy exponent predicted in mean field theory\cite{Salje2014,Friedman2012,BenZion2011,Dahmen2017,Dahmen2009a,Tsekenis2013}.\par
\begin{table}[h]
	\begin{center}
		\begin{tabular}{||c|c|c|c|c|c||} 
			\hline&&&&&\\[0.5ex]
			Systems &PZT B&PZT F&PZT S& MFT&AT \\[0.5ex]
			\hline\hline&&&&&\\
			Exponent, $\alpha$ & $1.73\pm0.04$&$1.64\pm0.04$&$1.61\pm0.04$&$1.33$&$1.65$ \\ [0.5ex]
			\hline
		\end{tabular}
	\end{center}
	\caption{Summarised exponents of PZT samples B, F and S from this project and theoretical energy exponent from mean field theory (MFT)\cite{Salje2014,Friedman2012,BenZion2011,Dahmen2017,Dahmen2009a,Tsekenis2013} and from avalanche theory (AT)\cite{SaljePredict}. The results of the project work were showed in \textbf{\Cref{sec:latexpw}}. The uncertainties displayed in the results were the standard deviations, $\sigma$ of the estimated exponents.}
	\label{tbl:intro_sumresult}
\end{table}
In this report, the experimental details were summarised in \textbf{\Cref{sec:expdet}}. Our initial work (\textbf{\Cref{sec:iniexpw}}) was not fruitful, and we reviewed our methodologies in \textbf{\Cref{sec:interlude}}. We then shifted our emphasis towards avalanche theory using analytical methods from avalanche statistics such as the Maximum-likelihood analysis, these methods were reviewed in \textbf{\Cref{sec:analyticalmet}}. 
The results from our later experimental avalanche works were reported in \textbf{\Cref{sec:latexpw}} with discussions in \textbf{\Cref{subsec:ResDis}}.
\newpage
\section{Experiment Details and Procedures}\label{sec:expdet}
\subsection{Ferroelectric Samples}\label{subsec:fesamples}
The samples used in this project works are two types of commercially available lead zirconate titanate (PZT) ceramics, with materials labelled PIC 151 and PIC 255, from PI Ceramic Lederhose, Germany\cite{PiezoTech}. Two PIC 151 samples, named B and F, were used for the duration of the project. These ceramics were used in Verdier et al.'s work \cite{Verdier2005} for fatigue studies. According to \cite{Verdier2005}, the ceramics with material PIC 151 has the chemical composition\cite{Verdier2005} Pb$_{0.99}$[Zr$_{0.45}$Ti$_{0.47}$(Ni$_{0.33}$Sb$_{0.67}$)$_{0.08}$]O$_3$. \\\par
The PIC 255 is a newer PZT sample, labelled S, that was recently ordered for this project. The actual composition is proprietary and therefore undisclosed; but from the product brochure\cite{PiezoTech}, it was designed to have a higher coercive field $E_c$ compared to PIC 151. PZTs are usually modified by doping to introduce defects into their crystal structure\cite{Damjanovic1998}. Acceptor doping, for example, stabilises the domain structure, leading to high $E_c$ and harder\footnote{Thus given the term `hard' doped PZTs.} polarisation switching\cite{Damjanovic1998}.

\subsection{Hysteresis Apparatus Set-up}
\begin{figure}[h]
	\centering
	\includegraphics[width=0.8\textwidth]{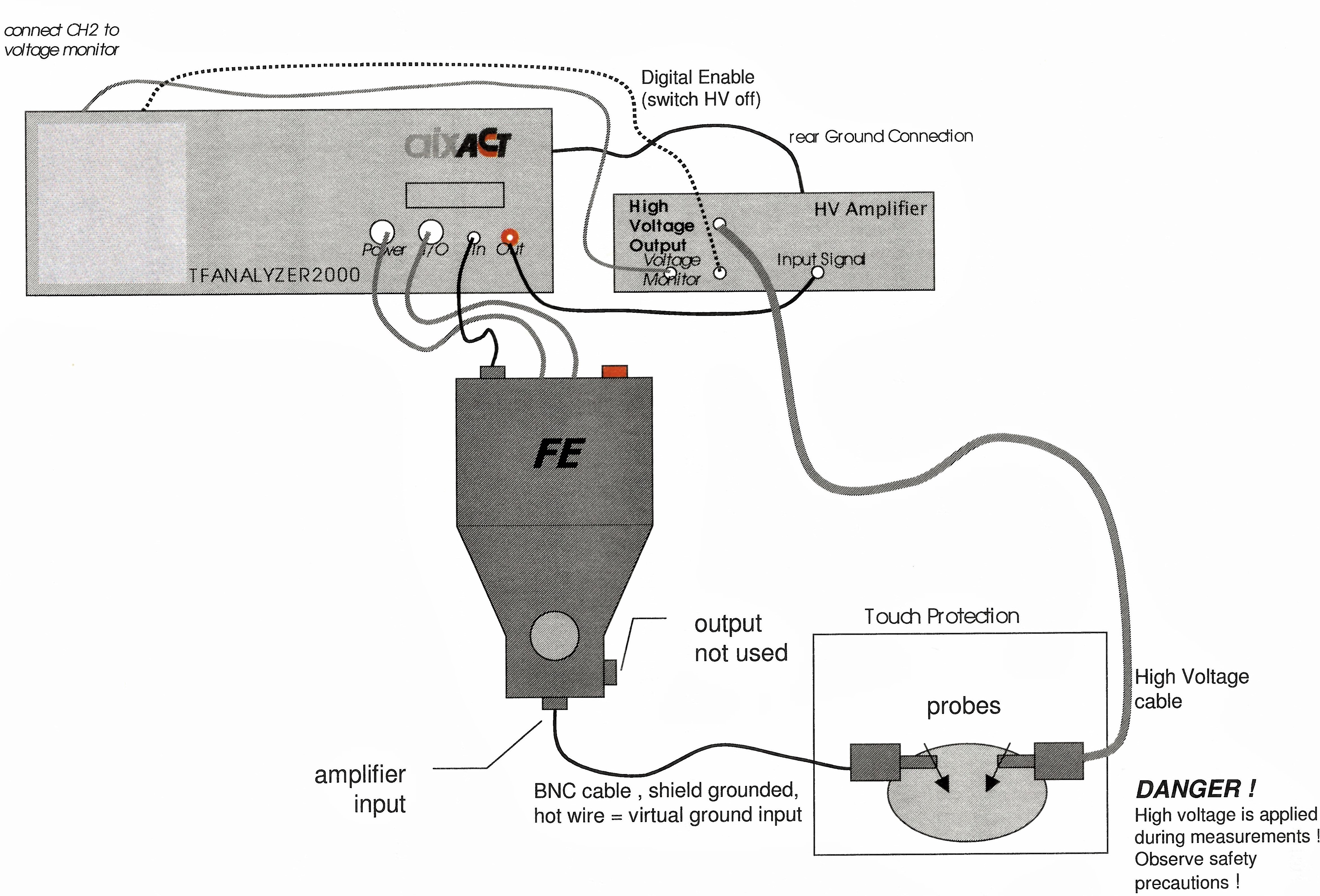}
	\caption{The apparatus set-up taken from aixACCT TF-Analyzer 2000's manual\cite{TfManual}. TF Analyzer 2000 comes with an operating system and its designated hysteresis software where all the measurements were made. The high voltage (HV) amplifier is an external module that enables voltage pulses as high as 40 kV\cite{TfManual} to be applied to a sample via the probes. The TF-Analyzer 2000 is a modular apparatus; attaching an FE-module (labelled FE) allows ferroelectric loops to be investigated\cite{TfManual}.}
	\label{fig:app_diagram}
\end{figure}
The measurements were conducted using the aixACCT TF Analyzer 2000\cite{tfanalyzerweb,TfManual} with a high voltage set-up (shown in \Cref{fig:app_diagram}). The external high voltage amplifier allows voltage pulses ranging from 200 V to 40 kV\cite{TfManual} to be applied to the sample via the probes. The TF-Analyzer 2000 system is actually modular; attaching an FE-module (labelled FE in the figure) will allow hysteresis loops to be measured\cite{TfManual}. The FE-module can be replaced by three other modules to investigate different electroceramic behaviours\cite{TfManual}.\\\par 

In our experiments, triangle or trapezoidal electric field/voltage pulses were applied through a PZT sample via the probes, and the current-time responses were measured. In the context of a hysteresis measurement, the change in polarisation, $\delta P$ can be obtained by integrating the current $I$ with respect to time $t$\cite{Griffiths2014}:
\begin{equation}
\delta P= \int I~dt
\end{equation}
These voltage pulses were designed using a manual waveform generator from the TF Analyzer 2000 software (details in \textbf{\Cref{apn:MWGandDataAc}}) by providing the target voltage values at specific timestamps. As the coercive field $E_c$ of ferroelectrics changes with the voltage ramp-rate\cite{Scott1996,Viehland2000}, the time taken for different maximum voltages were calculated and specified to keep the ramp rate constant.

\subsection{Two-Pulse Measurement}\label{sec:app_two-pulse-measurement}
\begin{figure}[t]
	\begin{multicols}{2}

	\includegraphics[width=0.5\textwidth]{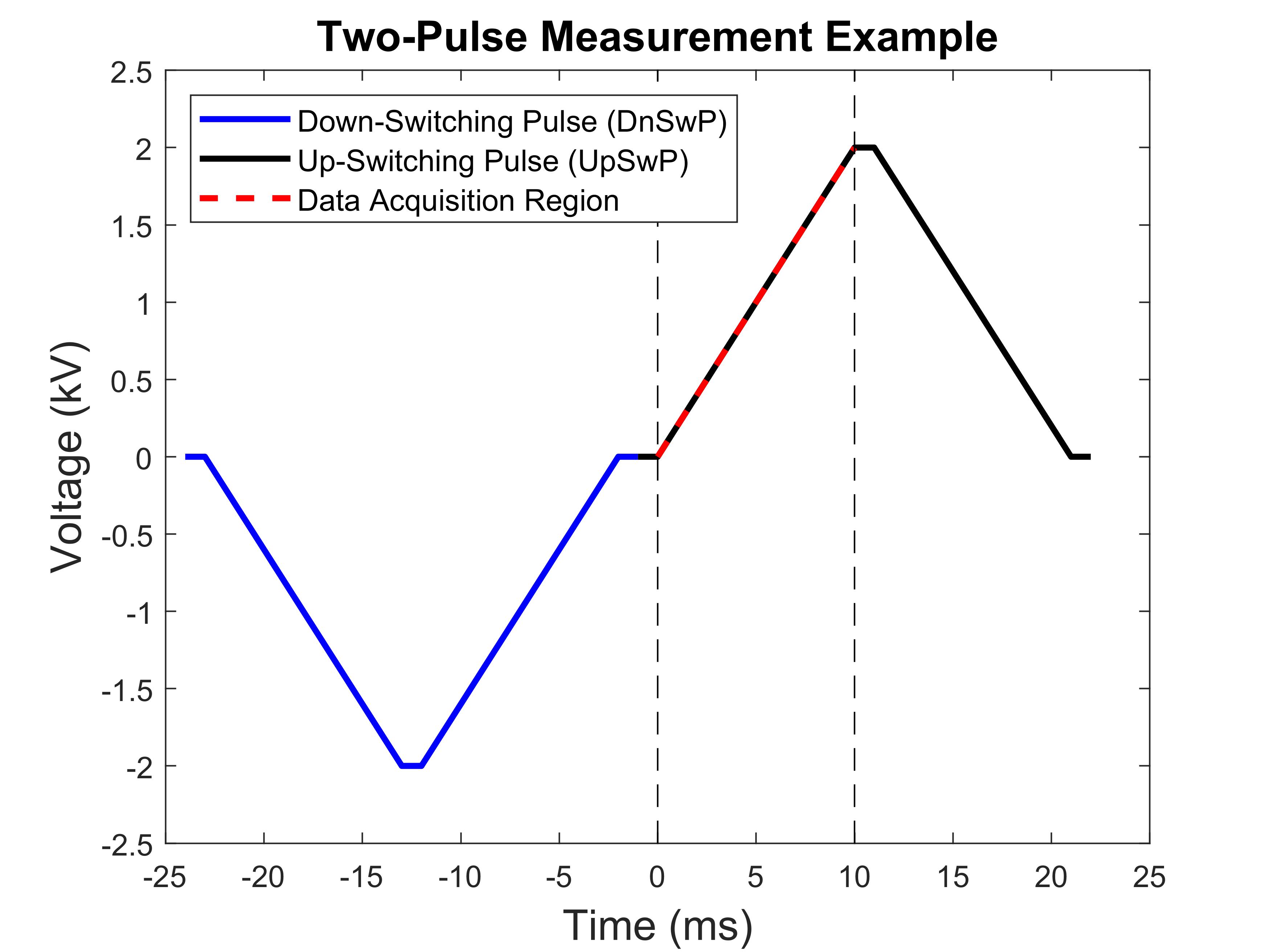}
	\centering

	\caption[width=0.3\textwidth]{An example of a two-pulse measurement waveform. Both the blue down-switching pulse (DnSwP), black up-switching pulse (UpSwP) and the red-dashed data acquisition region can be varied as required (see \textbf{\Cref{apn:MWGandDataAc}}). In this example, DnSwP and UpSwP are both trapezoidal and antisymmetrical, with ramp-rate $=200$ kV\textsuperscript{$-1$} and $\left|V_{max}\right|= 2$ kV. The sample was allowed to relax for 1 ms between the two pulses. The two vertical black-dashed lines indicate the start ($t=0$ ms) and end ($t=10$ ms) times of the data-acquisition region. The data before $t = 0$ ms are not recorded and the waveform after $t=t_\textrm{end} = 10$ ms will not be generated.}
\label{fig:app_meastwopulse}
	\end{multicols}

\end{figure}
The \textbf{two-pulse measurement} was the basis of the experiment work in this project. A triangular or trapezoidal negative/down-switching pulse (DnSwP) was generated across the ferroelectric ceramic, followed by a positive/up-switching pulse (UpSwP) counterpart where the data were usually acquired (see \Cref{fig:app_meastwopulse}).\\\par

Since the voltage waveform was generated manually, there was a lot of freedom with regards to how the pulses were generated. A schematic of a trapezoidal antisymmetrical voltage waveform is given in \Cref{fig:app_meastwopulse} where the magnitude of the voltage maximum $\left|V_\textrm{max}\right|$ is $2$ kV and the ramp-rate is $200$ kV\textsuperscript{$-1$}. In this figure, the data from the ramping region of UpSwP (red-dashed line sandwiched between two vertical black lines) will be acquired up to a maximum of 1000 points spaced evenly in between. However, there are some drawbacks: no data will be recorded before the first vertical black line ($t = 0$ ms) and the voltage pulse \textbf{will not follow through} for $t>t_{\textrm{end}} = 10$ ms (described in \textbf{\Cref{apn:MWGandDataAc}}). This leads to a few complications that needed to be addressed:
\begin{enumerate}
\item	\textbf{If the region before $t=0$ ms is not recorded, how do we know if the voltage pulse is generated as intended?}\label{enuopt:nopulseafter}

This can easily be determined by decreasing the magnitude of DnSwP while keeping UpSwP constant. As the magnitude decreases, fewer up-domains will be switched down and the resulting current peak from following UpSwP will decrease. If DnSwP is then set to $0$, no current peak from UpSwP will be observed, leaving only electrical noise. 

\item	\textbf{If the pulses do not continue beyond $t_\textrm{end}$, what actually happens at $t>t_\textrm{end}$?}

Without a second voltage measurement apparatus, it is impossible to tell. The voltage supply is assumed to be immediately cut off. Testing with an external voltmeter is not a good idea either since the voltages supplied are in the kV range, which poses a health hazard if not executed professionally. However, this effect will not cause any difficulties in taking the measurements as long as the region of interest is recorded.

\item \textbf{How do you ensure that the up-switching voltage pulse is fully generated?}

Taking \Cref{fig:app_meastwopulse} as an example: instead of the ramping region, one can set the recording region to encompass the whole pulse, from $t=0$ to $t=t_\textrm{end} = 21$ ms (the additional $1$ ms arises from the $1$ ms plateau). The full current spectrum of 1000 points in $21~\mu$s-intervals will then be recorded. But this comes with a price: the time taken for data acquisition is now larger, essentially reducing our sampling rate. 

\item \textbf{Why is the sampling rate reduced?}

From the software, the maximum number of allowed points for data acquisition is 1000 points\footnote{1001 including the point at $t=0$ s.}; if the recording region takes a total of 1 ms, 1000 points of data in 1 $\mu$s-intervals will be taken, giving a sampling rate of 1 MHz. The ramping region in \Cref{fig:app_meastwopulse} takes 10 ms; the time interval between two acquired data points will be $10~\mu$s, giving a sampling rate of 100 kHz. The full pulse takes 21 ms, which reduces the sampling rate to approximately 48 kHz.
\item \textbf{What is the maximum sampling rate?}
From the manual\cite{TfManual}, the maximum sampling rate is 1 MHz (1000 points in $1~\mu$s-intervals recorded in 1 ms). Going beyond 1 MHz will result in data points recorded with repeated entries. 

\end{enumerate}
To clarify question \ref{enuopt:nopulseafter} above, imagine the recording region is set at the downwards ramping region ($-24 \textrm{ ms} \leq t \leq -14 \textrm{ ms}$) of DwSwP in \Cref{fig:app_meastwopulse} with multiple readings taken. If the sample is previously poled upwards, a negative current switching peak will be observed in the \textbf{first} measurement. But the next consecutive measurements will yield only \textbf{electrical noise} because the UpSwP is not followed through and the sample is not switched as expected.\\\par
Hence to conclude, care must be taken when setting the data acquisition region. One can make sure the pulses are generated as intended by encompassing the full voltage waveform but this will result in a reduced sampling rate during data acquisition.\\\par

\newpage

\section{Initial Experimental Work}\label{sec:iniexpw}
Our work initially involved applying triangular or trapezoidal, down-switching pulses (DnSwPs) and up-switching pulses (UpSwPs), respectively, to PZT samples (as detailed in \textbf{\Cref{sec:app_two-pulse-measurement}}) with maximum voltage $\left|V_\textrm{max}\right|$ ranging from 1 to 3 kV\footnote{Corresponding to fields of 1 to 3 kVmm$^{-1}$ for a sample with a thickness of 1 mm.} with a ramp-rate that varies between 200 and 500 kVs\textsuperscript{$-1$}. As mentioned by Rudyak (1971) in \cite{Rudyak1971}, Barkhausen jumps of 90$^\circ$ degree domain walls will require a larger critical electric field than their 180$^\circ$ counterparts. As the domains in lead zirconate titanate (PZT) consist of both $180^\circ$ and $90^\circ$ walls\cite{Okayasu2012,Roelofs2002}, our initial hypothesis was, by introducing a high enough electric field, the Barkhausen jumps triggered by $90^\circ$ domain walls could be observed.\\\par
These early works allowed us to take a step back and think critically on how we should approach our project (see \textbf{\Cref{subsec:nstepfor}}) and to develop new methodologies for our later experimental stage \textbf{\Cref{sec:interlude}}. An example of our early results is summarised in this section. 
\subsection{Two Pulse Experiment Example}\label{subsec:tpulex}
DnSwPs and UpSwPs were applied in succession to PZT (PIC 151) sample B and the current-time response during UpSwPs was recorded. Sample B had a thickness of $0.93\pm0.01$ mm with gold electrodes on both sides with an area of $28.5\pm0.1$ mm$^2$ (radius = $3.01\pm0.01$ mm)\footnote{Sample B was categorised in greater detail in the later section (\textbf{\Cref{subsec:srun_pztb}}).}. \par
\begin{figure}[h]
	\centering
	\begin{subfigure}{0.48\textwidth}
		\includegraphics[width=1.0\textwidth]{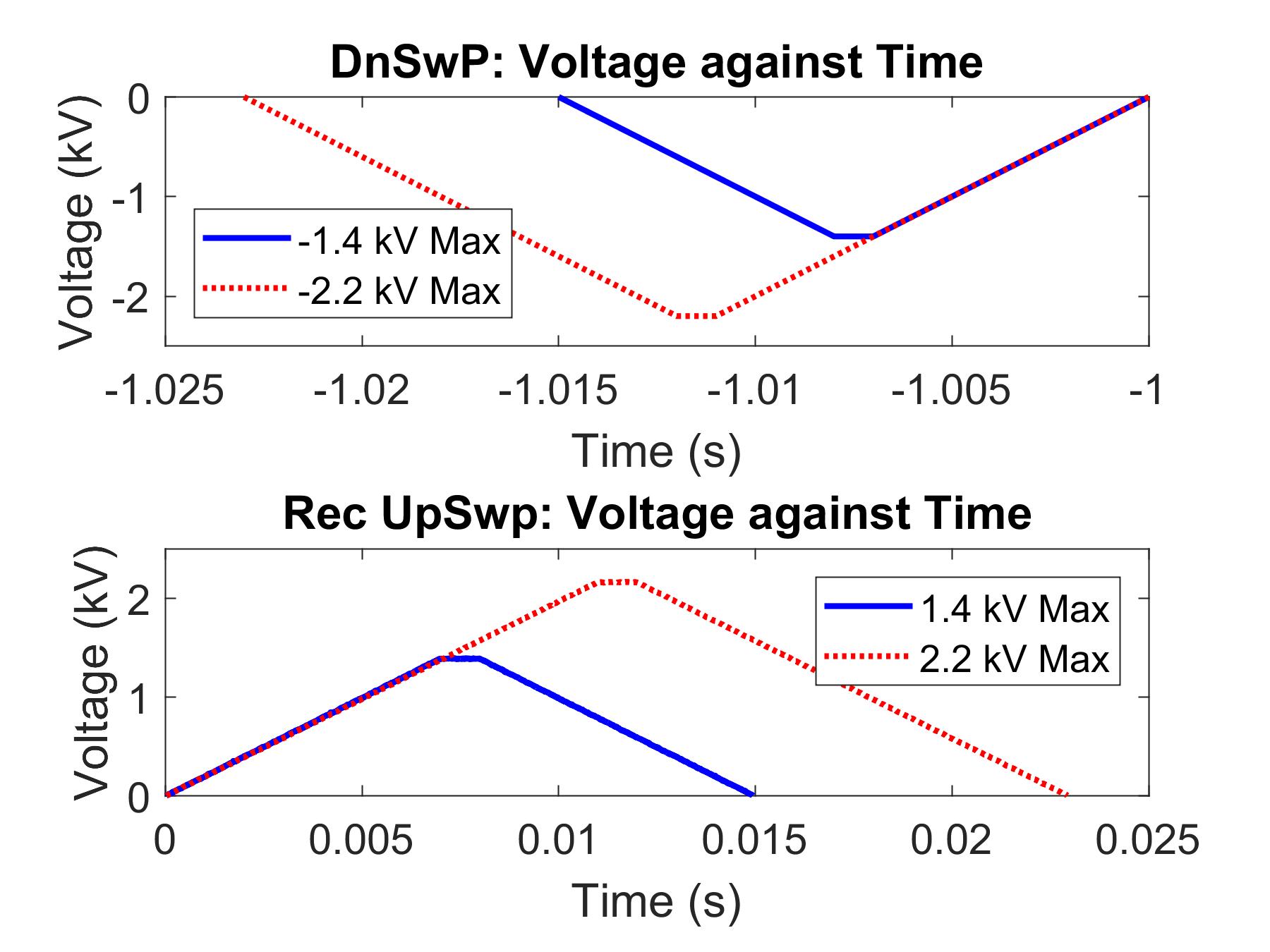}
		
		\subcaption{Voltage waveforms of DnSwP (Top) and UpSwP (Bottom). Only data from UpSwP were acquired.}	
		\label{subfig:ew_vvst}
	\end{subfigure}
	\hfill
	\begin{subfigure}{0.48\textwidth}
		\includegraphics[width=1.0\textwidth]{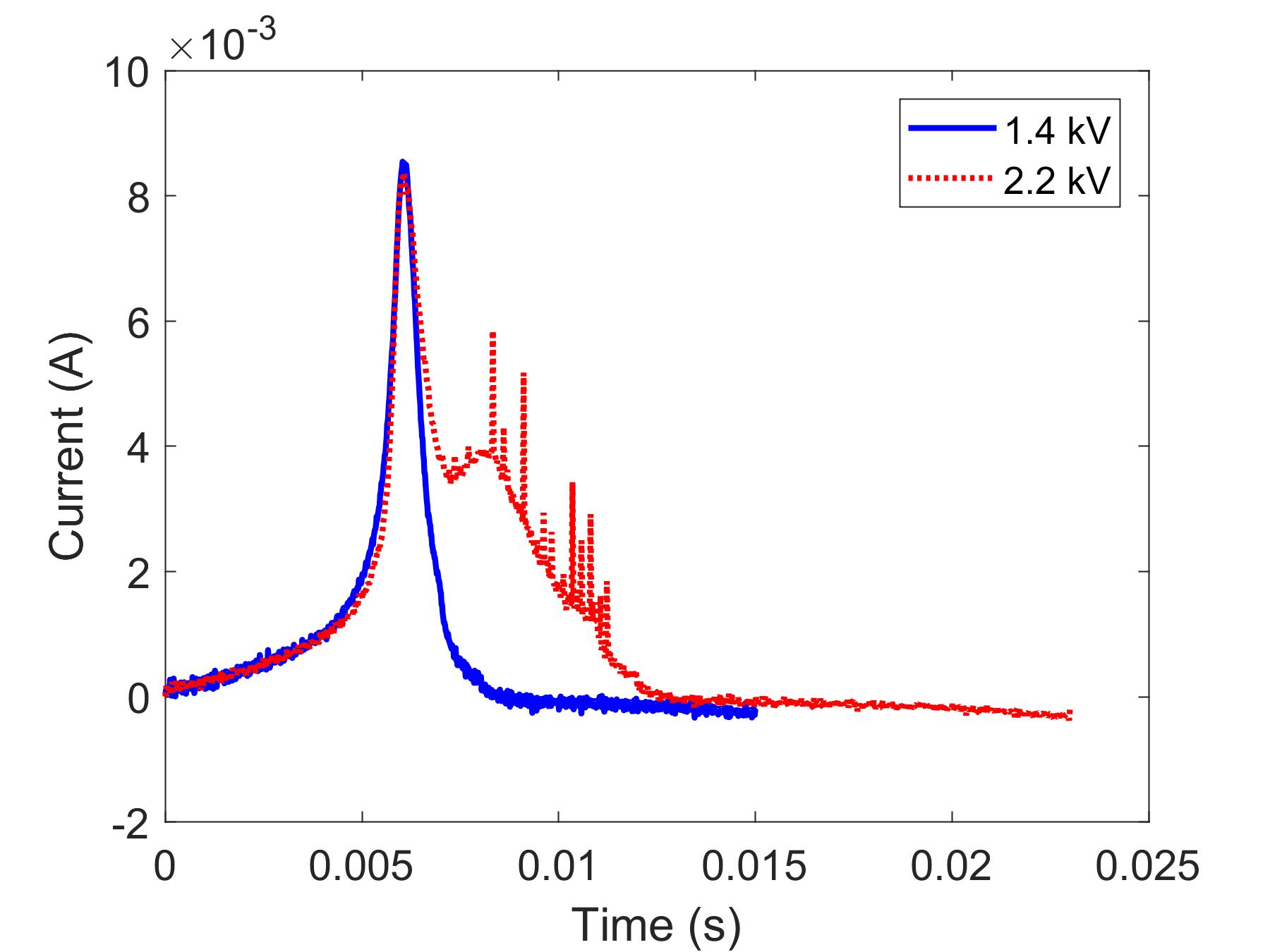}
		
		\subcaption{Current response during UpSwP for $V_\textrm{max}=1.4$ kV (blue) and $2.2$ kV (red dotted).}
		\label{subfig:ew_ivst}
	\end{subfigure}
	\caption{(a) shows the trapezoidal down-switching pulses (DnSwP) and up-switching pulses (UpSwP) for two measurement examples. With a rate of $200$ kVs$^{-1}$, the pulses were generated at a maximum magnitude of $\left|V_\textrm{max}\right|=1.4$ kV (blue) and $2.2$ kV (red dotted) respectively and were held constant for $1$ ms. Note the timeline of the pulses; the DnSwPs were applied up to time $t=-1$ s and the sample was allowed to rest for one second. The current responses were then recorded at $t=0$ s for the whole duration of the UpSwPs, shown at (b). The current responses for the higher maximum voltage exhibited some intense spikes and a bump. These intense spikes were hypothesised to be Barkhausen pulses from $90^\circ$ domain switchings, based on discussion in the literature.}
	\label{fig:ew_experiment}
\end{figure}
An example of the two pulse measurement is shown at \Cref{fig:ew_experiment}, with DnSwPs and UpSwPs of two different maximum magnitudes ($1.4$ kV as blue and $2.2$ kV as red dotted) plotted at the top and bottom of \Cref{subfig:ew_vvst}. The pulses were trapezoidal with a plateau of 1 ms and were antisymmetrical with respect to each other. \\\par

The timeline of \Cref{subfig:ew_vvst} shows that the DnSwPs were applied up to time $t=-1$ s (as a notation as the recording time starts at $t = 0$ s), and the sample was allowed to relax for one second. The UpSwPs were then applied at $t=0$ s and the current data points were acquired through the duration of the UpSwPs. The ramp-rate of the pulses was $200$ kVs$^{-1}$; therefore the $2.2$ kV pulse took a longer time to complete. As the number of data points acquired is fixed, the current response from $1.4$ kV will have a finer resolution compared to $2.2$ kV.\\\par

The current responses from the $1.4$ kV UpSwP (blue) and $2.2$ kV UpSwP (red dotted) is displayed as \Cref{subfig:ew_ivst}. The current plot for $2.2$ kV presents some massive current spikes as the UpSwP is increasing towards a very high voltage. These spikes are evident in the interval after the current peak, when the applied voltage exceeds the coercive voltage $V_c$. In addition, we analysed the noise spectrum (\textbf{\Cref{apn:noisean}}) and compared the results to other non-ferroelectrics. The noises generated from non-ferroelectrics (see \Cref{subfig:ew_perov_noise}) were oscillatory and resembled ``echoes", while these current spikes from PZT were random (see \Cref{fig:ew_three_fft}). So, these current spikes did fit the picture that we hypothesised; we had applied a voltage pulse that was large enough to induce Barkhausen jumps from the $90^\circ$ domains! \\\par

\begin{figure}[h]
	\begin{multicols}{2}
		
		\centering
		\includegraphics[width=0.5\textwidth]{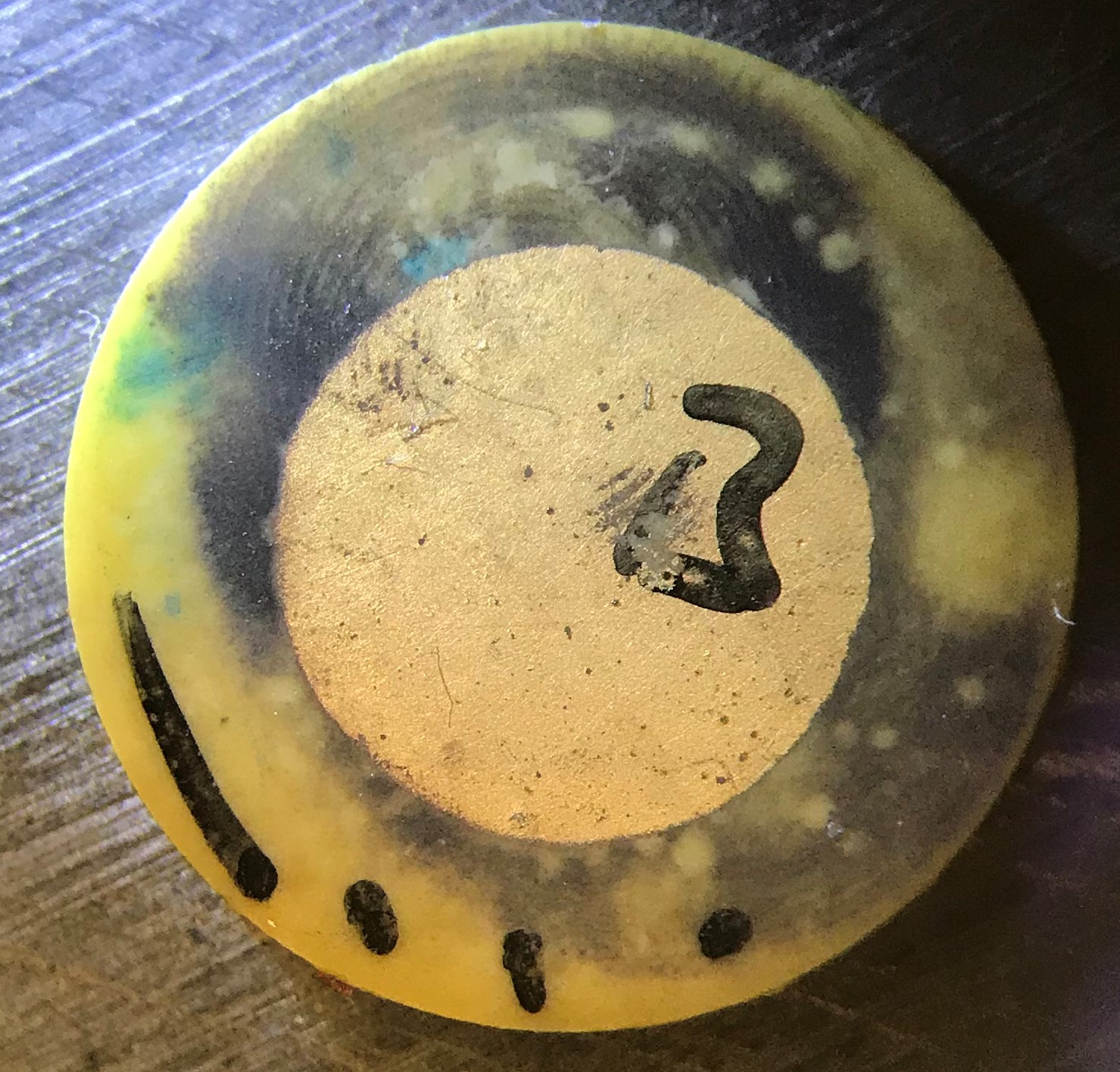}
		\caption{PZT sample B after multiple measurements using fast and high voltage pulses. The black/scorched patches resembled the non-ferroelectric layer that was generated when the PZT sample was fatigued\cite{Verdier2005}. The surface layer could be sanded off and the sample could be annealed in order to refurbish it, but it was decided that the high voltage and ramp-rate methodology was not practical and time-consuming. The radius of the electrode is $3.01\pm0.01$ mm..}
		\label{fig:ew_ruin_pzt}
	\end{multicols}
	
\end{figure}

However, neither the hypothesis nor the PZT samples lasted, as these high voltage pulses easily denatured our PZT samples. The samples were fatigued very quickly, forming a black non-ferroelectric layer on the surface\cite{Verdier2005}, as shown in \Cref{fig:ew_ruin_pzt}. Though these samples could easily be refurbished by polishing off the modified layer and by annealing\cite{Verdier2005}, we began to ponder the question: did the ability to observe Barkhausen noise come at the cost of destroying our samples? We conclusively decided that applying fast and large voltage pulses to our samples was not practical and we took a step back to review our methodology (elaborated in the next section (\textbf{\Cref{sec:interlude}})). 
\newpage
\section{Interlude}\label{sec:interlude}
As mentioned in the introduction (\textbf{\Cref{sec:introduction}}), Barkhausen pulses typically occur in the steepest region of the polarisation against electric field/voltage curve\cite{Rudyak1971}, which is when the electric voltage reaches the coercive voltage $V_c$ (the point where current $I$ is at its maximum).  Although when categorising ferroelectrics, it is more appropriate to address the electric field, such as $E_c$\footnote{since coercive field $E_c$ for each chemically identical PZT will be dimension independent.}; in this report the voltage is emphasised more often instead as both DnSwPs and UpSwPs applied to the samples are generated in terms of voltage, and are mostly identical throughout the experimentation. The thickness for each sample, however,  may vary slightly\footnote{The average thickness is 0.97 $mm$ with a standard deviation of $\pm0.04$ mm.}. \\\par

As seen in previous experimental works \textbf{\Cref{subsec:tpulex}}, the noises were only evident in the region where $V>V_c$ (see \Cref{subfig:ew_ivst}). This raises some concerns which are summarised below:
\begin{enumerate}
	\item If, according to Rudyak (1971)\cite{Rudyak1971}, Barkhausen noise is most intense at region $V\leq V_c$; why were these pulses observed only at $V>V_c$?
	\item If, according to Chynoweth (1958)\cite{Chynoweth1958} and Rudyak (1971)\cite{Rudyak1971}, Barkhausen pulses were generated in ferroelectrics by domain nucleation; shouldn't the domains be nucleated and/or propagated as $V$ increases from $V=0$  towards $V_c$ and not when $V>V_c$? 
	\item If the pulses at $V>V_c$ were indeed caused by the nucleation of new domains, then what was generating the current switching peak for the region $V\leq V_c$? 
	\item During the initial experimentation period, the samples were degrading very quickly (as shown in \Cref{fig:ew_ruin_pzt}) and were sent to anneal very often; were the Barkhausen pulses (or the experimentation method used to observe them) destructive?
\end{enumerate}
Though our early work had given us some insights with regards to noises in a ferroelectric current signal, the experimental results did not match the findings by Rudyak and Chynoweth; these noises might not be Barkhausen. Thus, we re-evaluated our strategy and designed new methods to investigate these pulses.

\subsection{The Next Step Forward}\label{subsec:nstepfor}
We did some literature research on more recent works with regards to Barkhausen noise and discovered that a lot of research had been conducted, both theoretically and experimentally, on systems that generate crackling noise. From earthquakes to magnetic avalanches (ferromagnetic Barkhausen noise)\cite{Dahmen2009}, systems that exhibit crackling noises are statistically similar \cite{Salje2014}. This gave rise to a new idea: since domain mechanics of ferroelectrics are comparable to their ferromagnetic cousins, do they exhibit the same statistical similarities? Hence, we decided to look into \textit{Avalanche Analyses}.\\\par
According to Salje and Dahmen (2014)\cite{Salje2014}, avalanche experiments ``must be slow" in order to observe these crepitations. Therefore, with \textit{slow}\footnote{low ramp-rate.} voltage pulses, the experiments were redesigned and conducted on two PZT samples B and F (material PIC 151 from PI Ceramic Lederhose, Germany) and a newly ordered PZT sample, labelled S, (material PIC 255 also from PI Ceramic Lederhose, Germany). The changes are detailed below for clarity:
\begin{enumerate}
	\item The maximum voltage range was reduced from $3$ kV to $1$ kV (for PZT samples B and F) and $1.5$ kV (for PZT sample S). 
	\item The ramp-rate rate was decreased from $200$ kVs\textsuperscript{$-1$} to $40$ Vs\textsuperscript{$-1$} ($5000$ times slower) for samples B and F and $60$ Vs$^{-1}$ for sample S. 
	\item Both UpSwP and DnSwP pulses were now triangular and antisymmetrical. Both of them were applied individually instead of via a single waveform file.\label{enuopt:sepwav}
	\item The sampling rate was decreased to\footnote{For example: $1000$ points $\times 40$ Vs\textsuperscript{$-1$}$/1$ kV $= 40$ Hz.} $40$ Hz. \label{enuopt:lowsamprate}	
\end{enumerate}
To reiterate \cref{enuopt:sepwav} above: In contrast to previous waveforms where a DnSwP was followed by a UpSwP and the upwards ramping region (from V$=0$ to V$_{\textrm{max}}$) was recorded,  both up-switching and down-switching pulses were now applied separately. This was to ensure that the samples were really switched by reading the current responses from both pulses instead of only relying on the current response of the UpSwP. \\\par
However, as mentioned in \cref{enuopt:lowsamprate}, the drawback was these pulses now took a longer time to reach the voltage maximum, thus reducing the sampling rate to a smaller frequency ($40$ points per second). The drawback of the low sampling rate was the jerk peaks may overlap in time\cite{Friedman2012}. However, there are two arguments we made to justify the use of a lower sampling rate:
\begin{itemize}
	\item The \textbf{shape} of the current curve is assumed to be almost identical as the ramping rate decreases. The general current response behaviour is assumed to not deviate from a fast ramp to a slow ramp. 
	\item Reducing the rate reduces a lot of ringing noise/echoes at low current signals. This generates a better current response, giving us more statistically relevant data points.
\end{itemize}
\subsubsection{Observations on the Change of Methodology}
The first effect we noticed going from a high to low ramping rate was the smoothing of the current response with no ringing noise present at the low current region\footnote{The ringing noise can be seen at \Cref{subfig:ew_ivst} in the region $t<0.005$ s.}. The current generated had also decreased in magnitude which is consistent in theory since polarisation\cite{Griffiths2014} is now changing over a larger time interval:
\begin{equation}
\textrm{Current, }I=\frac{dP}{dt}
\end{equation}
Assuming the total polarisation change, $\delta P$ of a sample is constant for all ramp-rates; a sample takes a longer time, $dt$ to achieve a small change in polarisation, $dP$ and thus decreasing the magnitude of $I$. In addition, the coercive voltage $V_c$ or field $E_c$\footnote{the voltage or electric field where current $I=I_\textrm{max}$.} decreases as the ramp-rate decreases. This observation is consistent with theory\footnote{Proposed by Viehland and Chen (2000)\cite{Viehland2000}, at lower ramp-rates, more time is allowed for small oppositely polarised domains to nucleate, which allow polarisation reversal to occur at a lower electric field.}\cite{Scott1996,Viehland2000}. 
\newpage
\section{Analytical Methods}\label{sec:analyticalmet}
\subsection{Jerks}\label{subsec:jerks}
\begin{figure}[h]
\centering
\begin{subfigure}[t]{0.48\textwidth}
\includegraphics[width=1\textwidth]{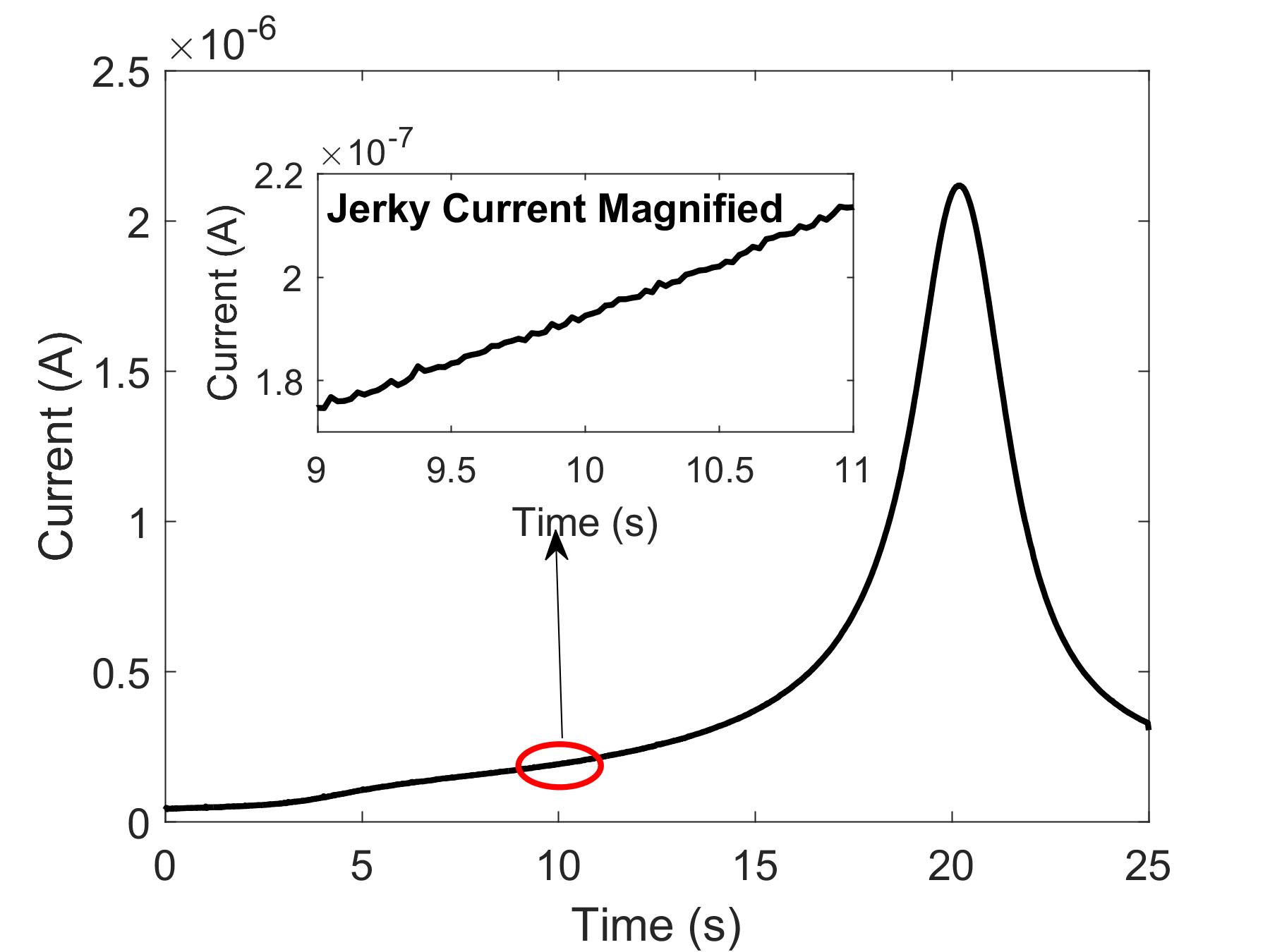}
\caption{Jerky current, $I$ from PZT B}
\label{subfig:pztb_run2_ivst}
\end{subfigure}
\hfill
\begin{subfigure}[t]{0.48\textwidth}
\includegraphics[width=1\textwidth]{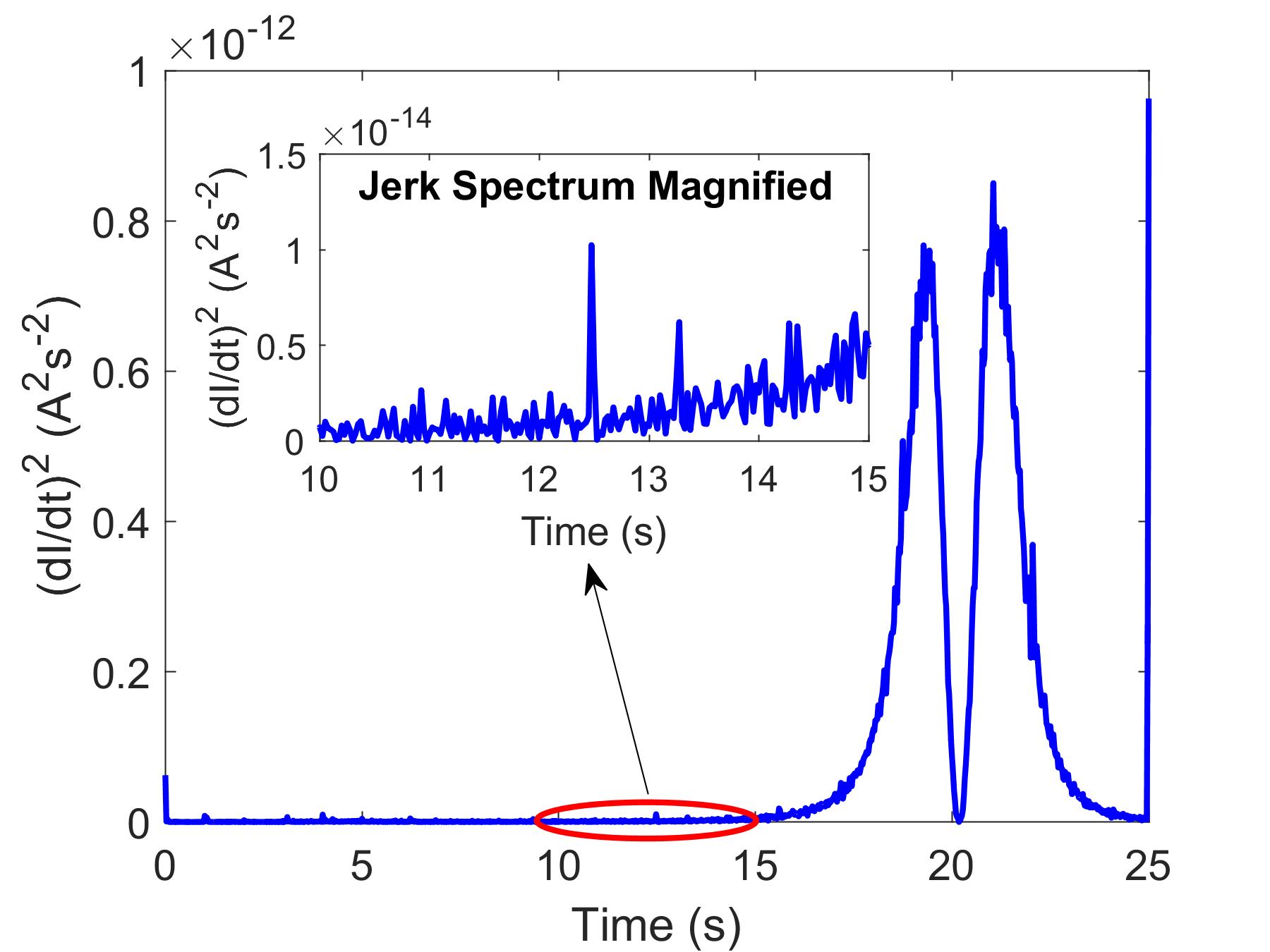}
\caption{Jerk spectrum, $(dI/dt)^2$ with baseline}
\label{subfig:pztb_run2_jvst}
\end{subfigure}
\caption{(a): A current $I$ response example taken from sample PZT B under a slow ramp-rate ($40$ Vs\textsuperscript{$-1$}) to $1$ kV. The curve may look smooth, but as the inset shows, it is actually ``jerky". (b): The ``jerkiness" is more evident when the current curve is differentiated with respect to time (and squared). This plot is called the jerk spectrum with baseline; the peaks of this curve (after removing the baseline) are later defined as \textit{jerks} and are used for statistical analysis.}
\label{fig:jerkycurr}
\end{figure}

As mentioned previously, the ramp-rate was now greatly reduced and the resulting current response was neither noisy nor filled with random peaks. An example measurement of a slow ramp-rate is shown as \Cref{subfig:pztb_run2_ivst}. However, a closer examination of the current-time response will show that the curve is not smooth but exhibits a jerky motion. To fully extract these \textit{jerks}, the first time-derivative of the current $I$ spectrum was taken and then squared, resulting in a jerk spectrum with a baseline (\Cref{subfig:pztb_run2_jvst}).\par
The baseline mentioned above is actually the two humps of the jerk spectrum (\Cref{subfig:pztb_run2_jvst}) between times, $t=15$ s and $t=25$ s; one can see there are multiple jerky peaks superimposed onto them. The baseline is removed (described in \textbf{\Cref{apn:RemBase_pztf}}) and the resulting peaks are defined as \textit{jerks}, $J$. These peaks play a crucial role in analysing Barkhausen noise as Barkhausen noises adhere to avalanche theory\cite{Salje2014,Dahmen2009}. This leads to our new hypothesis:

\begin{quote}
	If these jerks peaks did arise from Barkhausen pulses, they should show statistical similarities with other avalanche systems. These jerk peaks should obey power-law.
\end{quote}
\begin{equation}\label{eq:jerkpowerlaw}
P(J)=\frac{\alpha-1}{J_\textrm{min}}\left(\frac{J}{J_\textrm{min}}\right)^{-\alpha}
\end{equation}
where $J$ is jerks, $\alpha$ is a critical exponent and $J_\textrm{min}$ is the lower-bound normalisation condition \cite{Clauset2007,Salje2017}.\\\par
In avalanche studies, jerks are defined in multiple ways, even for the same system\cite{BenZion2011,Salje2014,He2016}. In He et al. (2016) \cite{He2016} ferroelastic switching simulations, the jerks were defined as the total change in potential energy, the energy drop and the shear stress drop. In crystal plasticity experiments, the jerks were defined as the square of the velocity of \textit{slip} avalanches squared, $(dv/dt)^2$, which took the form of an energy\cite{Salje2014,Dahmen2017,Friedman2012,Tsekenis2013}.\\\par 
In our case, the jerks were defined as peaks of the square of the \textit{slew-rate}\cite{Charalambous2018}, $(dI/dt)^2$. The slew-rate $dI/dt$, is a term commonly found in electronics to describe how fast can an input waveform change when it passes through an op-amp before the signal is distorted\cite{Carter2013}. The slew-rate is usually defined as the time derivative of voltage $dV/dt$\cite{Carter2013} but it can also be defined as $dI/dt$\cite{Charalambous2018}. Taking the square of the slew rate allow the peaks in our jerk spectrum to proxy the jerks defined in crystal plasticity and acoustic emission experiments, which is $(dv/dt)^2$ as mentioned.

\subsection{Power-law Histogram}
\subsubsection{Linear Binning}
Knowing that the distribution of jerks may take the form of a power-law, the most straight-forward way to extract the critical exponent from a jerk spectrum is to bin the magnitude of the observed peaks into bins of constant linear width. The number of counts (or frequency) of the jerks $J$ with various magnitudes can then be divided by the total jerk counts and can now be approximated as a probability distribution $P(J)$ \cite{White2008}. Taking the natural logarithm\footnote{All the $\log$ functions in the figure labels are natural logarithms, $\ln$.} of both the magnitude of the jerks (x-axis) and the probability (y-axis), a linear regression can be fitted through the data points to retrieve the exponent as the gradient of the fitted line\footnote{Empty bins however must be excluded to prevent taking $ln(0)$, which is undefined.}. 
\subsubsection{Logarithmic Binning}
On the other hand, one can take natural logs of the data and then proceed to bin the log-transformed data to form a distribution, this is known as logarithmic binning. This reduces the number of zero and low-count entries since the \textit{linear} bin width now increases linearly with the magnitude of $J$ \cite{White2008}, binning more low-count data as the bin width increases. However, there is a little subtlety while using this binning method, as one will find the linear fitted straight line will estimate a gradient, $m$ of the expected exponent, $\alpha + 1$ and not $\alpha$ \cite{White2008,Bonnet2001}! This is due to the bin width increasing linearly with the jerk magnitude and can easily be remedied by normalising the distribution (discussed by White, Enquist and Green (2008) in \cite{White2008}).
\subsubsection{Disadvantages of Linear Regression Fitting}
\label{sssec:disadvLinFit}
However, as argued by Clauset, Shalizi and Newman\cite{Clauset2007} in 2007, fitting a linear regression through a log-log plotted distribution is not necessarily a good idea, even if the transformed distribution exhibits a straight-line behaviour. The full details can be found in Clauset et al.'s \textit{Power-Law Distribution in Empirical Data} (2007) Appendix A\cite{Clauset2007} while the relevant reasonings can be summarised as follows:
\begin{enumerate}
\item 	The standard error calculated by the linear regression formula is inapplicable as the noise in the logarithmic dependent variable ($\ln{P(J)}$ in this case) is not Gaussian. The formula for error is only true when the assumption, where the dependent variables fitted exhibit independent Gaussian noise, holds.
\item	High coefficient of determination, $R^2$ value of the linear fit cannot be trusted. Shown in Clauset et al.'s 2007 paper\cite{Clauset2007}, \textit{non-power-law distributed} histograms \textit{can resemble} a power-law distribution over many orders of magnitude, thus providing a large $R^2$ value. Although low $R^2$ values are a valid reason to show that a distribution does not follow power-law, the converse is not true. 
\end{enumerate} 
With the reasonings above in mind, linear regression fitting of the jerk distributions in the upcoming Avalanche Results Section will only function as a guide to the readers. In order to determine fully if a distribution of the data (jerk peaks in this report) truly obeys power-law, a maximum-likelihood analysis must be performed, which is detailed in the next section (\textbf{\Cref{subsec:MLAnalysis}}).
\subsection{Maximum-Likelihood Analysis}\label{subsec:MLAnalysis}

The Maximum-Likelihood (ML) analysis estimates\footnote{also known as Maximum-Likelihood Estimate (MLE).} how likely a parameter, the exponent $\alpha$ from the power-law model, had generated an experimental jerk spectrum\cite{Clauset2007,Rice2007}. To perform an ML estimate, the data must be independent, identically distributed (i.i.d.) and have a joint frequency function \cite{Clauset2007,Rice2007,Salje2017}. The probability or \textit{likelihood} of these jerk data points that resembles a power-law model is\cite{Clauset2007}:
\begin{equation}\label{eq:MLfunction}
p(J|\alpha)=\prod_{i=1}^{N}\frac{\alpha-1}{J_\textrm{min}}
\left(\frac{J_i}{J_\textrm{min}}\right)^{-\alpha}
\end{equation}
where $p$ is the probability, $J$ are discrete jerk peaks and $J_\textrm{min}$ is the lower bound normalisation condition for the power-law. To find the best value of $\alpha$ that generates the jerk spectrum, the likelihood function must be maximised with respect to $\alpha$. Shown in \textbf{\Cref{apn:MLderive}}, the conventional way is to log-transform \cref{eq:MLfunction} into \textit{log-likelihood}, $\mathcal{L}$ (\cref{eq:MLDerive}), and then solve $\frac{d\mathcal{L}}{d\alpha}=0$ for parameter $\alpha$\cite{Clauset2007,Rice2007}. The resulting estimated exponent that \textit{most likely} generated the jerk spectra will then be:
\begin{equation}\label{eq:estiexponent}
\hat{\alpha}=1+N\left[\sum_{i=1}^{N}\ln{\frac{J_i}{J_\textrm{min}}}\right]^{-1}
\end{equation}
where $\hat{\alpha}$ is the convention used to denote that the exponent is an estimate \cite{Clauset2007,Rice2007,Salje2017}. ML analysis is not confined solely to power-law distributions; other distributions with a joint density or frequency function can be analysed, such as \textit{Poisson} and \textit{normal} distributions\cite{Rice2007}. The standard error of $\hat{\alpha}$ is:
\begin{equation}
\textrm{Standard error, }\sigma=\frac{\hat{\alpha}-1}{\sqrt{N}} + \textrm{higher-order terms}
\end{equation}
where the higher-order terms are positive\cite{Clauset2007}.
As $N\rightarrow\infty$, $\sigma\rightarrow0$ and $\hat{\alpha}\rightarrow\alpha$; the estimation of the true exponent $\alpha$ becomes more reliable as the number of entries increases\cite{Clauset2007}.
\subsubsection{Performing ML Fitting}

When analysing experimental jerk signals, not all data points are power-law distributed; due to the limitations of the apparatus, low signals may be under-counted due to saturation effects\cite{Clauset2007,Salje2017}. This creates a problem in using \cref{eq:estiexponent} because the value for lower bound $J_\textrm{min}$ is unknown\cite{Salje2017}!\\\par
Putting this problem aside for now and assuming that $J_\textrm{min}$ is known; to properly estimate $\hat{\alpha}$, any jerk signals $J$ below $J_\textrm{min}$ must be discarded. Thus in a sample jerk spectrum with $N$ jerk peaks $J$ where only $N'$ number of  the peaks are greater than $J_\textrm{min}$ and obeys a power-law, \cref{eq:estiexponent} can then be transformed to:

\begin{equation}\label{eq:correstiexponent}
\hat{\alpha}=1+N'\left[\sum_{i=1}^{N'}\ln{\frac{J_i}{J_\textrm{min}}}\right]^{-1}\textrm{, where } J_i>J_\textrm{min} \textrm{ and } N'<N
\end{equation}

Now, back to the unknown $J_\textrm{min}$ problem, one can guess the value of $J_\textrm{min}$ when performing the ML estimate but will run the risk of\cite{Clauset2007}:
\begin{enumerate}
\item Underestimating $J_\textrm{min}$, performing a power-law ML estimate to non-power-law $J$ data.
\item Overestimating $J_\textrm{min}$, discarding valid data points and increasing statistical error of $\hat{\alpha}$.
\end{enumerate}

Thus to fully utilise the ML estimation, a computational algorithm of \cref{eq:correstiexponent} is iterated on the jerk data by a wide-range of cut-off values, $J_0$ (that replaces $J_\textrm{min})$\cite{Clauset2007,Salje2017}. The algorithm is described in detail in \textbf{\Cref{apn:MLalgo}}, but to summarise: $J_\textrm{min}$ is replaced by a series of $J_0$ and an exponent is estimated for each of them. This results in a series of estimated exponents, $\hat{\alpha}$ which is plotted against the natural-log transformed, $J_0$ (see \Cref{subfig:ml_sing}). This method is called \textit{lower-bound estimation}\cite{Clauset2007} but is conventionally understood as performing the Maximum Likelihood Estimation/Method/Fit \cite{Salje2017,Salje2014,Jiang2017,Gallardo2010}. 

\subsubsection{Interpreting ML Fit} 
\begin{figure}[h]
	\centering
	\begin{subfigure}[t]{0.48\textwidth}
		\includegraphics[width=1\textwidth]{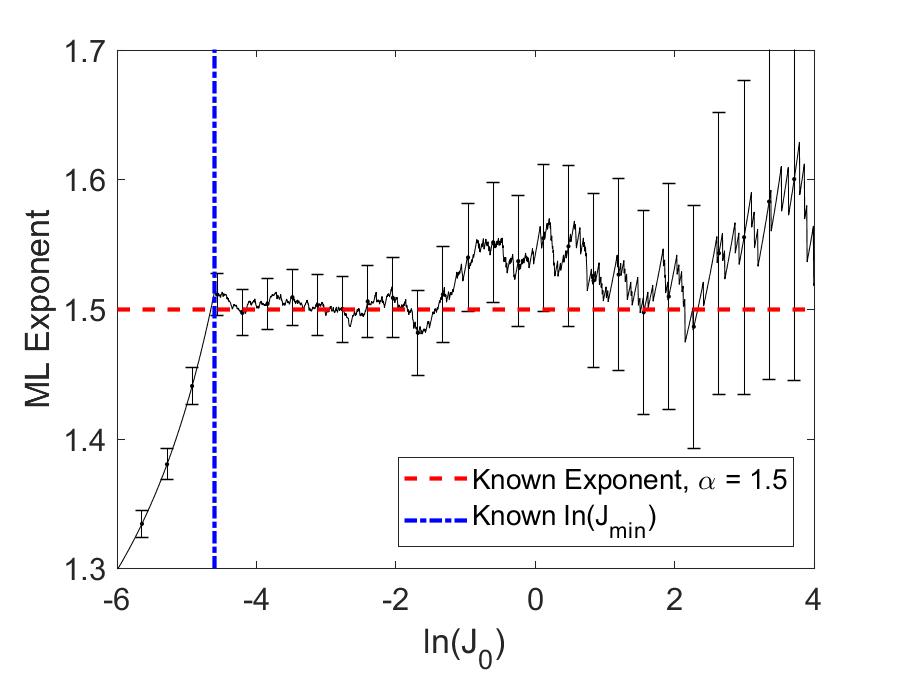}
		\subcaption{ML Estimate with known exponet $\alpha=1.5$ and lower-bound $J_\textrm{min}=0.01$}
		\label{subfig:ml_sing}
	\end{subfigure}\hfill
	\begin{subfigure}[t]{0.48\textwidth}
		\includegraphics[width=1\textwidth]{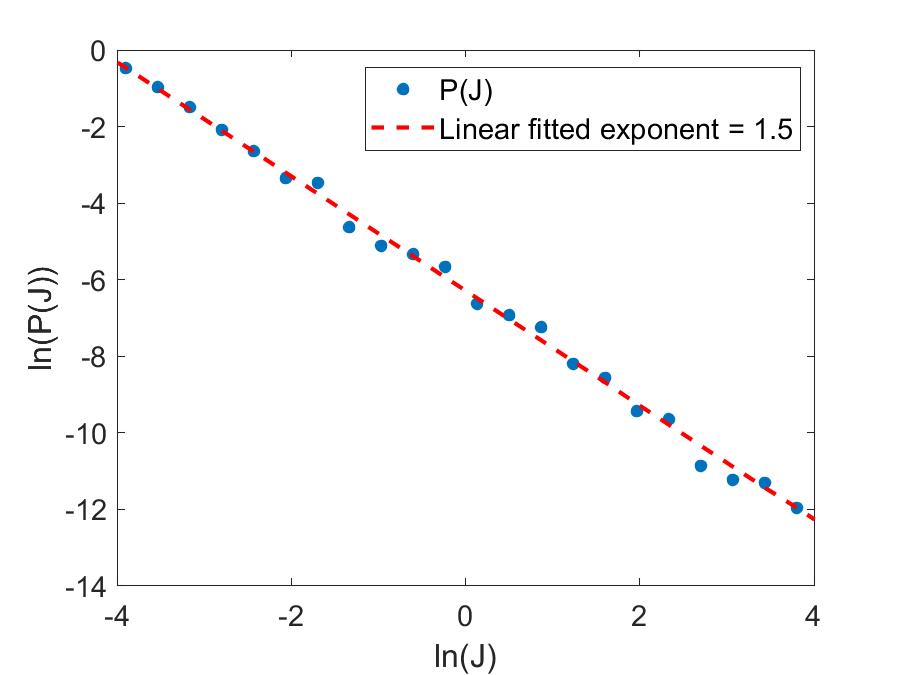}
		\subcaption{Log-log fit of P(J) vs J}
		\label{subfig:loglog_sing}
	\end{subfigure}
	\caption{(a) is a Maximum-likelihood fit of the perfect power-law randomly generated jerks with 1000 points. Each $J_0$ value tells us the most likely exponent that generates the jerk spectrum. As $J_0$ equals $J_\textrm{min}$ (blue dot-dashed line), a kink is observed at the known exponent value 1.5 (red dashed line in error range), and the curve plateaus over a few decades. As $J_0$ increases, there exists lesser relevant data, the errors become larger and the ML curve fluctuates away from the known exponent. (c) is a simple linear regression fit of the normalised probability distribution the simulated jerks. }
	\label{fig:perfpowlaw_comb}
\end{figure}
An ML fit example using a perfect power-law randomly generated jerk spectrum is shown as \Cref{subfig:ml_sing} with conditions $\alpha = 1.5 $ and $J_\textrm{min}= 0.01$. The power-law spectrum is generated by the inversion sampling method\cite{Vassiliev2017}, shown in \textbf{\Cref{apn:MCgensingpowlaw}} along with the spectrum as \Cref{fig:simjerks_sing}.\\\par 

The estimated exponent (a.k.a ML curve) takes the form of an increasing curve when $J_0<J_\textrm{min}$ and then plateaus at the most-likely exponent when the condition $J_0=J_\textrm{min}$(blue dot-dashed line) is met. After a kink at $J_0 = J_\textrm{min}$, the plateau extends over a few decades  around the known $\alpha = 1.5$ (red dashed line) within error range. The plateau then fluctuates away from the known exponent as $J_0$ further increases, due to less statistically relevant data being fed to the iterative algorithm, consequently increasing the error in the exponent estimate.\\\par
The origin of the decrement as $J_0$ decreases below $J_\textrm{min}$ can be shown mathematically (in \textbf{\Cref{apn:MLdecJ0Jmin}}, \cref{eq:MLdecrementshow}); as $J_0$ traverses below $J_\textrm{min}$, the estimated exponent, $\hat{\alpha}$ decreases due to an increasing $\ln{J_\textrm{min}/J_0}$ in reciprocals\cite{Salje2017}. Also, the number of error bars \textbf{do not} reflect the number of $J_0$ iterations; they were evenly spaced out to prevent a cluttering of error bars.\\\par
However, not all ML curves exhibit plateaus exactly as above, especially in experimental settings where high or low cut-offs and saturation events might come into play \cite{Salje2017}. In the next section, the ML curve and the log-log probability distribution are put together to demonstrate that various properties of a crackling system can be unveiled by analysing them in parallel.

\subsection{ML Application in Real Systems}\label{subsec:MLapps}
\Cref{fig:perfpowlaw_comb} represents an ideal case of a power-law jerk spectrum. In experimental settings, ML curves and log-log binning plots are more complicated but are the keys to understanding an avalanche system. Many subtleties of a system can be uncovered using ML analysis, which is well described by Salje, Planes and Vives in \textit{Analysis of crackling noise using the maximum-likelihood method: Power-law mixing and exponential damping} (2017)\cite{Salje2017}. They are:
\begin{enumerate}
\item 	Mixture of two power laws, a system that involves two exponents, perhaps due to two different mechanics that are involved in the avalanche.
\item	Exponentially damped power-law, a power-law distribution that takes the form of
\begin{equation}
P(J)=J^{-\alpha} e^{-J/\Lambda}/J_\textrm{min}
\end{equation}	
where $\Lambda$ is the damping scale that has the same dimensions as $J$. Damping usually originates from experimental settings that suppress higher $J$ counts.
\item Exponentially pre-damped lower cut-off, a power-law distribution that takes the form of
\begin{equation}
P(J)=J^{-\alpha}\left(1- e^{-J/\lambda}\right)/J_\textrm{min}
\end{equation}
where $\lambda$ is another damping scale that has the same units as $J$. This introduces a smooth cut-off to better describe the distribution at the low $J$ limit. These usually arise from saturation effects from the apparatus\cite{Salje2017}.
\end{enumerate} 
The mixture of two power laws is discussed in \textbf{\Cref{theory:exponentmixing}} to provide a deeper understanding of the ML method. For the other two cases, we would like to refer the readers to Salje et al.'s paper\cite{Salje2017} mentioned previously. The lower cut-off effect is observed in every jerk distribution. Take \Cref{subfig:pztb_loglogbin} from the results section (\textbf{\Cref{subsec:srun_pztb}}) for example, the straight-line behaviour extended until $\ln(J)=-34$ and discontinued as a horizontal line due to saturation effects. In terms of ML, the lower cut-off effect removes the hard kink (\Cref{subfig:pztb_mlfit} and \Cref{fig:pztb_MLcomb} as examples) seen in perfect power-law systems, this makes the lower bound $J_\textrm{min}$ harder to determine.
\newpage
\section{Results from Later Experimental Work}\label{sec:latexpw}
\subsection{PZT sample B: Single Run}\label{subsec:srun_pztb}
Described in \textbf{\Cref{subsec:fesamples} and \Cref{subsec:tpulex}}, PZT sample B was $0.93\pm0.01$ mm thick (not including electrode), and had circular gold electrodes of area $= 28.5\pm0.1$ mm\textsuperscript{$2$} (radius $= 3.01\pm0.01$ mm) on both sides. The sample was annealed at least twice before conducting the jerk experiments. For a ramp-rate of $40$ Vs\textsuperscript{$-1$}, it had a coercive voltage $V_c$ of $795\pm2$ V or in terms of field, $E_c=855\pm9$ Vmm\textsuperscript{$-1$}.\\\par 
\begin{figure}
	\centering
	\begin{subfigure}{0.48\textwidth}
		\includegraphics[width=1.0\textwidth]{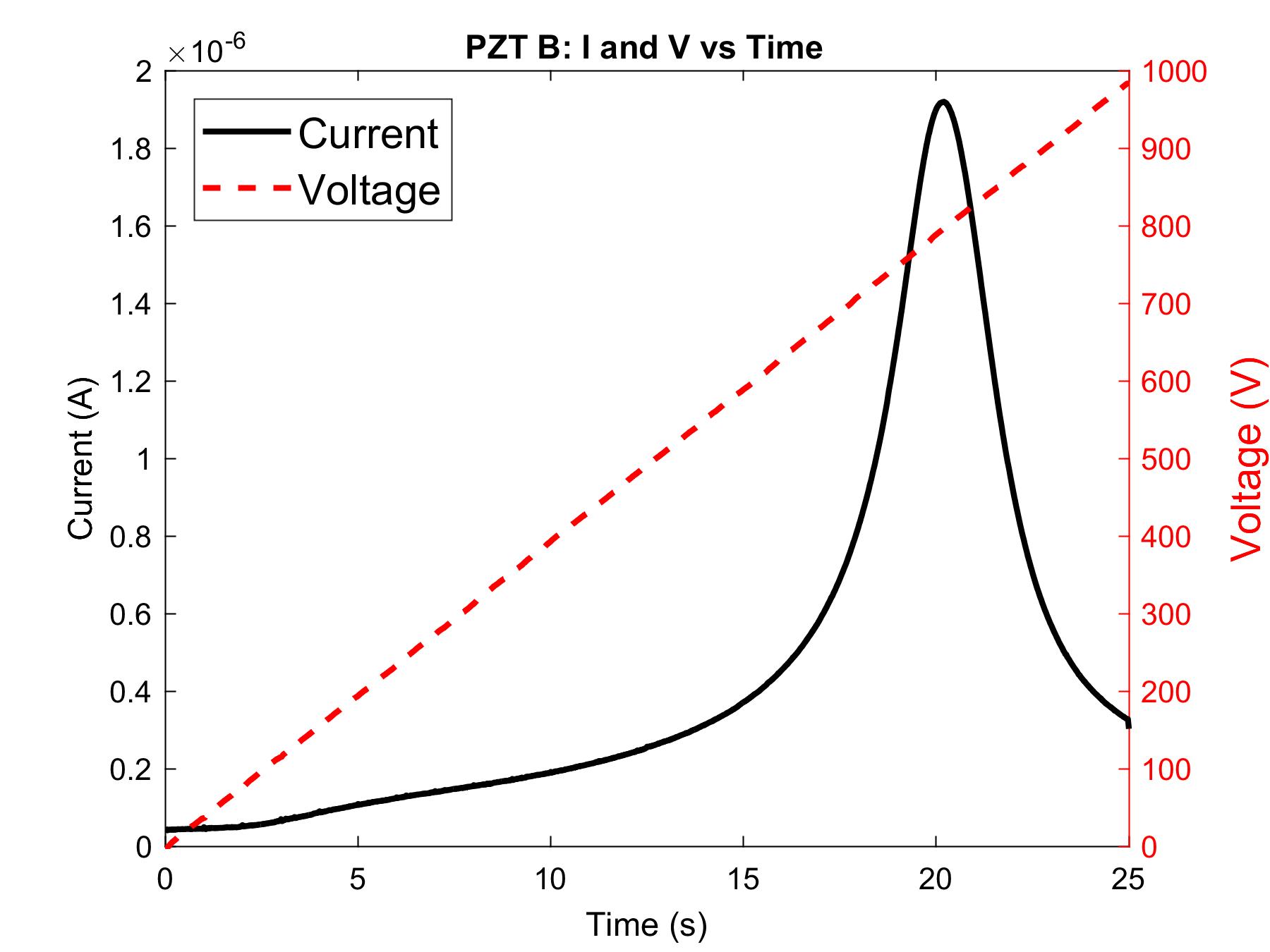}
		
		\subcaption{The current $I$ (black) response of PZT sample B after a voltage $V$ (red dashed) of ramp rate $=40$ Vs\textsuperscript{$-1$} was applied to it.}	
		\label{subfig:pztb_ivvst}
	\end{subfigure}
	\hfill
	\begin{subfigure}{0.48\textwidth}
		\includegraphics[width=1.0\textwidth]{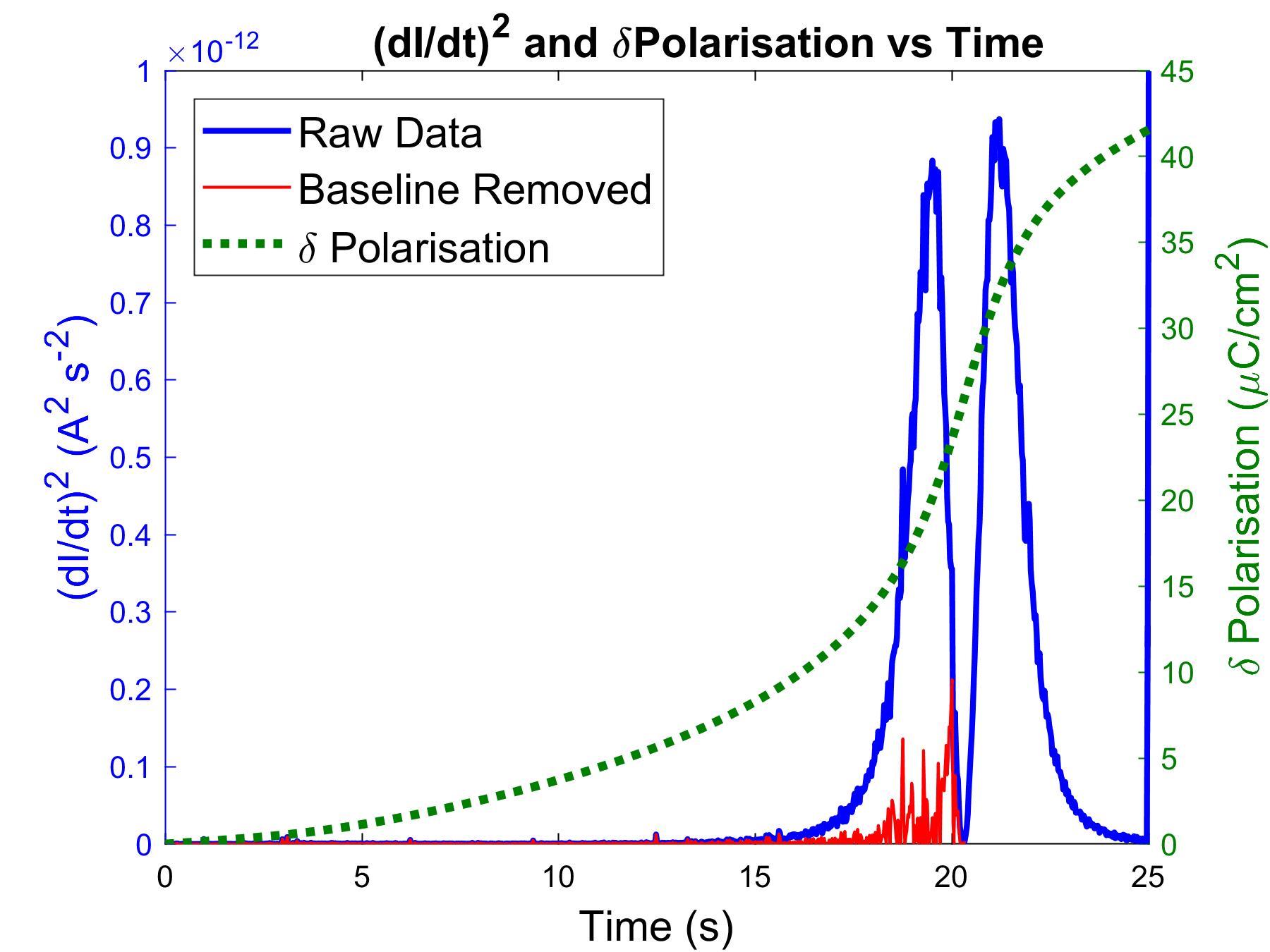}
		
		\subcaption{The square of first time derivative, $(dI/dt)^2$ (blue) with baseline removed (red) and polarisation change, $\delta P$ (green dotted) plotted agaisnt time.}
		\label{subfig:pztb_jdpvst}
	\end{subfigure}
	\begin{subfigure}{1\textwidth}
		\centering
		
		\includegraphics[width=0.9\textwidth]{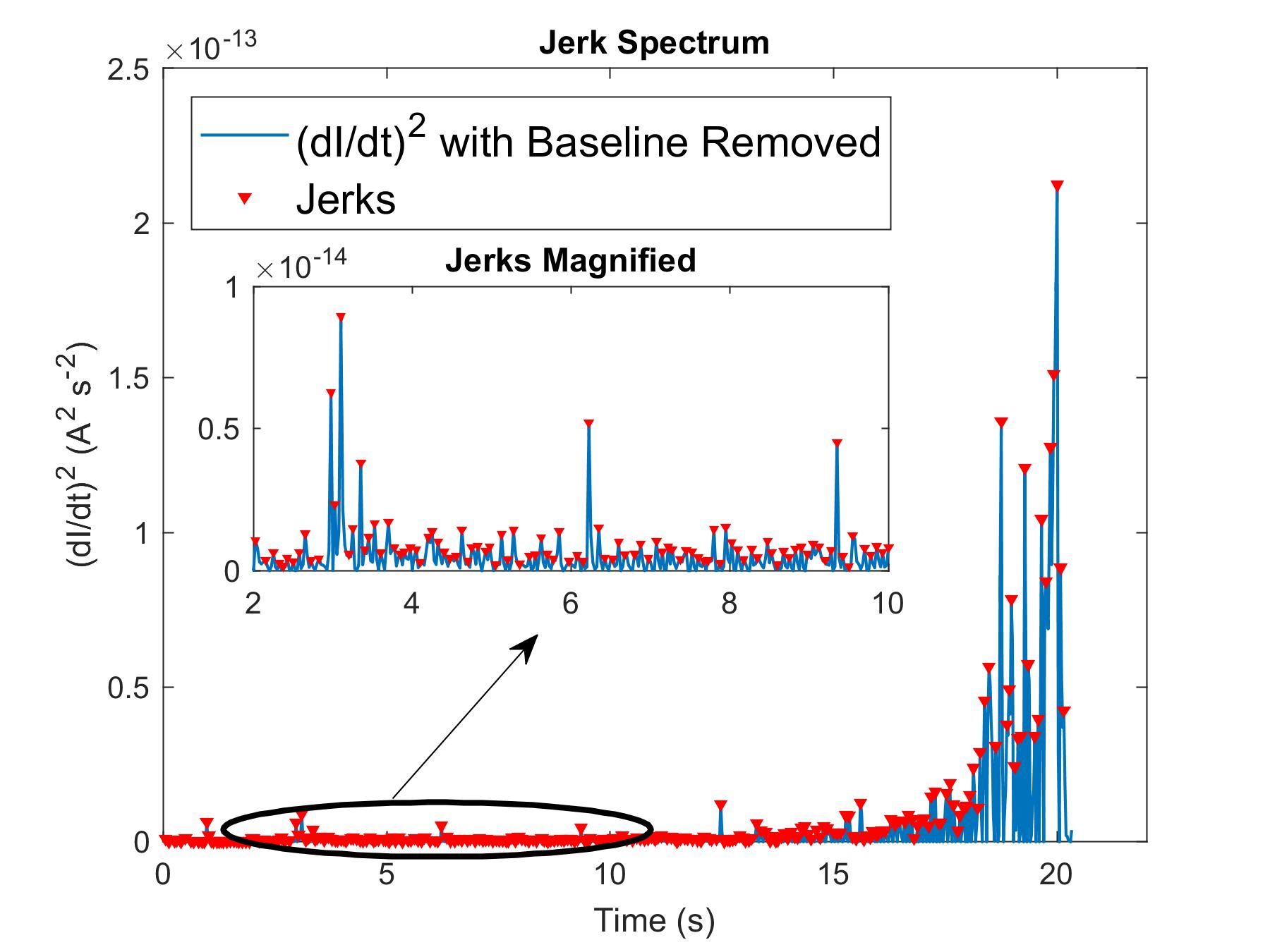}
		\subcaption{Jerk spectrum: baseline removed $(dI/dt)^2$ with peaks defined as jerks $J$. Inset is the magnified jerk spectrum at region $2\textrm{ s}\leq t\leq 10\textrm{ s}$ with same units on both axes. }
		\label{subfig:pztb_jerkspectrum}
	\end{subfigure}
	
	\caption{A single measurement from PZT sample B. The smooth looking current $I$ curve in (a) was actually jerky, as shown after taking the square of its time derivative, $(dI/dt)^2$ (blue) in (b). As mentioned by Rudyak in \cite{Rudyak1971}, Barkhausen noise is most intense in the steepest region of polarisation (against field). Hence, the baseline up to the steepest region of $(dI/dt)^2$ was removed and the peaks (defined as jerks $J$ (red labels at (c)) were used for statistical analysis. The graph with baseline removed is conventionally called the jerk spectrum.}
	\label{fig:pztb_firstrun}
\end{figure}
\begin{figure}
	\centering
	\begin{subfigure}{0.48\textwidth}
		\includegraphics[width=1.0\textwidth]{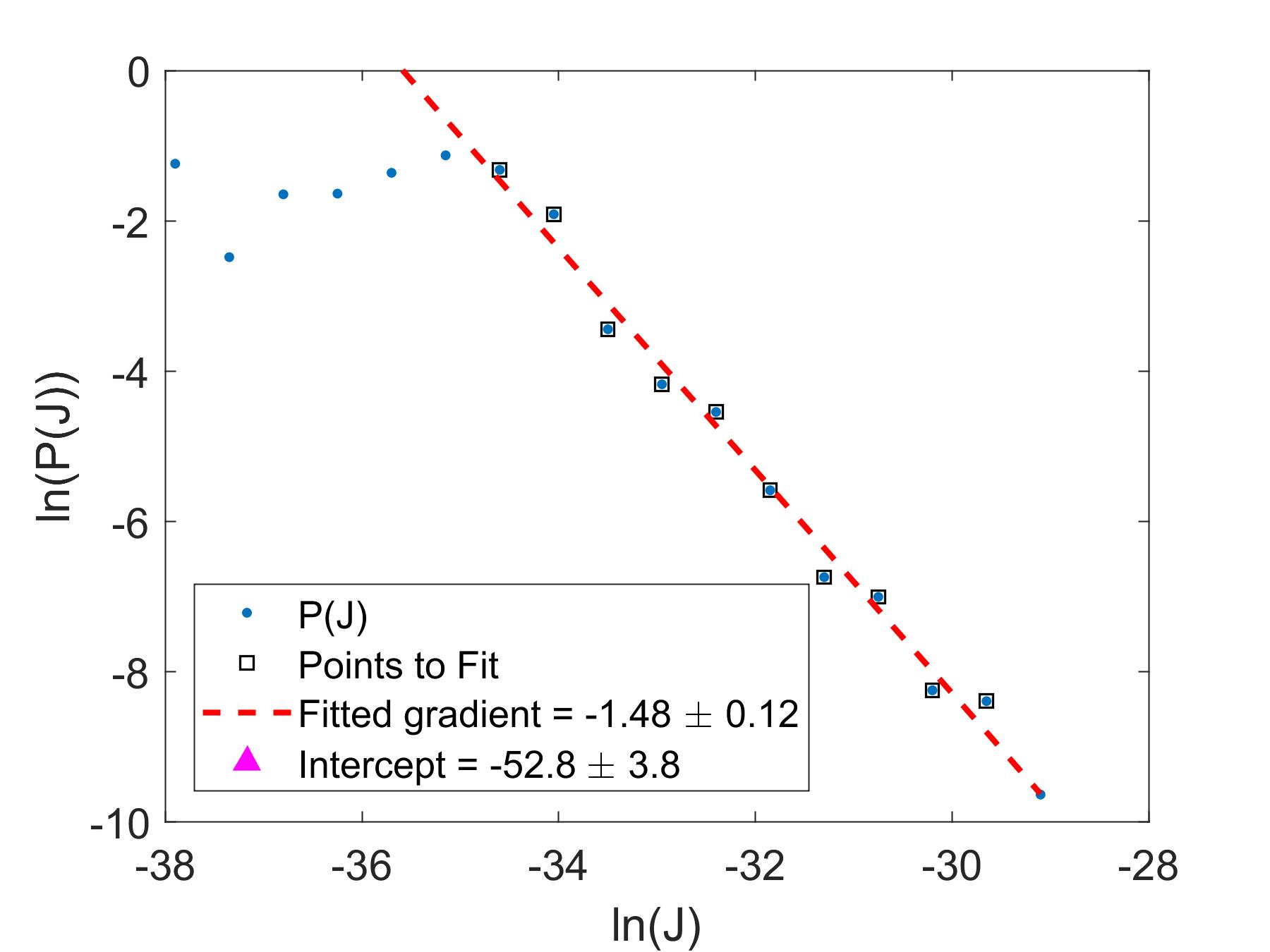}
		\subcaption{Total jerk peaks: 252 points. Log-log binning and fitted with exponent $=1.5$}
		\label{subfig:pztb_loglogbin}	
	\end{subfigure}
	\hfill
	\begin{subfigure}{0.48\textwidth}
		\includegraphics[width=1.0\textwidth]{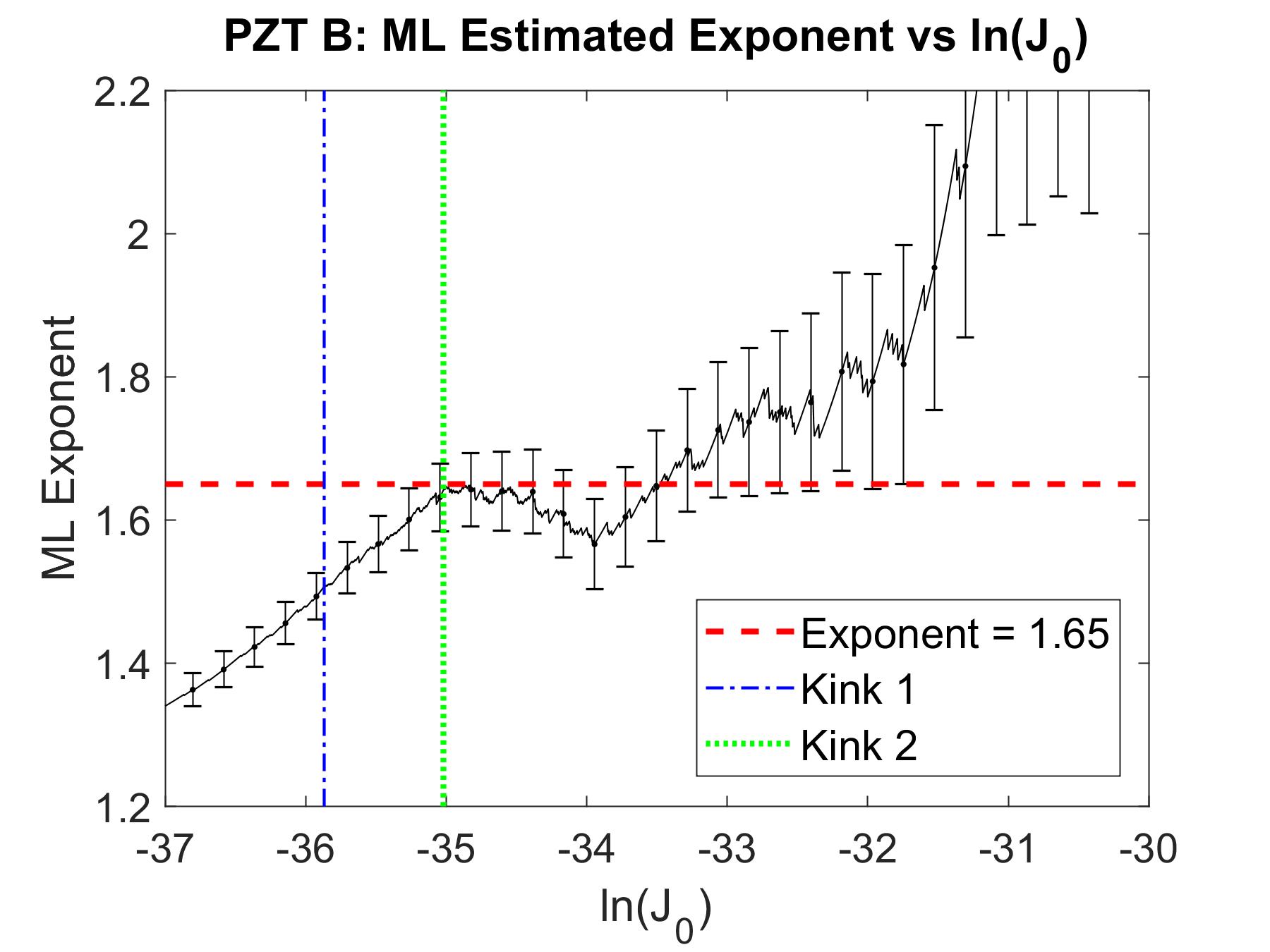}
		\subcaption{Corresponding maximum-likelihood (ML) fit of 252 jerk peaks. An estimated exponent of $1.65$.}
		\label{subfig:pztb_mlfit}
	\end{subfigure}
	\caption{The jerk peaks from \Cref{subfig:pztb_jerkspectrum} (252 in total) were analysed statistically with the methods detailed in \textbf{\Cref{sec:analyticalmet}}. As mentioned before, due to the pitfalls in log-log linear regression fit, the fitting in (a) only functions as a guide, which shows a straight line behaviour of gradient $=-1.5$. Do note that the error $\pm0.12$ is not valid because the error for linear regression does not apply in log-transformed data (see \textbf{\Cref{sssec:disadvLinFit}}). The ML fit in (b) shows a power-law behaviour with a lower cut-off effect which extends around two decades worth of jerk peaks $(-35<\ln(J_0)<-33)$. The exponent obtained was $1.65$ within uncertainty range except for the dip at $\ln(J_0)\approx-34$. The dip may indicate that there is an exponent mixing effect\cite{Salje2017} (as explained in \textbf{\Cref{theory:exponentmixing}}). As expected, the fit deviates as $\ln(J_0)$ increases due to lack of statistically relevant data.}
\end{figure}

An example of a current-time response of PZT B is shown in \Cref{subfig:pztb_ivvst}. The current response looked smooth but is actually jerky when magnified, as previously indicated in \Cref{subfig:pztb_run2_ivst} in \textbf{\Cref{subsec:jerks}}. This current-time response was then jerk analysed using the methodology described  below:
\begin{enumerate}
	\item The first time derivative of the current response curve (black in \Cref{subfig:pztb_ivvst}) was taken and then squared, $(dI/dt)^2$ (blue in \Cref{subfig:pztb_jdpvst}).
	\item The background of the time derivative was removed using the Piecewise Cubic Hermite Interpolating Polynomial (PCHIP) function (briefly described in \textbf{\Cref{apn:RemBase_pztf}}), leaving jerky peaks (\Cref{subfig:pztb_jerkspectrum}) in the spectrum.
	\item The height of the peaks (now defined as jerks $J$) were extracted and subjected to two major analyses:
	\begin{enumerate}
		\item Log-log binning and linear regression fit (\Cref{subfig:pztb_loglogbin}):
		\begin{itemize}
			\item The jerks were binned in logs and a straight-line behaviour was observed between $\ln(J) \approx -35$ and $-29$.
			\item The gradient of the straight line was $-1.5$. The error calculated ($\pm0.12$) is not valid since the formula for error analysis in linear regression does not apply in log-log fitting\footnote{the common pitfalls of log-log fitting is described in \textbf{\Cref{sssec:disadvLinFit}} or \cite{Clauset2007}.\label{fn:loglogpitfall}}.
		\end{itemize}
		\item Maximum-Likelihood Fit (\Cref{subfig:pztb_mlfit}).
		\begin{itemize}
			\item Due to the low cut-off effect (\textbf{\Cref{subsec:MLapps}}), the kink for the true $J_\textrm{min}$ is hard to determine\footnote{The errors for $J_\textrm{min}$ arise from error propagation in logs $\Delta\ln(x)\approx(\Delta x)/x$ while the errors in natural logarithms are due to the uncertainty of the kink itself being determined by eye.} \cite{Salje2017}. There are two possibilities 
			\begin{itemize}
				\item	kink $1$ (blue dot-dashed), $\ln(J_\textrm{min}) = -35.8\pm0.2$. $J_\textrm{min}=2.6\pm0.5\times10^{-16}~\textrm{A}^2\textrm{s}^{-2}$.
				\item	kink $2$ (green dotted), $\ln(J_\textrm{min}) = -35.0\pm0.1$. $J_\textrm{min}=6.2\pm0.6\times10^{-16}~\textrm{A}^2\textrm{s}^{-2}$.
			\end{itemize}	
			\item  A plateau that extends for two decades $(-35<\ln(J_0)<-33)$ was observed with exponent $= 1.65\pm0.05$ at the initial kink.
		\end{itemize}
	\end{enumerate}
\end{enumerate}
As mentioned in \textbf{\Cref{sssec:disadvLinFit}}, the exponent extracted from the log-log fitting (\Cref{subfig:pztb_loglogbin}) must be taken with a grain of salt. Though a straight-line behaviour is observed (with exponent $= 1.5$), the jerk spectrum may not be power-law distributed\textsuperscript{\ref{fn:loglogpitfall}}. Thus we need to rely on the maximum-likelihood analysis.\\\par

Looking at the ML analysis (\Cref{subfig:pztb_mlfit}) for a single run of PZT sample B, the ML curve displays a plateau that spans two decades $(-35<\ln(J_0)<-33)$ with an exponent of $= 1.6$ and then deviates as $\ln(J_0)$ increases due to a lack of relevant data. The plateau is a good indicator to show that this jerk spectrum is power-law distributed. Comparing this to the perfect power-law ML fit that has an obvious kink, it is hard to determine the location of $J_\textrm{min}$ (where the power-law distributed jerks starts) in the ML curve due to saturation or lower cut-off effects\cite{Salje2017} (as remarked in \textbf{\Cref{subsec:MLapps}}). The two possible $J_\textrm{min}$ values are $2.6\pm0.5\times10^{-16}~\textrm{A}^2\textrm{s}^{-2}$ and $6.2\pm0.6\times10^{-16}~\textrm{A}^2\textrm{s}^{-2}$. \\\par

Four more measurements were taken (spectra and analyses in \textbf{\Cref{apn:pztb_2_5}}) and each individual distribution of the jerks were shown to obey a power-law. These single run measurements show that the jerk spectra are power-law distributed via ML fitting and exhibit avalanche statistics which is consistent for Barkhausen noise\cite{Dahmen1996,Salje2014}. However, a larger sample size is needed to increase statistical significance. Hence, we decided to analyse a \textit{grand} jerk spectrum of multiple readings combined (after being normalised by the mean) from PZT sample B. If these spectra are statistically similar, combining them will increase the confidence in both log-log binning and ML fit. \\\par

Most importantly, although much research in avalanche theory has been carried out, such as ferromagnetic \cite{Dahmen1996,Dahmen2009}, ferroelastic switching \cite{He2016,Zhao2014} and crystal plasticity compressions\cite{Friedman2012,Baro2013}, this discovery is exciting as it is the first time electrical measurement of Barkhausen pulses are statistically analysed and shown to be consistent with avalanche theory. 

\newpage
\subsection{Normalised and Combined Jerks for PZT Sample B}\label{subsec:mrun_pztb}
\begin{figure}[h]
	\centering
	\begin{subfigure}[t]{0.48\textwidth}
		\includegraphics[width=1\textwidth]{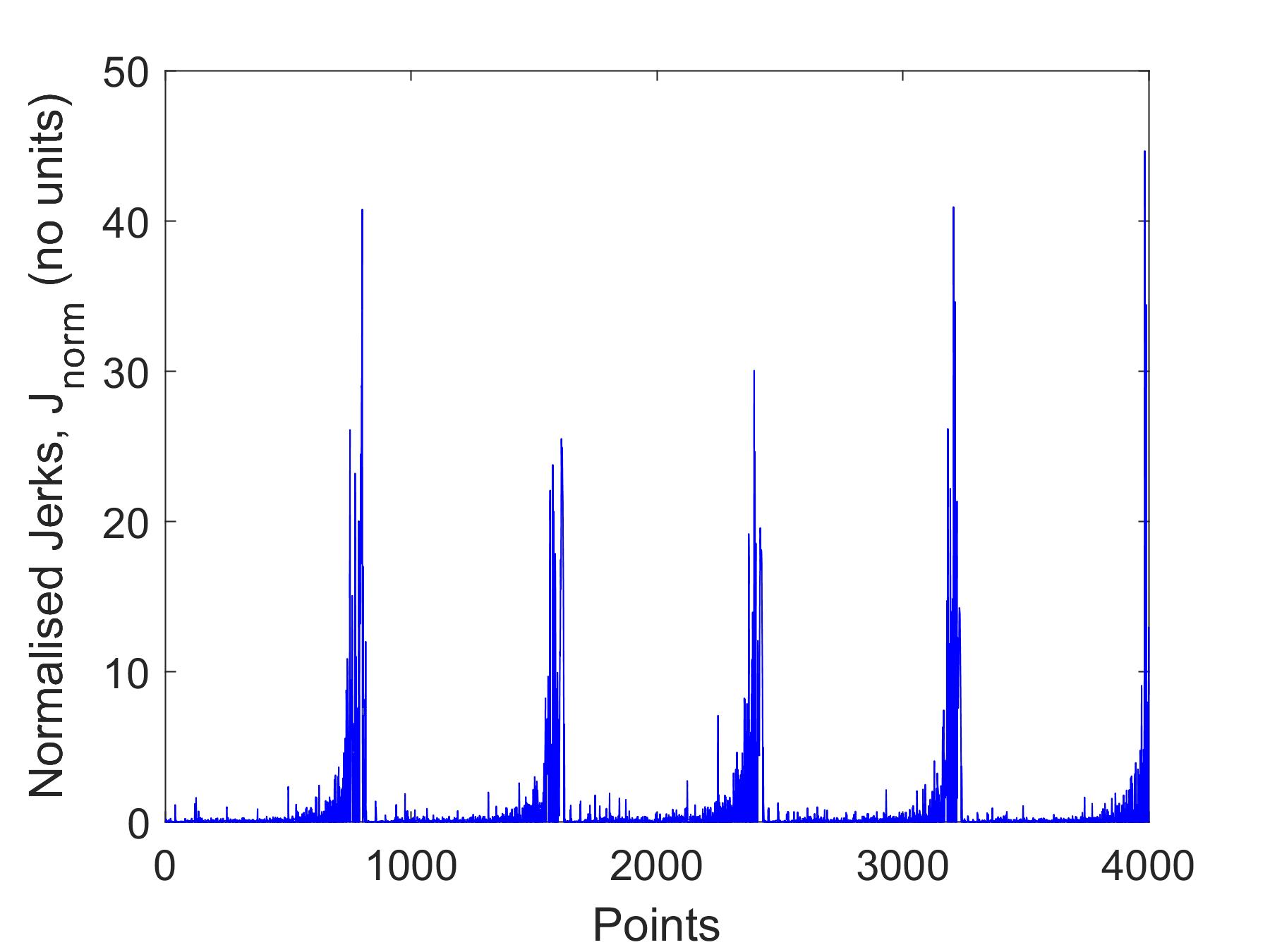}
		\caption{Normalised and combined jerk spectrum}
		\label{subfig:pztb_combinedjerks}
	\end{subfigure}\hfill
	\begin{subfigure}[t]{0.48\textwidth}
		\includegraphics[width=1\textwidth]{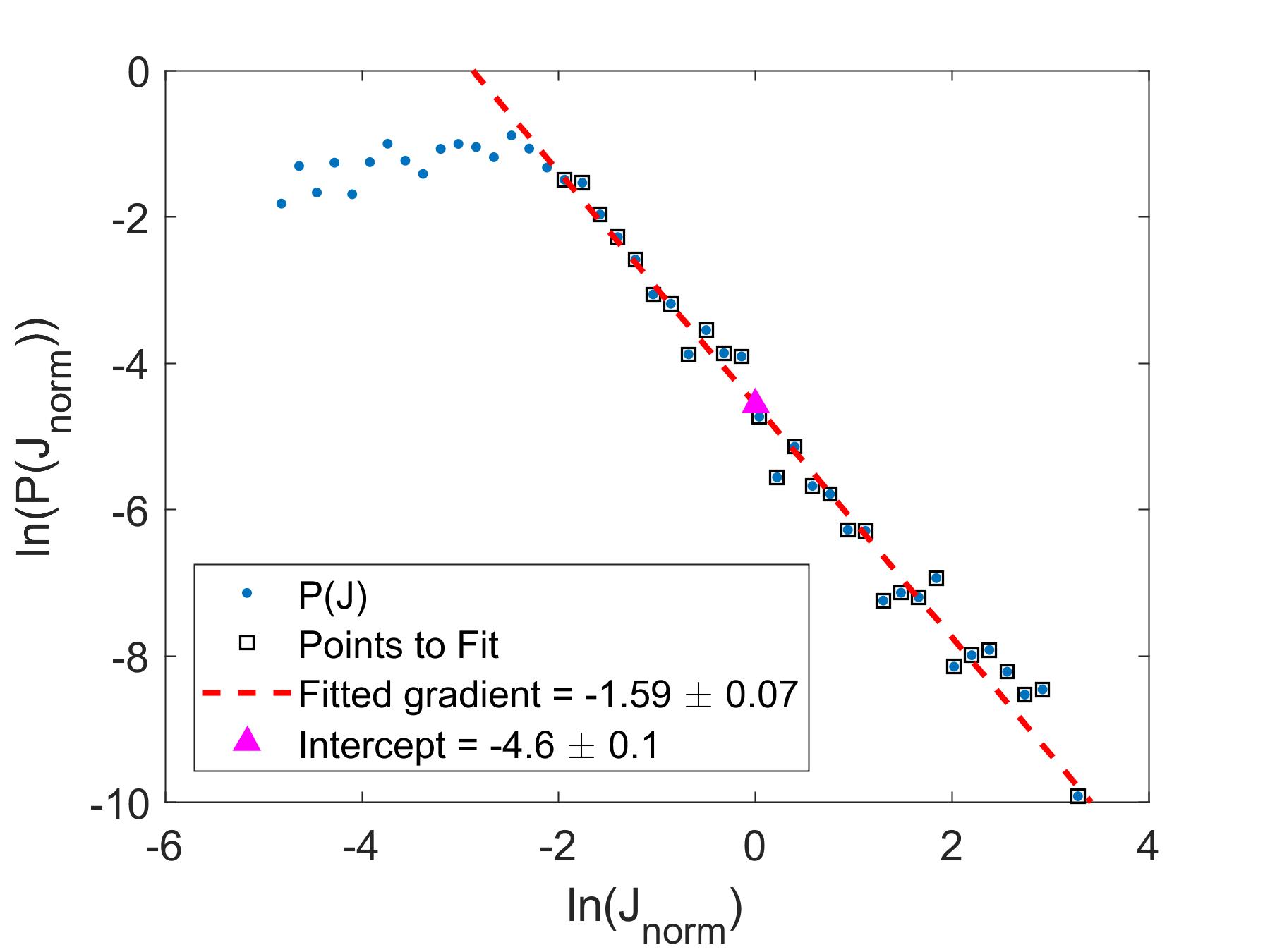}
		\caption{Log-log Fit of 1261 jerk peaks}
		\label{subfig:pztb_loglogcomb}
	\end{subfigure}\\
	
	\caption{(a): All 5 jerk spectra were combined via normalisation (by the mean) to form a \textit{grand} jerk spectrum. Note the change in magnitude of the jerk peaks and the lost of units. (b): Log-log fit for the distribution of 1261 jerk peaks that demonstrates a lower cut-off effect.}
	\label{fig:pztb_combjerklogfit}
\end{figure}

The jerk peaks of the five runs are normalised by the mean and combined as a \textit{grand} jerk spectrum (\Cref{subfig:pztb_combinedjerks}). Note that after normalisation, the normalised jerk peaks are now dimensionless and are rescaled in magnitude. Though there around 4000 points labelled, only the peaks from the spectrum (1261 in total) are used for the analyses.\par

The log-log binning in \Cref{subfig:pztb_loglogcomb} shows a straight-line behaviour with a low cut-off effect, as mentioned in \textbf{\Cref{subsec:MLapps}} and in \cite{Salje2017}. A probability distribution plot with a lower cut-off effect takes the form of a curve instead of a straight-line as $J_\textrm{norm}$ decreases. The fit gives an exponent of $1.59$.\\\par
\begin{figure}[h]
	\centering

		\includegraphics[width=0.7\textwidth]{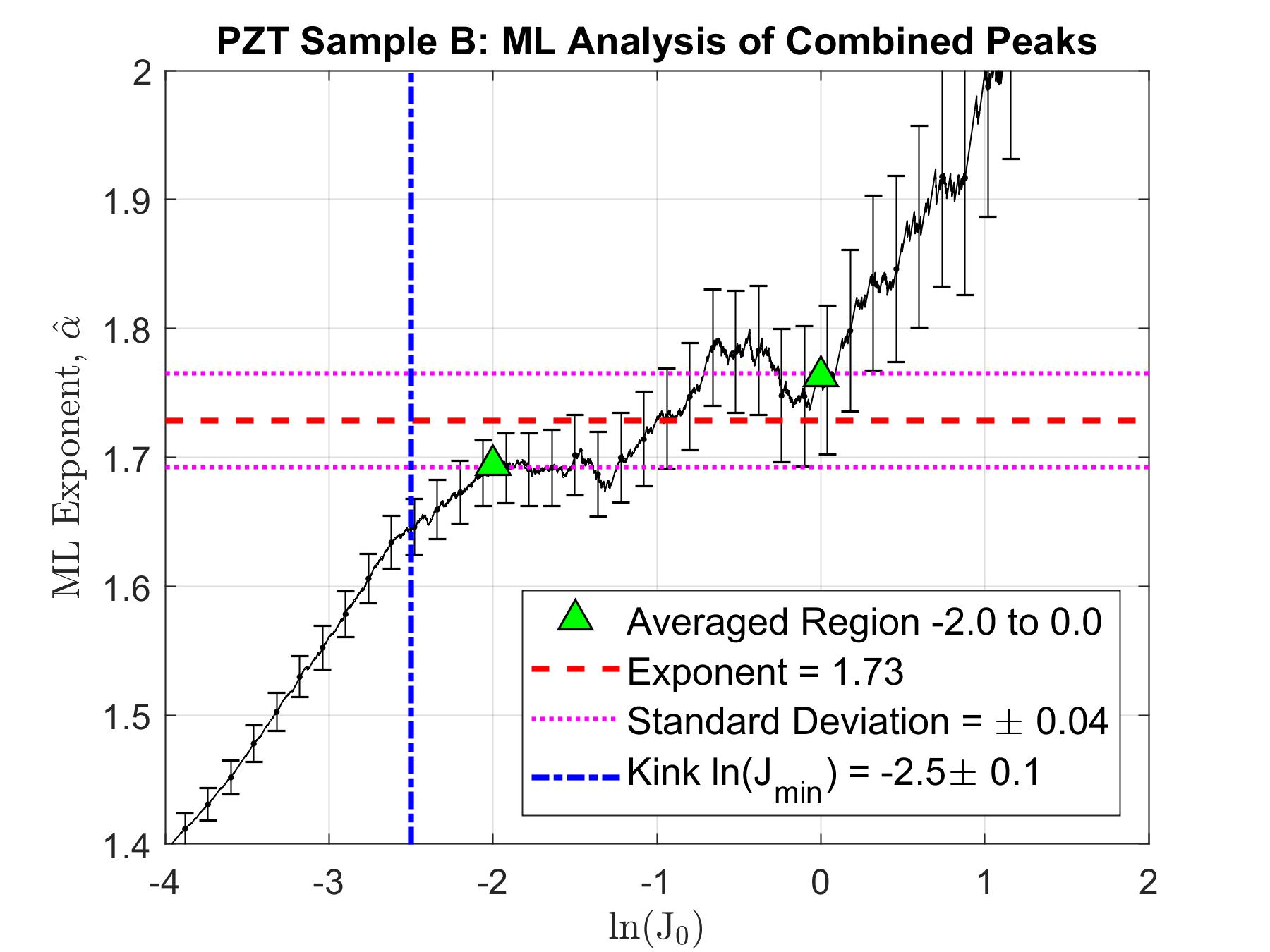}
		\caption{ML analysis of the combined jerks. The kink (blue dot-dashed line), $\ln(J_\textrm{min})=-2.5\pm0.1$, is not well defined and can only be estimated due to the low cut-off effect. The exponent plateau (red dashed line), $1.73$ is obtained by taking the mean from the region between two green triangle labels ($-2\leq\ln(J_0)\leq0$) with a standard deviation, $\sigma$ (pink dotted line) of $\pm0.04$.}
		\label{fig:pztb_MLcomb}

\end{figure}
Finally, with the ML analysis in \Cref{fig:pztb_MLcomb}, a plateau that spans for two decades can be observed. The averaged exponent (red dashed line) in the plateau region $2.0\leq\ln(J_0)\leq0.0$ (labelled as green triangles) is $1.73$ with standard deviation (pink dotted line), $\sigma=\pm0.04$; the plateau then deviates away as $\ln(J_0)>0$. Also due to the lower cut-off effect, the kink (blue dot-dashed line) for true $J_\textrm{min}$ is hard to determine. $\ln(J_\textrm{min})$ is estimated to be $-2.5\pm0.1$, thus the \textit{normalised} $J_\textrm{min}$ is $0.082\pm0.008$. This lower cut-off effect also caused a shorter plateau due to no contributions from the lower signals. \\\par

\subsubsection{Categorising PZT Sample B}\label{sssec:minicon_b}
Using both log-log linear regression fit and ML analysis, the combined jerks of sample B extracted from the current-response are concluded to exhibit power-law statistics with mean exponent of $1.73$ and $\sigma=\pm0.04$ but this spans only about two decades.\\\par
The normalised $J_\textrm{min}$ is $0.082\pm0.008$ and can be cross-checked with $J_\textrm{min}$ from the first run. The mean of the jerks from run 1, $\bar{J}_\textrm{Run 1}$ is $8.527\times10^{-15}~\textrm{A}^2\textrm{s}^{-2}$. Multiplying the mean with the normalised $J_\textrm{min}$ yields $7.0\pm0.7\times10^{-16}~\textrm{A}^2\textrm{s}^{-2}$, and taking the natural log of it gives $-34.9\pm0.1$. This shows that $J_\textrm{min}$ of kink $2$ from \Cref{subfig:pztb_mlfit} is within uncertainty range of $J_\textrm{min}$ extrapolated from the normalised grand jerk spectrum. The kinks can also be seen at $\ln(J_0)=-35$ for all the other ML curves (\Cref{fig:pztb_MLs_2_5}) in \textbf{\Cref{apn:pztb_2_5}}. Thus the average value\footnote{obtained by finding all the $J_\textrm{min}$ values of each run, and take the mean and standard error.} of $J_\textrm{min}$ for all runs is $\bar{J}_\textrm{min}=6.1\pm0.3\times10^{-16}$A$^{2}$s$^{-2}$.\\\par
Hence, for PZT sample B:
\begin{itemize}
	\item The coercive voltage $V_c$ at ramp-rate $= 40$ Vs$^{-1}$ is $795\pm2$ V. In terms of coercive field $E_c=855\pm9\textrm{ Vmm}^{-1}$. The thickness of the sample was $0.93\pm0.01$ mm.
	\item The critical exponent is estimated to be $1.73$ with $\sigma=\pm0.04$.
	\item The value for the \textit{normalised} $\ln(J_\textrm{min})=-2.5\pm0.1$.
	\item The average value of 5 runs for $J_\textrm{min}=6.1\pm0.3\times10^{-16}~\textrm{A}^2\textrm{s}^{-2}$; $\ln(J_\textrm{min})=-35.03\pm0.05$.
\end{itemize}
\newpage
\subsection{PZT Sample F}\label{subsec:an_pztf}

PZT sample F (PIC 151 from PI Ceramic, Lederhose, Germany\cite{PiezoTech}) is a very similar sample to PZT B from \textbf{\Cref{subsec:srun_pztb}}. It had a thickness of $0.97\pm0.01$ mm (not including electrodes) and was annealed at least once before conducting the experiments. The coercive voltage $V_c$ for $40$ Vs\textsuperscript{$-1$} was $872\pm1$ V. In terms of field, $E_c$, that would be $899\pm9$ Vmm\textsuperscript{$-1$}. Similar to PZT B (since they were prepared in the same way), sample F had circular gold electrodes of area $= 28.5\pm0.1$ mm\textsuperscript{$2$} (radius $= 3.01\pm0.01$ mm) on both sides.\\\par
\begin{figure}[h]
	\centering
	\begin{subfigure}{0.48\textwidth}
		\includegraphics[width=1\textwidth]{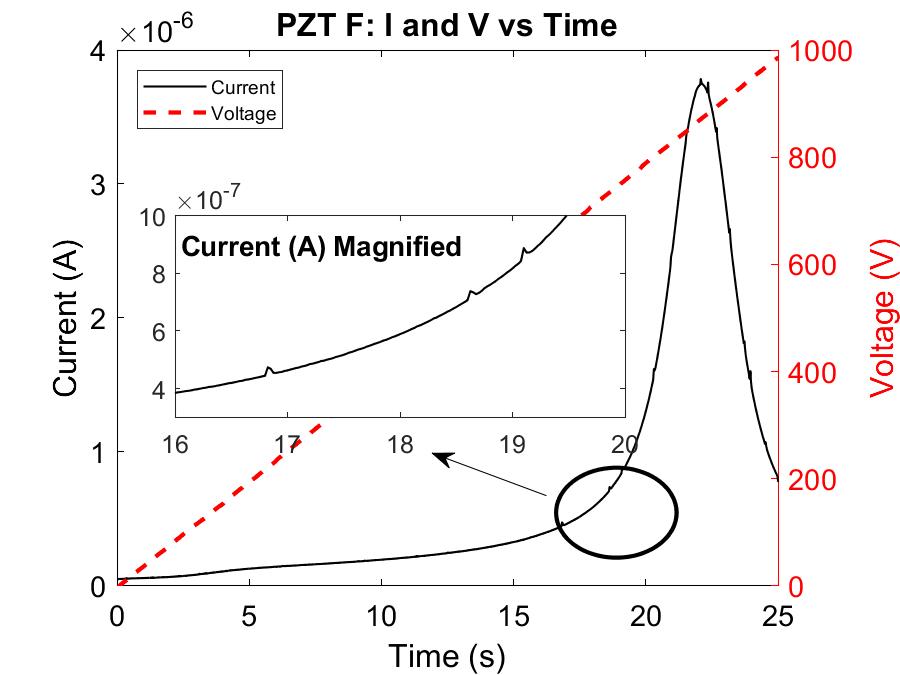}
		\caption{Run 1: Current and Voltage against time. Inset: current magnified at region 16 s to 20 s.}
		\label{subfig:pztf_IVvsT}
	\end{subfigure}\hfill
	\begin{subfigure}{0.48\textwidth}
		\includegraphics[width=1\textwidth]{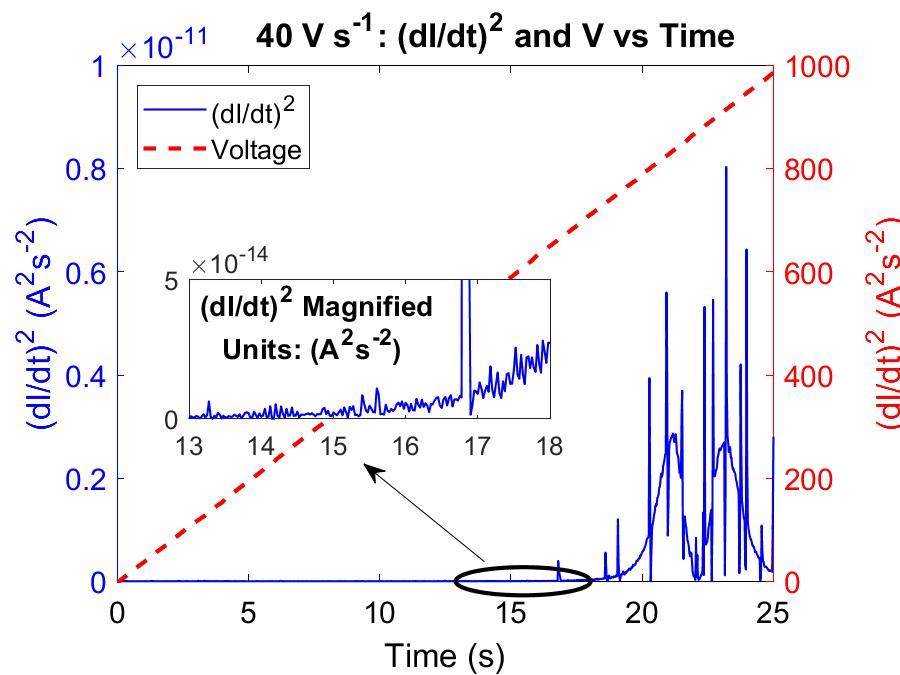}
		\caption{Jerk spectrum with baseline. Inset: jerk spectrum magnified at region 13 s to 18 s}
		\label{subfig:pztf_JVvsT}
	\end{subfigure}
	
	\caption{(a): This is the current(and voltage)-time response of PZT sample F which had been subjected to more stress in its history. The usual jerky current response had incorporated some spikes, causing major jerk peaks in the jerk spectrum (b). These huge jerks may not play a role in an avalanche system\cite{He2016}. As seen in the inset of (b), the magnified $(dI/dt)^2$ curve still shows jerk peaks.}
	\label{fig:pztf_run1}
\end{figure}
As mentioned in \textbf{\Cref{subsec:srun_pztb}}, these samples were at least 10 years old and they were used for fatigue experiments\cite{Verdier2005}, thus the quality of each sample varies. Throughout the project, these samples were annealed to ensure that fatigue effects were removed \cite{Verdier2005}, yet certain samples were subjected to more experimentations than the others. PZT sample F was an example of a sample that had underwent more tests in its history. \\\par
In \Cref{subfig:pztf_IVvsT}, the current-time response showed tiny spikes as an electric field was being applied to sample F. The origin of these spikes in the current are currently unknown and may or may not play a role in avalanche statistics. As seen in the jerk spectrum (\Cref{subfig:pztf_JVvsT}), these spikes create huge jerk peaks. We suspect these intense peaks may be spanning avalanches as seen in He et al.'s (2016) \textit{simulated ferroelastic} switching work\cite{He2016}. The statistical properties of spanning avalanches are sample size dependent and do not represent Barkhausen switching avalanches\cite{He2016}. In our case, the samples are almost the same size, but the robustness may be different since the samples had experience various degrees of fatigue. These little spikes do show a little resemblance to the current responses (\Cref{subfig:ew_ivst}) in our early works (\textbf{\Cref{subsec:tpulex}}) when we drove our samples through a very high electric field. In He et al.'s work, these intense avalanches were dismissed from statistical analyses\cite{He2016}. 

\subsubsection{Results of Combined 10 Runs with Spanning Peaks}\label{sssec:pztf_wispanpeaks}

\begin{figure}[h]
	\centering
	\begin{subfigure}{0.47\textwidth}
		\includegraphics[width=1\textwidth]{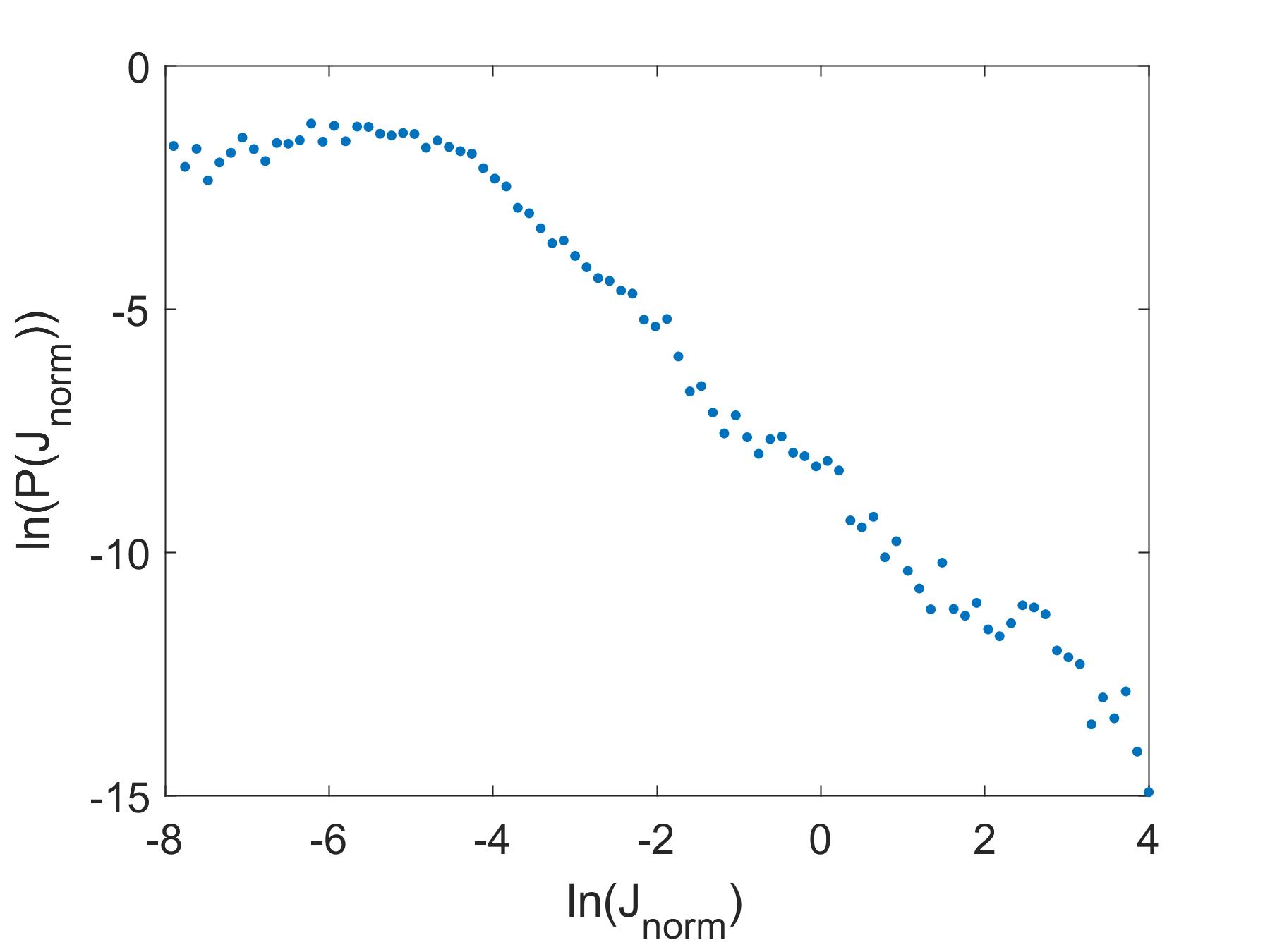}
		\caption{$P(J_\textrm{norm})$ against $J_\textrm{norm}$ of sample F.}
		\label{subfig:pztf_comb_loglognofit}
	\end{subfigure}\hfill
	\begin{subfigure}{0.47\textwidth}
		\includegraphics[width=1\textwidth]{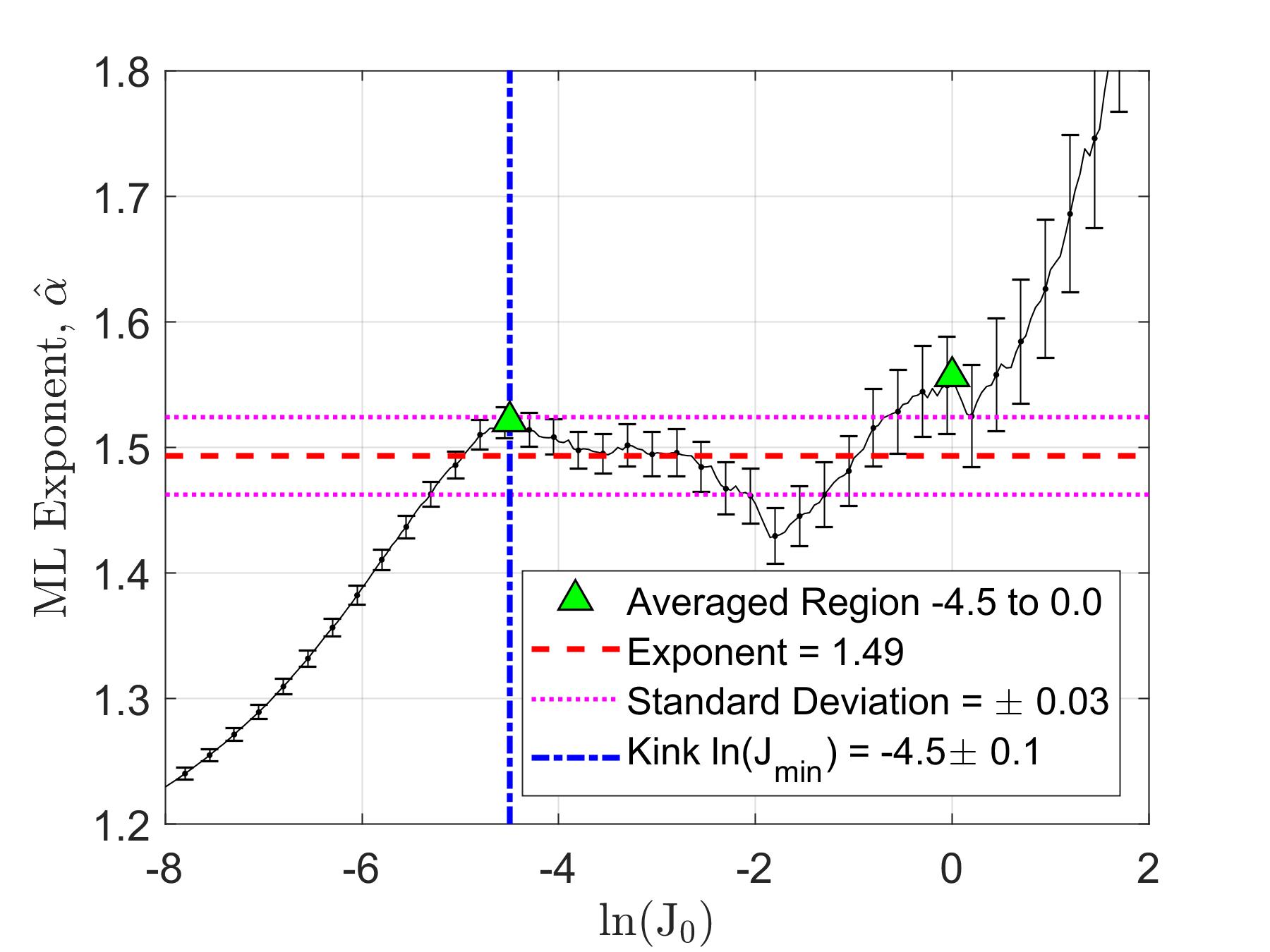}
		\caption{ML analysis of $2722$ jerk peaks from $10$ runs.}
		\label{subfig:pztf_MLcomb}
	\end{subfigure}\\
	\begin{subfigure}{0.47\textwidth}
		\includegraphics[width=1\textwidth]{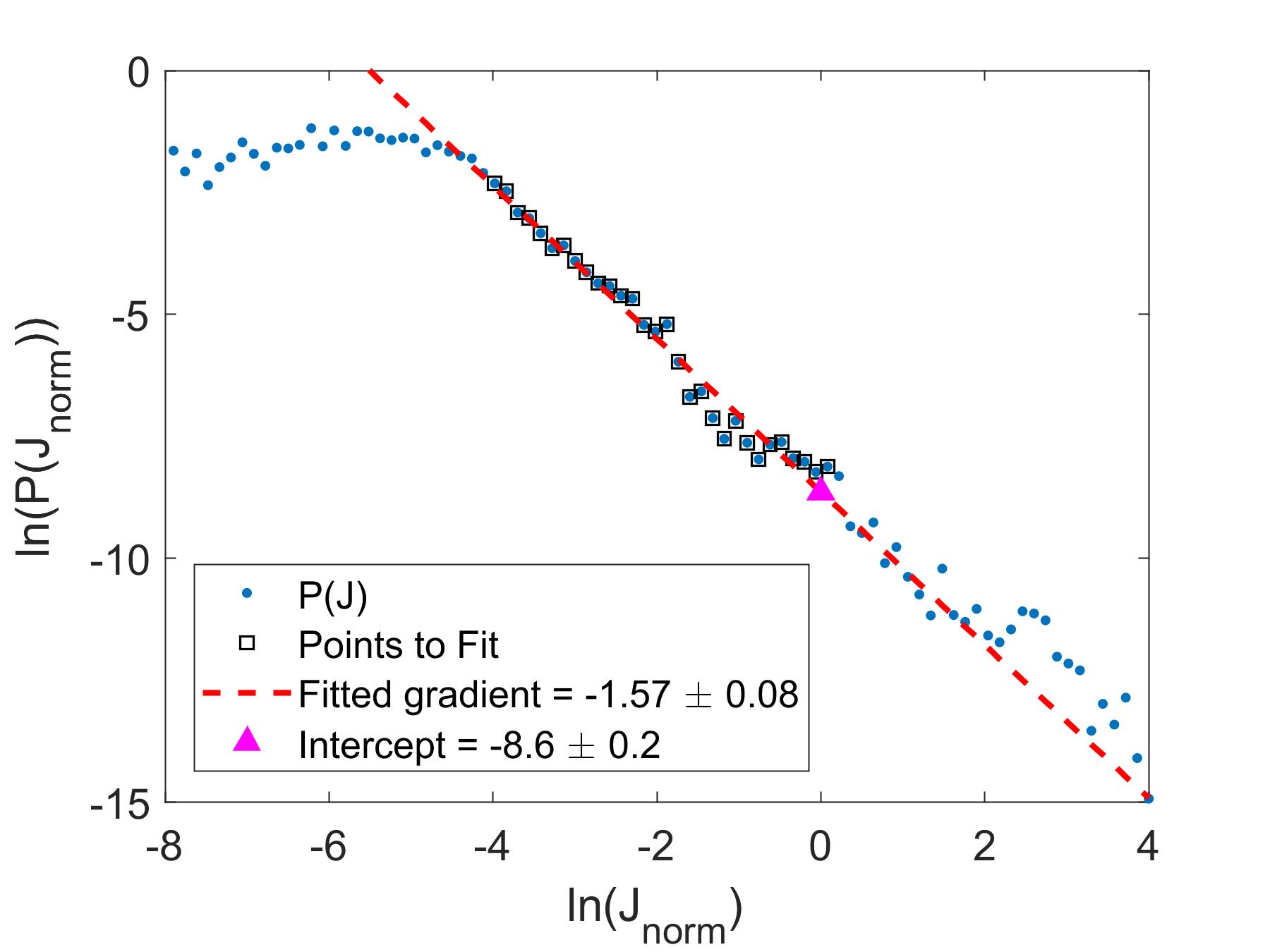}
		\caption{Linear fitted at region $-4\leq\ln(J_\textrm{norm})\leq0$.}
		\label{subfig:pztf_comb_loglogfit_1}
	\end{subfigure}\hfill
	\begin{subfigure}{0.47\textwidth}
		\includegraphics[width=1\textwidth]{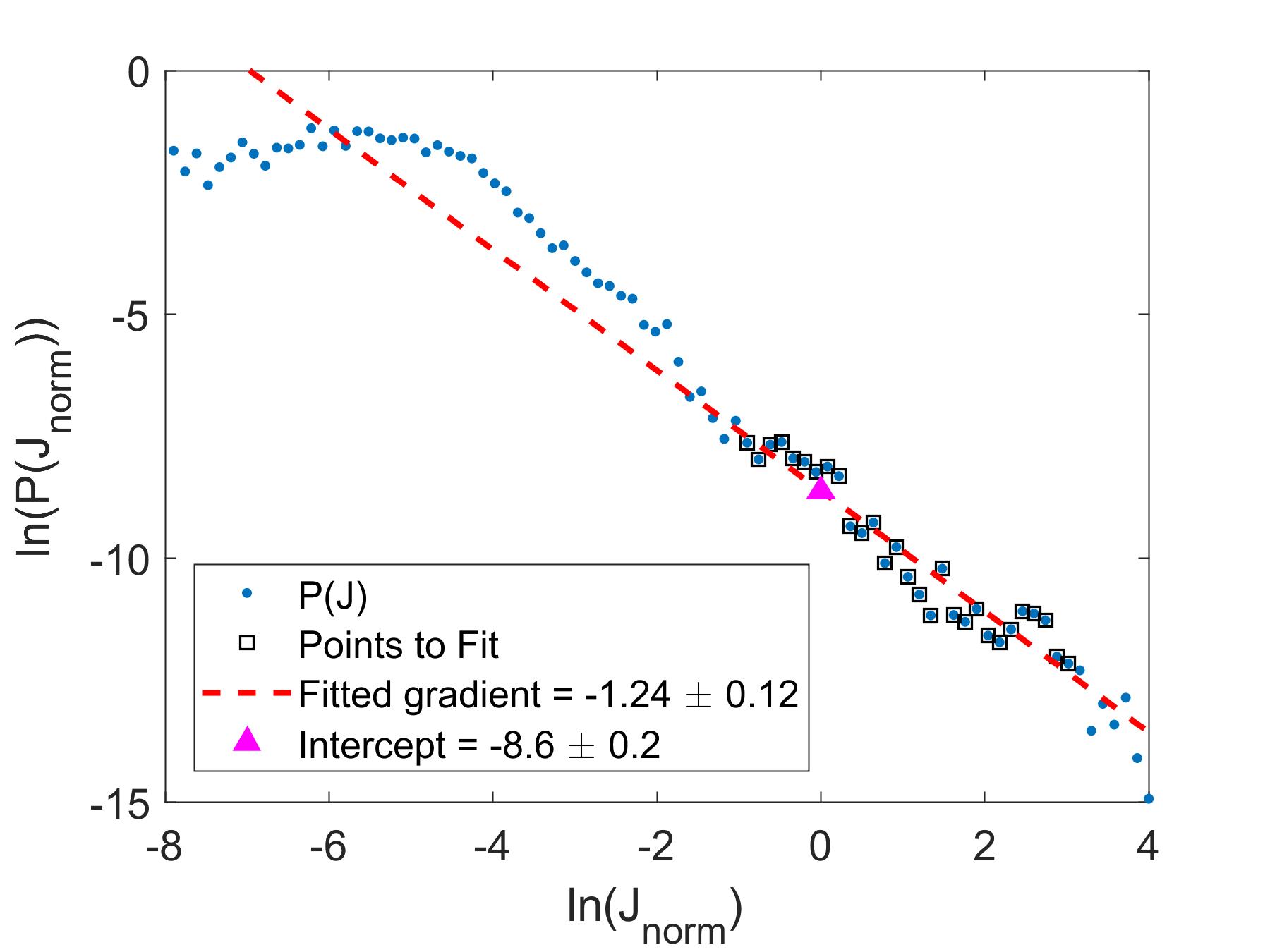}
		\caption{Linear fitted at region $-1\leq\ln(J_\textrm{norm})\leq3$.}
		\label{subfig:pztf_comb_loglogfit_2}
	\end{subfigure}
	
	\caption{A total of $2722$ normalised jerk peaks, $J_\textrm{norm}$ from $10$ measurements of sample F were log-log binned as probability distribution, $P(J_\textrm{norm})$ against $J_\textrm{norm}$ as (a). The distribution at $\ln(J_\textrm{norm})>2$ is mainly due to spanning peaks. (b) shows the ML analysis that was performed on the $2722$ jerk peaks. The ML curve shows some exponent mixing effect (described in \textbf{\Cref{theory:exponentmixing}}) due to the existence of larger peaks, which indicate a power-law distribution with a lower exponent (thus the dip in the ML curve). It is always safer to rely on ML, but one can be bold and fit (a) with two straight lines (c) and (d) to show the change in exponents (note the errors do not apply).}
	\label{fig:pztf_spanpeakscomb}
\end{figure}

As mentioned above, the large peaks may or may not play a role in the avalanche statistics\cite{He2016}. With the apparatus we were working with right now, it was not possible to work out how the peaks originated, thus it is best to perform the analyses with (\textbf{this section}) and without (\textbf{\Cref{sssec:pztf_wospanpeaks}}) the spanning avalanche peaks involved. The log-log binning and the ML estimate result of 10 runs are displayed as \Cref{subfig:pztf_MLcomb,subfig:pztf_comb_loglognofit}.\par
From the ML fit in \Cref{subfig:pztf_MLcomb}, the curve shows some exponent mixing effect (as demonstrated with simulated jerks in \textbf{\Cref{theory:exponentmixing}}) albeit very weak. This is caused by the inclusion of the large peaks. As larger peaks were being counted at a higher frequency, the most-likely exponent that generates them will be smaller, thus the estimated exponent decreases. When compared to the ML curve (\Cref{fig:pztb_MLcomb}) of sample B, the ML curve of sample F had a decrease in estimated exponent due to the influence of these large peaks that were absent for sample B.\\\par

Looking at \Cref{subfig:pztf_comb_loglognofit} one can argue that the distribution has only one exponent. However, if the exponent mixing effect from the ML fit was really true, one could try to fit the log-log distribution with two different straight-lines (\Cref{subfig:pztf_comb_loglogfit_1,subfig:pztf_comb_loglogfit_2}). Determining the point at which the first fit ends and the second fit begins is subjective. Thus, the fits in \Cref{fig:pztf_spanpeakscomb} function more as a guide, showing that there may be more than one straight-line behaviour in the log-log plot. It is always better to rely on the ML curves.
\subsubsection{Results of Combined 10 Runs without Spanning Peaks}\label{sssec:pztf_wospanpeaks}
\begin{figure}[h]
	\centering
	\begin{subfigure}{0.45\textwidth}
		\includegraphics[width=1\textwidth]{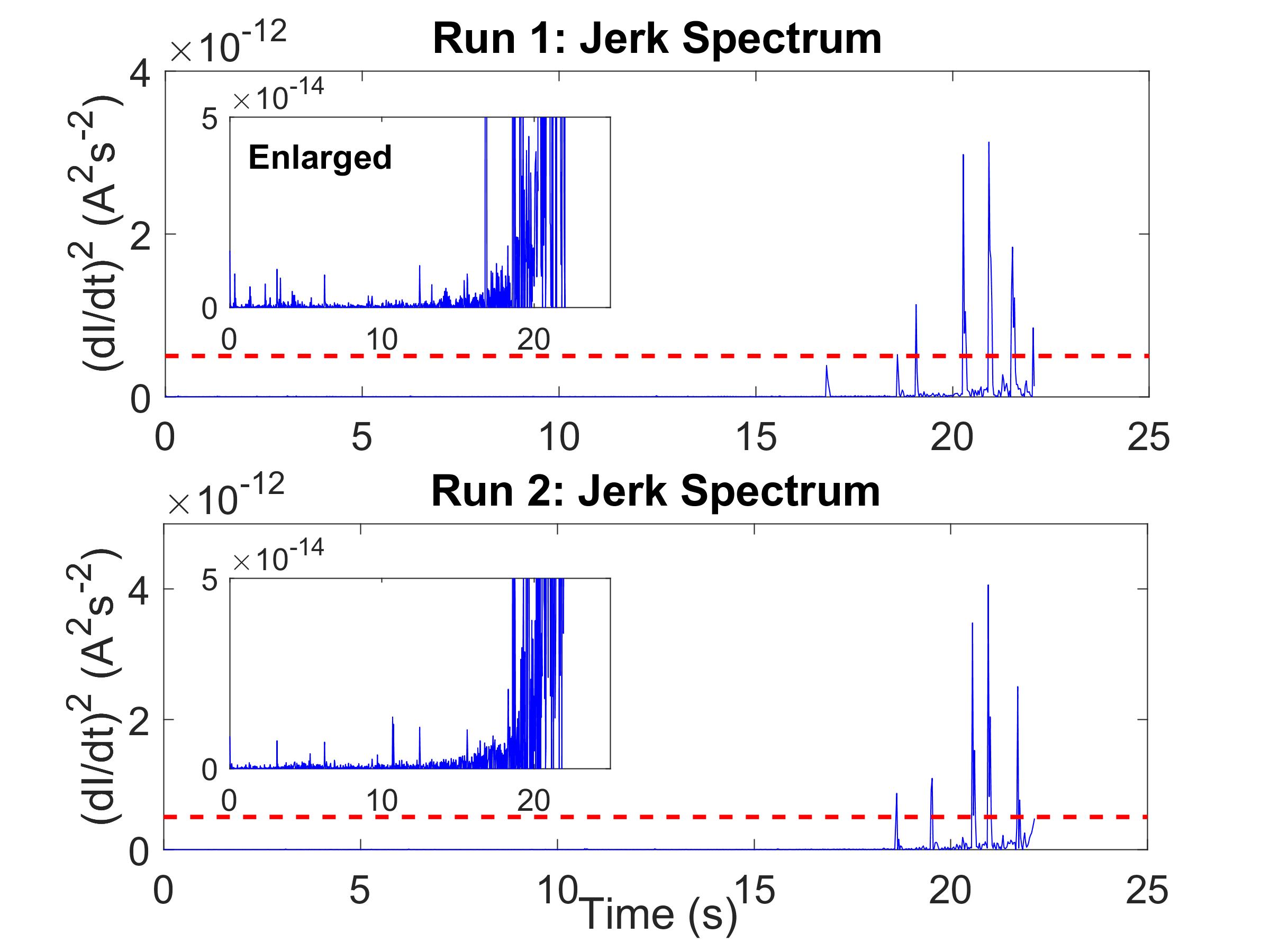}
		\caption{Spectra 1 and 2 with threshold.}
		\label{subfig:pztf_thresruns_1_2}
	\end{subfigure}\hfill
	\begin{subfigure}{0.45\textwidth}
		\includegraphics[width=1\textwidth]{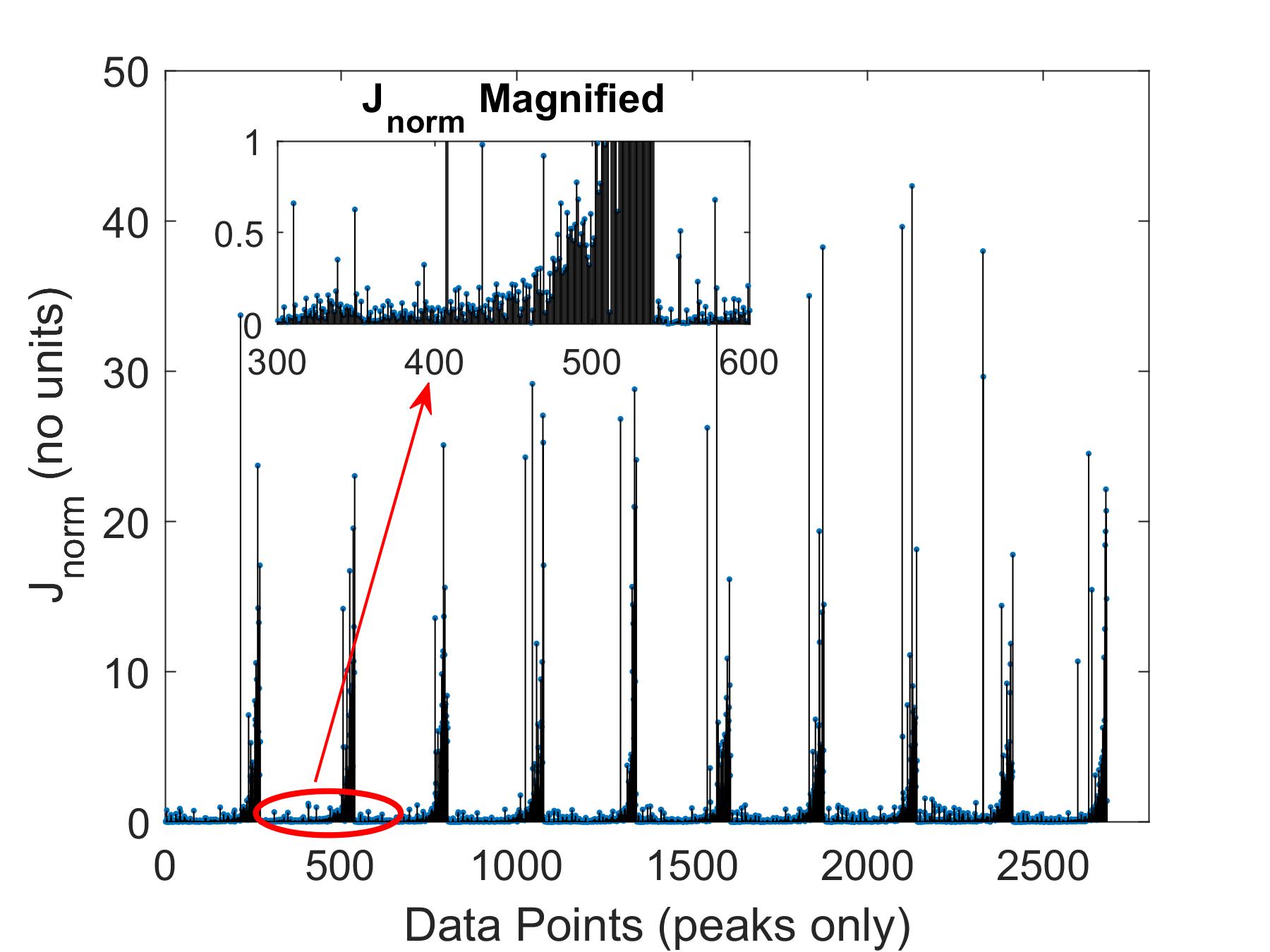}
		\caption{Combined 2681 jerks from 10 runs.}
		\label{subfig:pztf_rempeaksonly}
	\end{subfigure}\\
	\begin{subfigure}{0.47\textwidth}
		\includegraphics[width=1\textwidth]{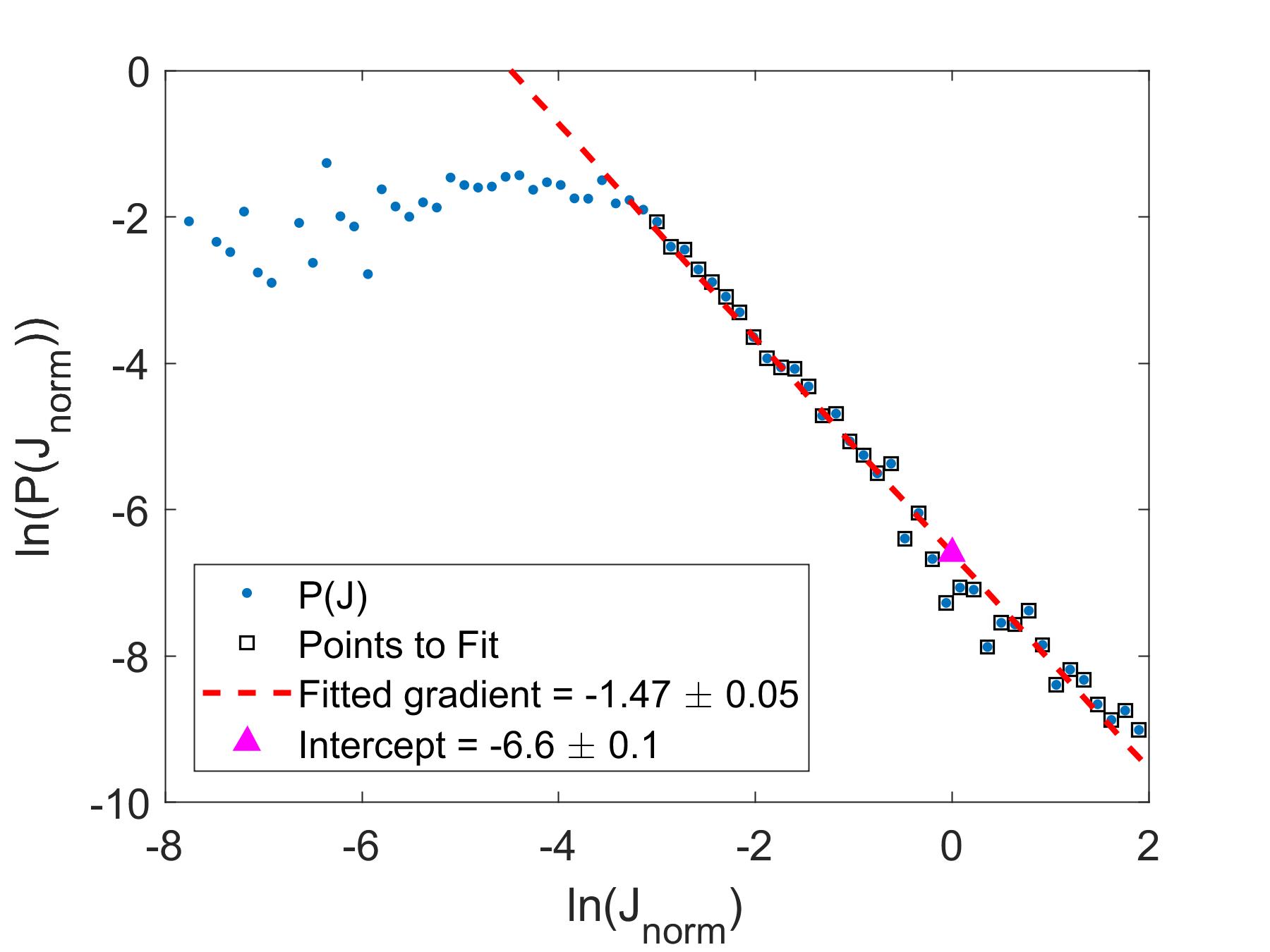}
		\caption{$P(J_\textrm{norm})$ against $J_\textrm{norm}$ with removed peaks.}
		\label{subfig:pztf_loglogpeaksrem}
	\end{subfigure}\hfill
	\begin{subfigure}{0.47\textwidth}
		\includegraphics[width=1\textwidth]{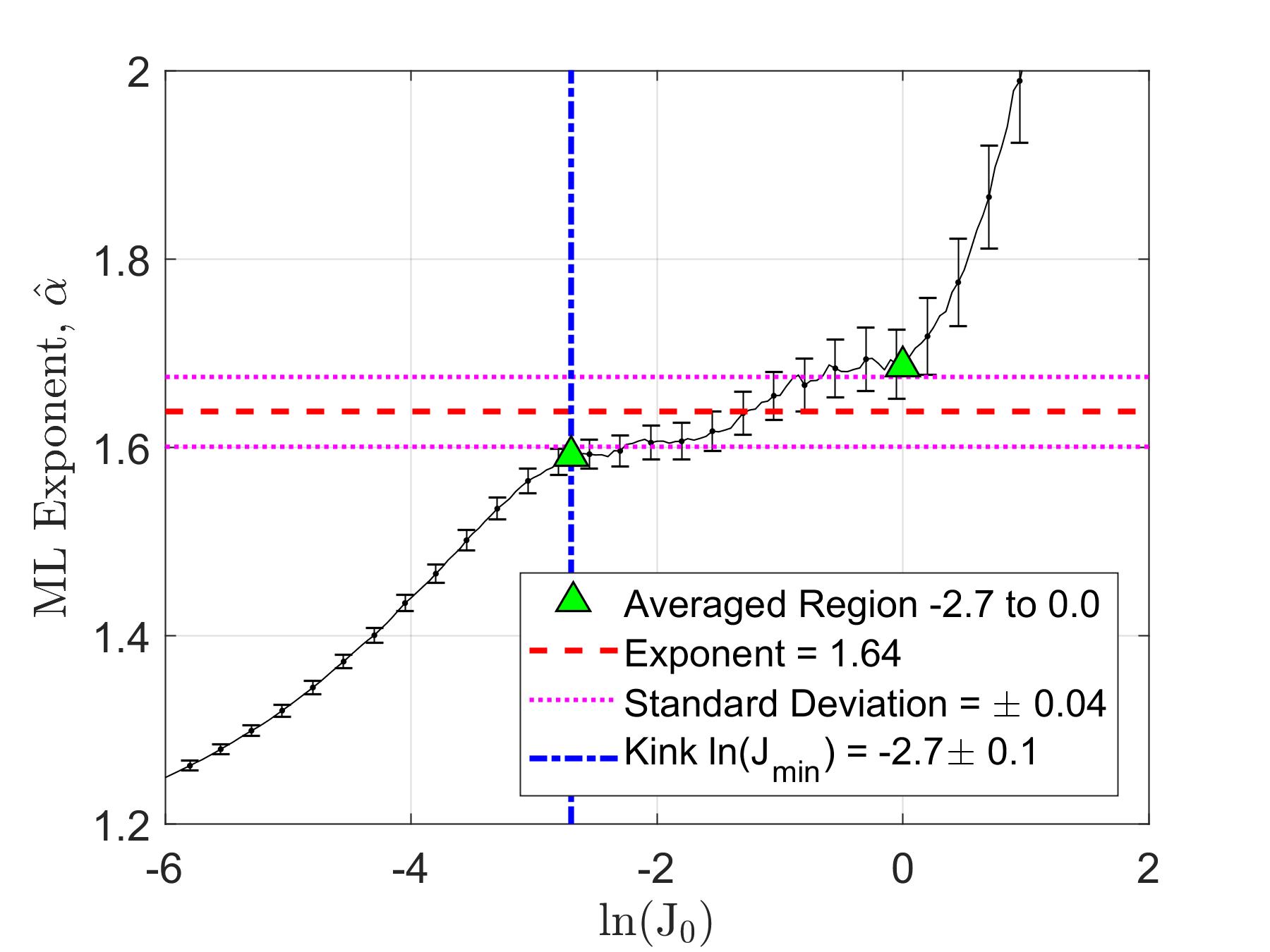}
		\caption{ML analysis of the $2681$ jerk peaks.}
		\label{subfig:pztf_MLpeaksrem}
	\end{subfigure}
	\caption{(a): Two examples of PZT sample F with a threshold of $0.5\times10^{-12}$ A$^2$s$^{-2}$ inserted; any spanning peaks that were above this threshold were removed. The spanning peaks were much larger compared to the usual jerks. (b): The jerk peaks were collated as a \textit{grand} spectrum with a region magnified in the inset. (c): Peaks from (b) were log-log binned and a straight line (with exponent $= 1.47$) was fitted. The ML in (d) shows a plateau that spans three decades (region between two green triangle labels) with an averaged exponent (red dashed line with standard deviation as pink dotted line) around $1.64\pm0.04$. A kink (blue dot-dashed line) can be observed at $\ln(J_\textrm{min})=-2.7\pm0.1$.}
\end{figure}

Similar to He et al.'s analysis\cite{He2016}, the large spanning peaks were now omitted by setting a threshold of $5\times10^{-13}$ A$^2$s$^{-2}$ for all the 10 runs from \Cref{sssec:pztf_wispanpeaks}. Any spanning peaks that were above the threshold (shown as the red dashed line in \Cref{subfig:pztf_thresruns_1_2}) were removed. Out of 10 measurements, the total jerk number was reduced from $2722$ to $2681$, with $41$ large peaks removed. The threshold was set just high enough that the ferroelectric switching jerk peaks were not affected. The curated jerks (\Cref{subfig:pztf_rempeaksonly}) were then analysed with the two methods: log-log binning in \Cref{subfig:pztf_loglogpeaksrem} and ML in \Cref{subfig:pztf_MLpeaksrem}.\\\par

Comparing the log-log fits (\Cref{subfig:pztf_comb_loglogfit_1,subfig:pztf_comb_loglogfit_2}) in the previous section with \Cref{subfig:pztf_loglogpeaksrem}, the latter has arguably a better straight-line fit with only a single exponent of $1.47$. However, to emphasize again, log-log fitting is highly subjective as one can argue that there may be a second exponent to be fitted at region $0\leq\ln(J_\textrm{norm})\leq2$. The ML estimate (\Cref{subfig:pztf_MLpeaksrem}) puts this case to rest by showing a plateau that spans three decades (region between two triangle labels) with a \textbf{single} exponent (red dashed line with standard deviation as pink dotted line) at about $1.64\pm0.04$, which concludes that the \textit{grand} jerk spectrum of PZT sample F is power-law distributed! A kink (blue dot-dashed line) can be observed at $\ln(J_\textrm{min})=-2.7\pm0.1$.\\\par
However, this raises a red flag; is the ML curve really reliable if the curve changes drastically (from \Cref{subfig:pztf_MLcomb} to \Cref{subfig:pztf_MLpeaksrem}) by removing 41 entries from the initial 2722 ($1.5\%$) peaks? This can be answered mathematically and with some forethought: as large peaks are more likely to arise from small exponents (the smaller the exponent, the higher the probability of creating a large peak), ML will try to estimate an exponent to describe where these large peaks come from, resulting in a lower exponent plateau overall. Keep in mind these spanning peaks are generally at least a magnitude larger than the switching jerks, they will take more weight in the ML algorithm. As $\ln(J_0)$ further increases, there are fewer switching peaks generated from the ``correct" exponent while these large peaks survived, causing the ML curve to dip to a lower ``false" exponent. Therefore, it is best to avoid samples that exhibit these abrupt current spikes to avoid the complications of curating the data. Nonetheless, both the ML curves (\Cref{subfig:pztf_MLcomb,subfig:pztf_MLpeaksrem}) convincingly show that the data from the jerk spectra are power-law distributed and consistent with Barkhausen noise.

\subsubsection{Categorising PZT sample F}\label{sssec:minicon_f}
The current response (\Cref{subfig:pztf_IVvsT}) of sample F showed little spikes that possibly resembled spanning avalanches \cite{He2016}. Taking these large peaks into account brings to light the effect of exponent mixing and the unsatisfactory log-log distribution plot (\Cref{fig:pztf_spanpeakscomb}). Following the convention of He et al.\cite{He2016} by omitting the spanning peaks from the analysis, a single plateau that spans three decades with a \textbf{single} exponent at $1.64$ with standard deviation, $\sigma$ of $\pm0.04$ is observed.\\\par
The lower bound $J_\textrm{min}$ for F can be worked out using the method from \textbf{\Cref{sssec:minicon_b}}. The averaged jerk peaks for run 1 was $1.134\times10^{-14}$ A$^{2}$s$^{-2}$, whereas the normalised $J_\textrm{min}=e^{-2.7}=0.067\pm0.007$. Taking the product, $J_\textrm{min}$ for Run 1 is $7.6\pm0.8\times10^{-16}$A$^{2}$s$^{-2}$. The averaged $J_\textrm{min}$ for 10 runs is $8.5\pm0.5\times10^{-16}$A$^{2}$s$^{-2}$.\\\par

To conclude for sample F:

\begin{itemize}
	\item The coercive voltage $V_c$ at ramp-rate $= 40$ Vs$^{-1}$ is $872\pm1$ V. In terms of coercive field $E_c=899\pm9\textrm{ Vmm}^{-1}$. The thickness of the sample was $0.97\pm0.01$ mm.
	\item The critical exponent is estimated to be $1.64$ with $\sigma = \pm0.04$.
	\item The value for the \textit{normalised} $\ln(J_\textrm{min})=-2.7\pm0.1$.
	\item The average value of 10 runs for $\bar{J}_\textrm{min}=8.5\pm0.5\times10^{-16}~\textrm{A}^2\textrm{s}^{-2}$; $\ln(\bar{J}_\textrm{min})=-34.70\pm0.06$.
\end{itemize}
\newpage
\subsection{PZT Sample S}\label{subsec:an_pzts}
PZT sample S was a brand new PZT sample (code: PRY+0111, material: PIC 255) that was ordered from PI Ceramic GmbH (Lederhose, Germany)\cite{PiezoTech}. PIC 255 is designed to be ``harder"/ have a higher $E_c$ compared to the two previous PIC 151 samples\cite{PiezoTech}. According to the product brochure from \cite{PiezoTech}, the product PRY+0111 has silver screen printed electrodes on both sides with a radius of $5.00\pm0.03$ mm (area $= 78.5\pm0.7$ mm\textsuperscript{$2$}) and a thickness of $1.00\pm0.05$ mm (including electrodes). For a ramp-rate of $60$ Vs$^{-1}$, the coercive voltage $V_c$ for sample S was $1237\pm5$ V; in terms of field, $E_c = 1240\pm60$ Vmm\textsuperscript{$-1$} (errors based on the tolerances  from the manufacturer). The ramp-rate was increased to $60$ Vs$^{-1}$ to match the data acquisition rate\footnote{Due to a higher $V_c$, a higher ramp-rate was needed to keep the measurement time and the sampling rate as a constant. A slower ramp will take a longer time to achieve $V_c$, thus decreasing the sampling rate.} ($40$ Hz) of the previous two measurements. \\\par

\begin{figure}[h]
	\centering
	\begin{subfigure}{0.48\textwidth}
		\includegraphics[width=1\textwidth]{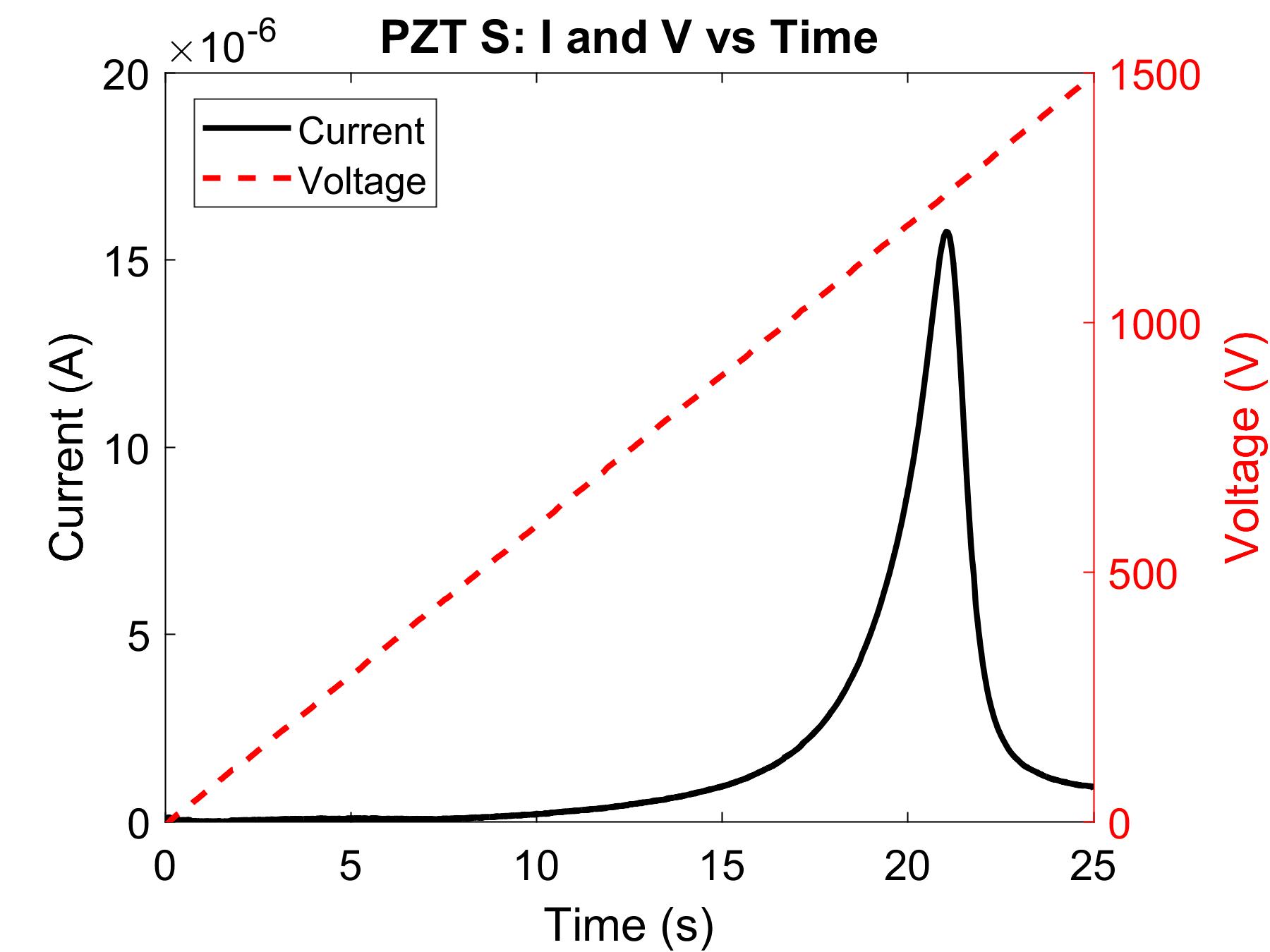}
		\caption{Run 1: Current and Voltage against time.}
		\label{subfig:pzts_IVvsT}
	\end{subfigure}\hfill
	\begin{subfigure}{0.48\textwidth}
		\includegraphics[width=1\textwidth]{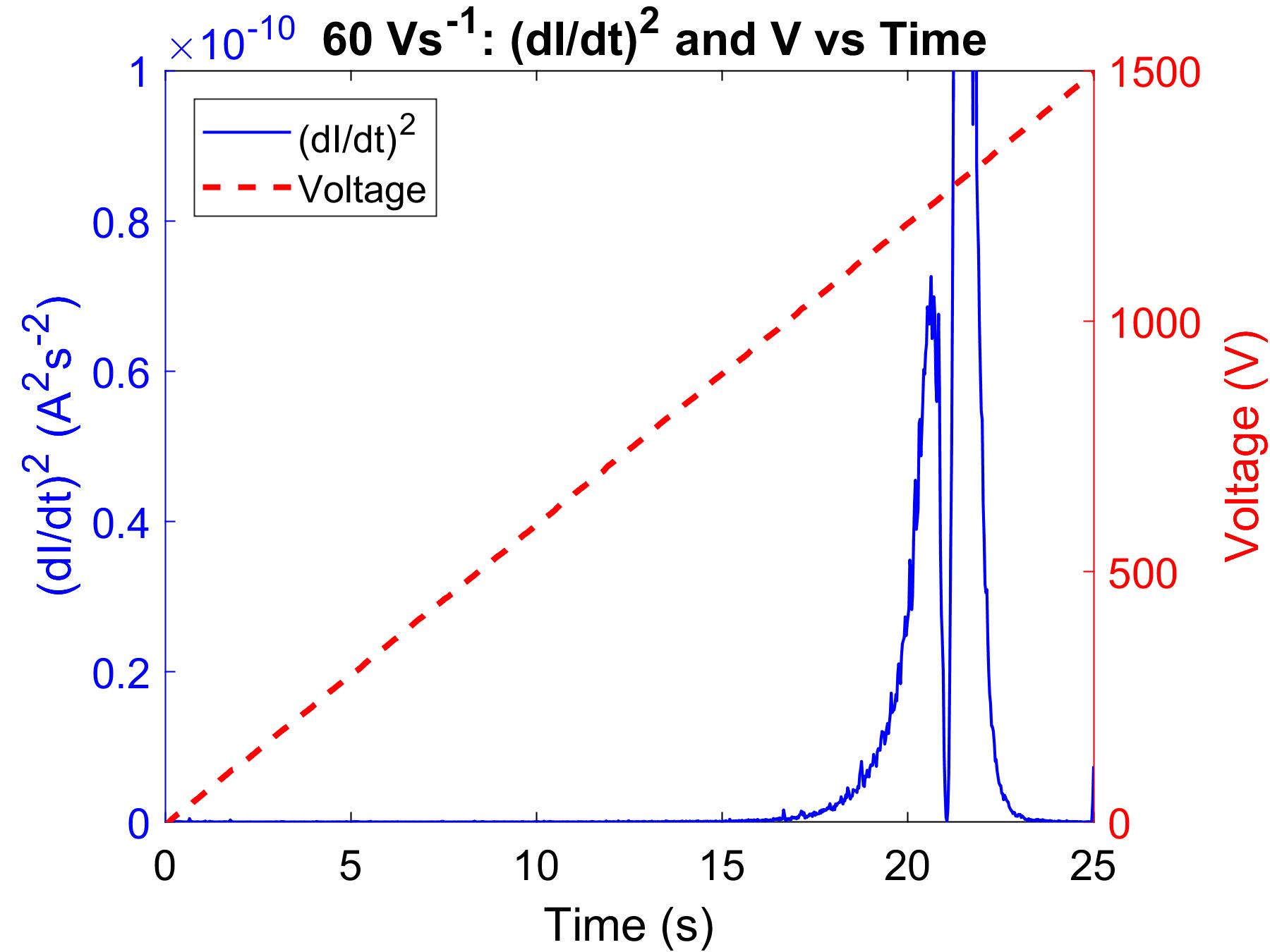}
		\caption{Jerk spectrum with baseline.}
		\label{subfig:pzts_JVvsT}
	\end{subfigure}
	
	\caption{(a): Current and voltage response of PZT sample S (material: PIC 255), which looked different when compared to samples (material: PIC151) B (\Cref{subfig:pztb_ivvst}) and F (\Cref{subfig:pztf_IVvsT}). (b): The resulting $(dI/dt)^2$ spectrum with baseline. The figure was enlarged to emphasise the jerk peaks around the first background maxima.}
	\label{fig:pzts_run1}
\end{figure}

\Cref{subfig:pzts_IVvsT} is the current and voltage responses of PZT sample S. The current spectrum has a different shape compared to the two former samples (\Cref{subfig:pztb_ivvst,subfig:pztf_IVvsT}). The resulting jerk spectrum (\Cref{subfig:pzts_JVvsT}) with baseline, $(dI/dt)^2$ is magnified to show the jerk peaks up until the baseline's first maxima (the steepest region of polarisation change, $\delta P$). Thus, the baseline's second maxima is cropped off. The analyses of this and four other jerk spectra of sample S are detailed in the next section.
\subsubsection{Results of Combined 5 Runs}\label{sssec:an_pzts}
\begin{figure}[h]
	\centering
	\begin{subfigure}{0.48\textwidth}
		\includegraphics[width=1\textwidth]{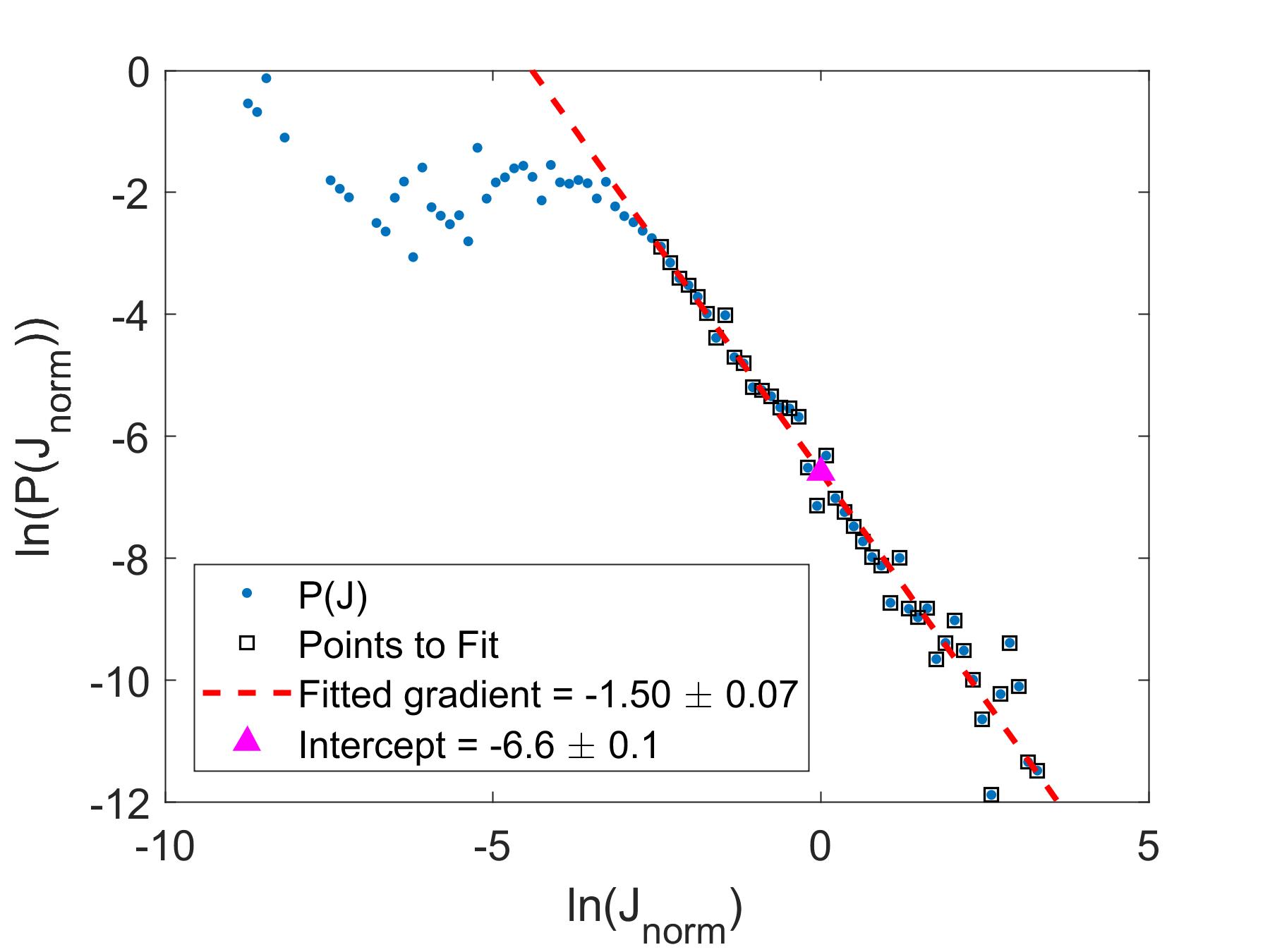}
		\caption{$P(J_\textrm{norm})$ against $J_\textrm{norm}$ of PZT sample S for 5 runs.}
		\label{subfig:pzts_loglogcomb}
	\end{subfigure}\hfill
	\begin{subfigure}{0.48\textwidth}
		\includegraphics[width=1\textwidth]{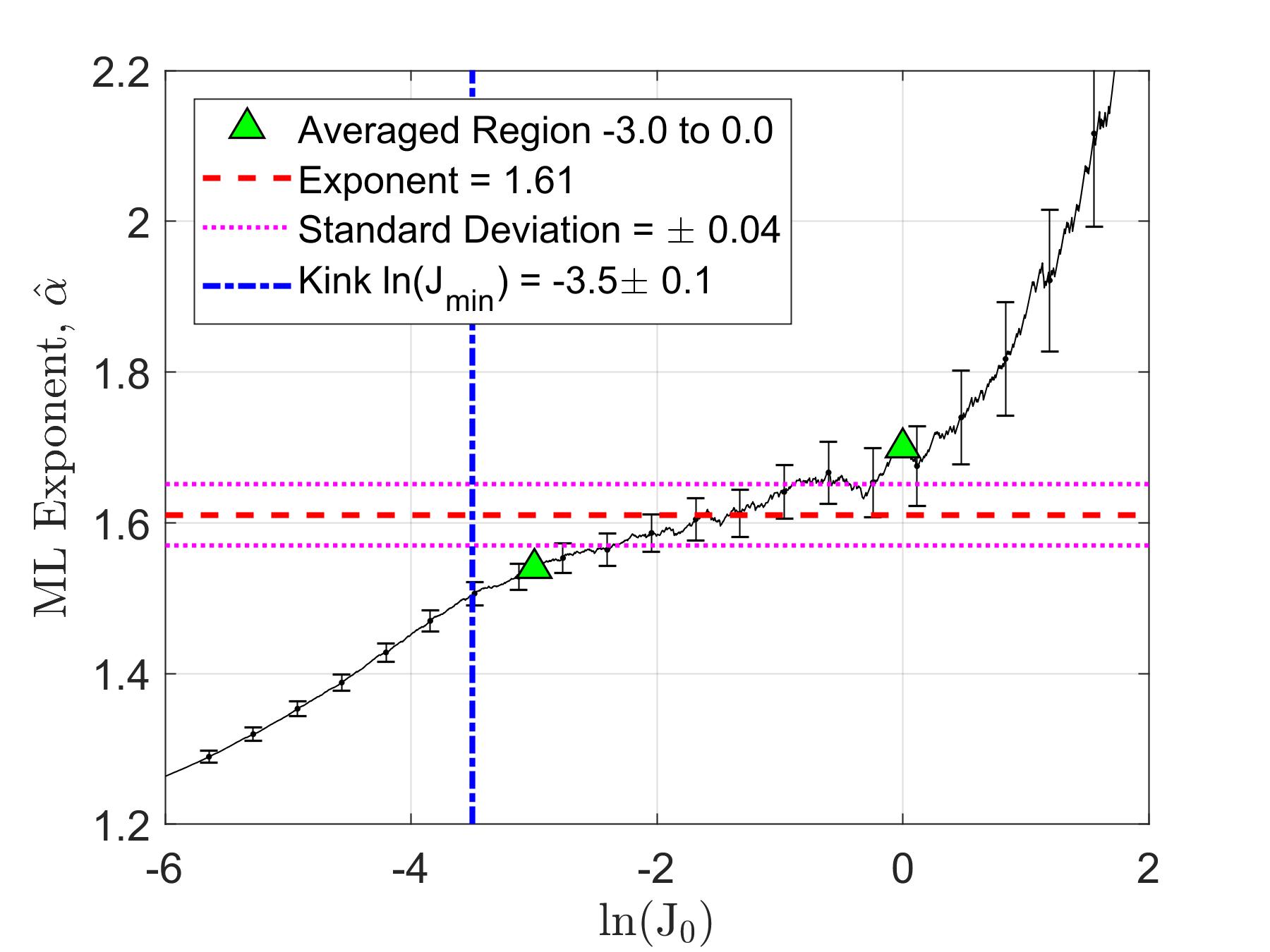}
		\caption{ML analysis of 1468 jerk peaks.}
		\label{subfig:pzts_MLcomb}
	\end{subfigure}
	\caption{(a): A total number of 1468 jerk peaks were log-log binned and the log-log probability distribution of the normalised jerk peaks, $P(J_\textrm{norm})$ against $J_\textrm{norm}$ was linear fitted with an exponent of $1.50$. (b): is the ML analysis of the jerk peaks, giving an averaged exponent (red dashed line) of $1.61$ with $\sigma$ (pink dotted line) $=\pm0.04$ that spans around three decades (region between two triangle labels). A kink (blue dot-dashed line) can be observed at $\ln(J_\textrm{min})=-3.5\pm0.1$.}
	\label{fig:an_pzts}
\end{figure}
A total of 5 measurements were taken on sample S, and as done previously, the jerk peaks were extracted after removing the baselines. These jerks were then normalised by the mean and combined for log-log binning (\Cref{subfig:pzts_loglogcomb}) and ML (\Cref{subfig:pzts_MLcomb}) analyses.\\\par

From the 5 measurements, a total of 1468 jerk peaks were extracted. The log-log linear regression, shown in \Cref{subfig:pzts_loglogcomb}, demonstrates a straight line behaviour with an exponent of $-1.50$. The ML analysis in \Cref{subfig:pzts_MLcomb} shows an averaged exponent (red dashed line) of $1.61$ with standard deviation (pink dotted line), $\sigma=\pm0.04$ that plateaus for about three decades (region between two green triangle labels). A kink (blue dot-dashed line) is observed at $\ln(J_\textrm{min})=-3.5\pm0.1$. 
\subsubsection{Categorising PZT sample S}\label{sssec:minicon_s}
The lower bound $J_\textrm{min}$ for S can be worked out as previously from \textbf{\Cref{sssec:minicon_b,sssec:minicon_f}}. For the jerks from run 1, the average was $4.596\times10^{-13}$ A$^{2}$s$^{-2}$, whereas the normalised $J_\textrm{min}=e^{-3.5}=0.030\pm0.003$. Taking the product, $J_\textrm{min}$ for Run 1 is $1.4\pm0.1\times10^{-14}$A$^{2}$s$^{-2}$. The averaged $J_\textrm{min}$ for 5 runs is $1.25\pm0.06\times10^{-14}$A$^{2}$s$^{-2}$.\\\par

To conclude for sample S:

\begin{itemize}
	\item Sample S (PIC 255) is compositionally different from samples (PIC 151) B and F. PIC 255 is designed to be harder than PIC 151.
	\item The coercive voltage $V_c$ at ramp-rate $= 60$ Vs$^{-1}$ is $1237\pm5$ V. In terms of coercive field $E_c=1240\pm60\textrm{ Vmm}^{-1}$. The thickness of the sample (given by manufacturer) was $1.00\pm0.05$ mm.
	\item The critical exponent is estimated to be $1.61$ with $\sigma = \pm0.04$.
	\item The value for the \textit{normalised} $\ln(J_\textrm{min})=-3.5\pm0.1$.
	\item The average value of $J_\textrm{min}$ for 5 runs is $1.25\pm0.06\times10^{-14}$A$^{2}$s$^{-2}$; $\ln(J_\textrm{min})=-32.00\pm0.04$.
\end{itemize}
\newpage
\subsection{Results Discussion}\label{subsec:ResDis}
The resulting estimated exponent, $\hat{\alpha}$ and the averaged lower bound of the three samples were tabulated in \Cref{tbl:results}. The values for $\bar{J}_\textrm{min}$ are out of range for all three samples, with the two PIC 151 samples B and F lying in the same order of magnitude. \\\par
\begin{table}[h]
	\begin{center}
		\begin{tabular}{||c|c|c|c|c||} 
			\hline&&&&\\[0.5ex]
			System & Esti. Exponent, $\hat{\alpha}$ & Std. Dev., $\sigma$ & $\bar{J}_\textrm{min}$ ($\times10^{-16}$ A$^2$s$^{-2}$) & $\Delta\bar{J}_\textrm{min}$ ($\times10^{-16}$ A$^2$s$^{-2}$)\\[0.5ex]
			\hline\hline
			PZT B & 1.73 & 0.04 & 6.1&0.3 \\ 
			PZT F & 1.64 & 0.04 & 8.5&0.5 \\
			PZT S & 1.61 & 0.04 & 125&6 \\
			\hline
			AT&1.65&\multicolumn{3}{|c|}{\cellcolor{gray}~}\\
			MFT&1.33&\multicolumn{3}{|c|}{\cellcolor{gray}~}\\
			
			\hline
		\end{tabular}
		
	\end{center}
	\caption{Table of estimated exponents, $\hat{\alpha}$ for three samples with their standard deviations, $\sigma$ and the averaged lower bounds of their power-law distribution, $\bar{J}_\textrm{min}$ with their standard error, $\Delta\bar{J}_\textrm{min}$. 
	AT is the theoretical prediction from avalanche theory\cite{SaljePredict} while MFT stands for the theoretically obtained energy exponent from mean field theory.}
	\label{tbl:results}
\end{table}

The values extracted for the jerk peaks were based on the first time derivative of the current squared ($(dI/dt)^2$), thus varying current responses taken from the measurements would yield different $\bar{J}_\textrm{min}$. There are two reasons for this:
\begin{itemize}
	\item As shown in \Cref{subfig:pztb_ivvst,subfig:pztf_IVvsT}, the current spectra for B and F (PIC 151) fell under the $10 \mu$A range while the range for sample S (PIC 255) fell under $100 \mu$A. Thus, sample S with a greater current response would have a larger $\bar{J}_\textrm{min}$ value. 
	\item The variation between B and F might be the result of various degrees of fatigue. A ferroelectric sample that is fatigued will experience a decrease in switched charge\cite{Scott1989}. Thus, the lower current generated by the more fatigued sample B will lead to a lower value of $\bar{J}_\textrm{min}$.
\end{itemize}
\par
The exponents evaluated were roughly close to each other at around 1.6 to 1.7. The values for both $\hat{\alpha}$ and $\sigma$ for each ML analysis depended highly on where the plateaus were defined and would be biased. Unlike ideal power-laws demonstrated in \textbf{\Cref{subsec:MLAnalysis}}, there was no definitive kink to determine where the plateau started due to the lower cut-off effect\cite{Salje2017}. Therefore, the exponents were concluded to be within range.\\\par

As previously remarked in \textbf{\Cref{subsec:jerks}}, there are many ways one can define the jerks in order to extract the critical exponents\cite{BenZion2011,Salje2014,He2016}. Crystal compression studies defined jerks as the velocity of \textit{slip} avalanche, squared\cite{Salje2014,Dahmen2017,Friedman2012}. The jerks would then be proportional to energy $\frac{1}{2}mv^2$\cite{Salje2014,Dahmen2017} and the probability distribution would yield an energy exponent. Since the jerk peaks in our experiments are not defined in terms of energy but in terms of the \textit{slew rate}, the exponents in \Cref{tbl:results} can only be treated as, at best, \textit{a proxy} for the energy exponent. \\\par

The critical exponents derived from mean-field theory (MFT) are also used commonly as reference points for researchers \cite{Salje2014,Friedman2012,BenZion2011,Dahmen2017,Dahmen2009a,Tsekenis2013} to compare their work. The MFT energy exponent, $\varepsilon= 4/3\approx1.33$\cite{Salje2014,Dahmen2017,Friedman2012} describes a theoretical avalanche system that has an infinite range interaction\cite{Friedman2012,Selinger2016,Dahmen1996}. If our estimated exponents, $\hat{\alpha}\approx1.66$ did resemble an energy exponent, the values of $\hat{\alpha}$ showed that the PZT systems were significantly different to the MFT regime, implying the switching interactions were more localised. Instead, our exponents are in excellent agreement with that of avalanche theory where $\alpha = 1.65$\cite{SaljePredict}.\\\par
Based on the power-law distribution of the jerk peaks, we hereby conclude that these Barkhausen noises were generated in an avalanche fashion. As a single cation is displaced along the axis of the applied field, the resulting polarisation change will trigger other nearby cations to be displaced as well, causing a displacement avalanche. This switching mechanism is analogous to the coupling interaction in a random field Ising model in ferromagnetic systems\cite{Salje2014,Dahmen2009}. 

\subsection{Future Work Ideas}
There are many ideas that can be implemented in the future to improve our analyses method and deepen our understanding. For instance, these experiments will benefit highly from modern scanning probe microscopy techniques such as atomic force microscopy (AFM)\cite{Potnis2011,Cheng2008}. Visualising the switching mechanisms of domain wall movement will increase our understanding of the origin of the jerky current response. \\\par

Besides, a more sophisticated hysteresis instrument with a higher sampling rate will significantly improve the resolution of the jerk spectra. The allocation of 1000 sampling points over a duration of 25 seconds is unsatisfactory in measuring these fine crepitations, as low sampling rates will cause an overlapping of jerks\cite{Friedman2012}.
An improvement in sampling rate will allow the jerks to be better defined. With more points per jerk, each peak can be numerically integrated to obtain their sizes and the distribution of them may lead to different critical exponents\cite{Salje2014,He2016,Friedman2012}. As jerks can be defined in various forms\cite{Salje2014,Dahmen2017}, it will be ideal to transform our \textit{slew-rate} definition into the form of an energy; obtaining a definitive energy exponent will allow our work to be compared with others.\\\par
There are other experiments that can be done in the near future. A system with tuned criticality will have its largest observed avalanche based on both its system size and external experimental parameters\cite{Salje2014,Friedman2012}. Compression experiments on nano-sized crystals showed that the greatest avalanche size is dependent on stress\cite{Friedman2012}; will the biggest jerks in our ferroelectric systems depend on the voltage ramp-rate? Are the criticality of these PZT systems tuned? \\\par 

The other option is to look into other ferroelectric systems and their jerk distributions. Categorising other ferroelectrics such as BTO will be the first step forward for this. Barkhausen noise was also found in the relaxor\footnote{A relaxor is a type of ferroelectric that has a frequency dependent permittivity\cite{Colla1998}.} regime of ferroelectric  PbMg$_{1/3}$Nb$_{2/3}$O$_3$ (PMN) where no macroscopic domains are present\cite{Colla2002}. It will be interesting to see, in other systems, if these crackling noises could be uncovered using a simple hysteresis apparatus and what their critical exponents may be.
\newpage
\section{Conclusion}
The project work in investigating Barkhausen noise involved two phases. Initial experiment works involved applying high voltage pulses with fast ramp-rates on PZT commercial ceramics. Although many irregular current spikes were observed in the current response, they were only evident at voltages greater than the coercive voltage. These large spikes did not fit the theories proposed by literature and the samples were easily fatigued.\\\par 
In later experiments, the voltage ramp-rate was greatly reduced and the jerks from the current responses were analysed statistically. The jerks were defined as the peaks of the slew-rate squared. From three different PZT samples albeit with a low sampling rate, we showed that the jerks obeyed power-law statistics and were consistent with Barkhausen noise \cite{Salje2014,Dahmen2009}. The distribution of Barkhausen noises obey a power-law and the determined exponents with an average of 1.66 were in excellent agreement with avalanche theory predictions of 1.65\cite{SaljePredict}. 
This observation, when compared with the expected MFT exponent 1.33\cite{Salje2014,Friedman2012,BenZion2011,Dahmen2017,Dahmen2009a,Tsekenis2013}, indicates that the polarisation switching interactions were more localised and short-ranged.\\\par 
The initial research carried out on the noise of ferroelectrics shows that these electrical crepitations in the current response can be statistically analysed. Via the power-law distributed jerks, we conclude that the switching processes in these PZT samples take the form of an avalanche and are consistent with Barkhausen noise and many other crackling systems. The discovery and categorisation of these exponents are crucial as they show that the domain switching mechanisms in ferroelectric ceramics demonstrate self-organised criticality (SOC)\cite{Bak1987,Bak1988,Bak1991}. Moreover, by comparing them with other systems, more properties with regards to the physics of polarisation switching can be unveiled.\\\par

\newpage
\pagenumbering{roman}
\phantomsection 
\addcontentsline{toc}{section}{Appendix} 
\appendix
\section{Manual Waveform Generator and Data Acquisition}
\label{apn:MWGandDataAc}
In this manual waveform generator window (\Cref{subfig:tf_wfg}), specific data acquisition time can be provided to record the current response at different regions of the ramp. At the bottom left of the waveform generator window, the data acquisition settings are specified by providing a starting time-stamp and the total time the current-time response was recorded. The number of points sampled will then be determined in the main measurement window, with a maximum of 1000 points, evenly distributed throughout the recording region. An example of using the waveform generator to record at $1$ kHz sampling rate is described below:
\begin{figure}[h]
	\centering
	\begin{subfigure}{0.60\textwidth}
		\includegraphics[width=1.0\textwidth]{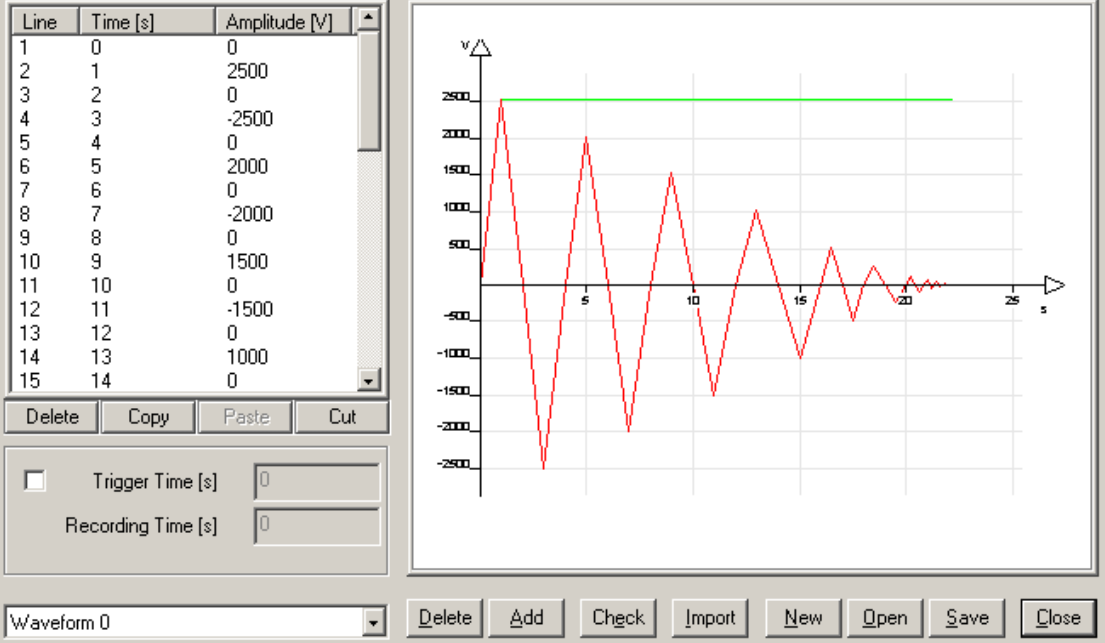}
		
		\subcaption{Waveform Generator Window.}	
		\label{subfig:tf_wfg}
	\end{subfigure}
	\hfill
	\begin{subfigure}{0.35\textwidth}
		\includegraphics[width=1.0\textwidth]{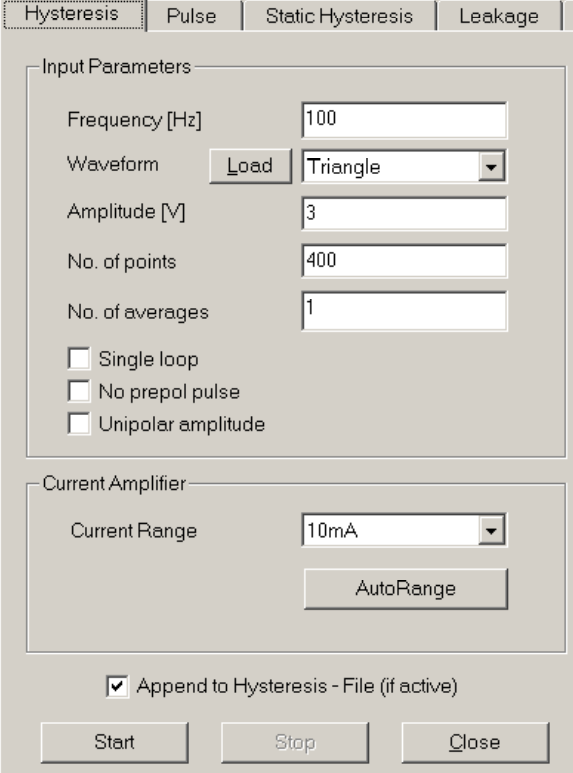}
		
		\subcaption{Hysteresis Measurement Dialog.}
		\label{subfig:tf_hysterdialog_crop}
	\end{subfigure}
\caption{(a): Waveform generator window that allow a specific voltage waveform to be designed and applied to a sample. (b): Hysteresis measurement dialog allow pre-set waveforms such as a sine and a triangle voltage pulse to be applied to the sample. The designed pulses from (a) can be loaded into (b) by pressing the Load button. The maximum number of points allowed is 1000 points. }
\end{figure}
\begin{enumerate}
	\item Design an intended waveform by specifying the timeline of the pulses with their pulse heights respectively.
	\item Tick the check-box next to the Trigger Time and specify the trigger time at the appropriate region of the waveform where one would like to investigate.
	\item Set a total recording time of 1 s.
	\item Save the waveform.
	\item On the main measurement dialog (\Cref{subfig:tf_hysterdialog_crop}), load the waveform using the load button and set the total number of points to 1000.

\end{enumerate}

~\par
With $1000$ points sampled in this $1$ s region, $\frac{1000}{1~\textrm{s}} = 1~\textrm{kHz}$ sampling rate can be achieved. However, there were some aspects of the software that were not straightforward and needed to be worked around. 
\begin{enumerate}
	\item From the manual, the maximum number of points for recording is $1000$\footnote{Technically it is $1001$ in the data output since the measurements include a data point at time t $= 0$ s followed by $1000$ points.}. This is a hard cap, providing any value above $1000$ will produce an error from the software.
	\item The maximum sampling rate provided in the manual is $1$ MHz. If one specifies a shorter recording time with $1000$ sampling points, a measurement can still be taken but the data file will consist of multiple points of the same value for each microsecond.
	\item The voltage pulse \textbf{will not} follow through once the pulse time exceeds the specified recording time-stamp. This had created a slight complication to the experiments conducted and was one of the pitfalls during the early periods of the project. For example, if one would like to examine the first five seconds a thirty second voltage ramp, the voltage supplied to the ferroelectric sample would not reach a maximum as intended and the sample would not fully switch. \label{enu:pulnofolthru}
\end{enumerate}

\section{Maximum-Likelihood}\label{apn:ML}
\subsection{Derivation}\label{apn:MLderive}
As shown in \cref{eq:MLfunction}, the ML function is:
\begin{equation}
p(J|\alpha)=\prod_{i=1}^{N}\frac{\alpha-1}{J_\textrm{min}}
\left(\frac{J_i}{J_\textrm{min}}\right)^{-\alpha}\tag{\ref{eq:MLfunction}}
\end{equation}
Log-transform the likelihood function to log-likelihood, $\mathcal{L}$:
\begin{equation}\label{eq:MLDerive}
\begin{split}
\mathcal{L}={}&\ln{p(J|\alpha)}\\
={}&\sum_{i=1}^{N}\left[\ln{(\alpha-1)}-\ln{J_\textrm{min}}-\alpha\ln{\frac{J_i}{J_\textrm{min}}}
\right]\\
=& N\ln{(\alpha-1)}-N\ln{J_\textrm{min}}
-\alpha\sum_{i=1}^{N}\ln{\frac{J_i}{J_\textrm{min}}}
\end{split}
\end{equation}
Maximising the function:
\begin{equation}
\frac{d\mathcal{L}}{d\alpha}=\frac{N}{\alpha-1}-\sum_{i=1}^{N}\ln{\frac{J_i}{J_\textrm{min}}}=0\notag
\end{equation}
Solve for the estimated exponent, $\hat{\alpha}$ to obtain:
\begin{equation}
\hat{\alpha}=1+N\left[\sum_{i=1}^{N}\ln{\frac{J_i}{J_\textrm{min}}}\right]^{-1}\tag{\ref{eq:estiexponent}}
\end{equation}
\subsection{Computational Algorithm}\label{apn:MLalgo}
\begin{enumerate}
\item A range of cut-off values, $J_0^{k=1,2,...n}$ is generated with a finite step size. The finer the step size the longer the computational time.
\item For the $k$\textsuperscript{th} iteration, remove all values of $J<J_0^{k}$.
\item Then, perform an iterative version of \cref{eq:correstiexponent}, as shown below, to obtain the $k$\textsuperscript{th} estimated exponent, $\hat{\alpha}_{k}$
\begin{equation}\label{eq:iterestiexponent}
\hat{\alpha}_k=1+N'_k\left[\sum_{i=1}^{N'_k}\ln{\frac{J_i}{J_0^{k}}}\right]^{-1}\textrm{, where } J_i>J_0^{k} \textrm{ and } N'_k<N
\end{equation}
\item After $n$ number of iterations, a plot of $\hat{\alpha}_{k=1,2...n}$ against $\ln(J_0^{k=1,2,...n})$ is generated.
\end{enumerate}
\subsection{Decrement as $J_0<J_\textrm{min}$}\label{apn:MLdecJ0Jmin}
In the perfect power-law ML fit (\Cref{subfig:ml_sing}), the estimated exponent decreases as $J_0$ descends below the lower-bound $J_\textrm{min}$. This phenomena can be observed in almost all ML fits and can be shown by rearranging \cref{eq:iterestiexponent}.\\

Since $J_0<J_\textrm{min}$, \cref{eq:iterestiexponent} can be expanded as follows:
\begin{equation}
\begin{split}
\hat{\alpha}&=1+N\left[\sum_{i=1}^{N}\ln{\frac{J_i}{J_{0}}}\right]^{-1}\\
&=1+N\left[\sum_{i=1}^{N}\ln{\frac{J_i\times J_\textrm{min}}{J_\textrm{min}\times J_{0}}}\right]^{-1}\\
&=1+N\left[
\underbrace{\sum_{i=1}^{N}\ln{\frac{J_i}{J_\textrm{min}}}}_\textrm{Part I}
+\underbrace{\sum_{i=1}^{N}\ln{\frac{J_\textrm{min}}{J_0}}}_\textrm{Part II}
\right]^{-1}\notag
\end{split}
\end{equation}
Part I is now a constant term; with some rearranging, let it take the form of an exponent, $\alpha_0$:
\begin{equation}
\textrm{Part I}\rightarrow
\sum_{i=1}^{N}\ln{\frac{J_i}{J_\textrm{min}}}=\frac{N}{\alpha_0-1}
\end{equation}
Part II is a summation of two constant terms, thus:
\begin{equation}
\textrm{Part II}\rightarrow
\sum_{i=1}^{N}\ln{\frac{J_\textrm{min}}{J_0}}=N\ln\frac{J_\textrm{min}}{J_0}
\end{equation}
Thus the equation is reduced to:
\begin{equation}\label{eq:MLdecrementshow}
\begin{split}
\hat{\alpha}&=N\left[\frac{N}{\alpha_0-1}+N\ln{\frac{J_\textrm{min}}{J_0}}\right]^{-1}
=\left[\frac{1}{\alpha_0-1}+\ln{\frac{J_\textrm{min}}{J_0}}\right]^{-1}
\end{split}
\end{equation}
Therefore, by inspection, as $J_0$ decreases below $J_\textrm{min}$, $\ln{J_\textrm{min}/J_0}$ increases, thus reducing $\hat{\alpha}$.
\newpage
\subsection{Mixture of two Power-laws}\label{theory:exponentmixing}
\begin{figure}
	\centering
	\begin{subfigure}[t]{0.48\textwidth}
		\includegraphics[width=1.0\textwidth]{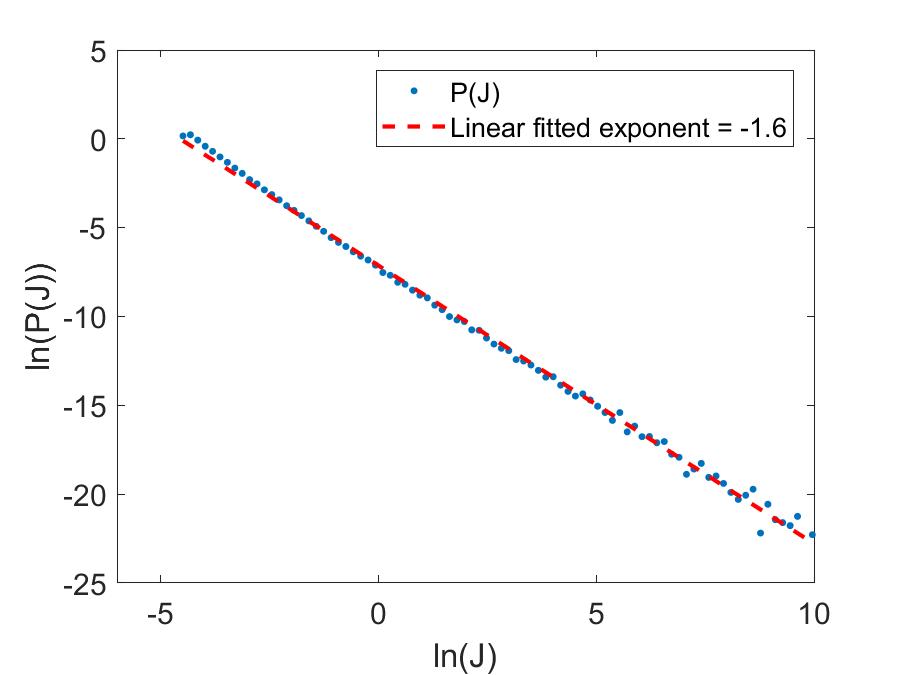}
		\subcaption{Log-log fit: $40:60$ mixture of exponents $\alpha=1.5$ and $\beta=2.0$.}
		\label{subfig:pjj_4060}	
	\end{subfigure}
	\hfill
	\begin{subfigure}[t]{0.48\textwidth}
		\includegraphics[width=1.0\textwidth]{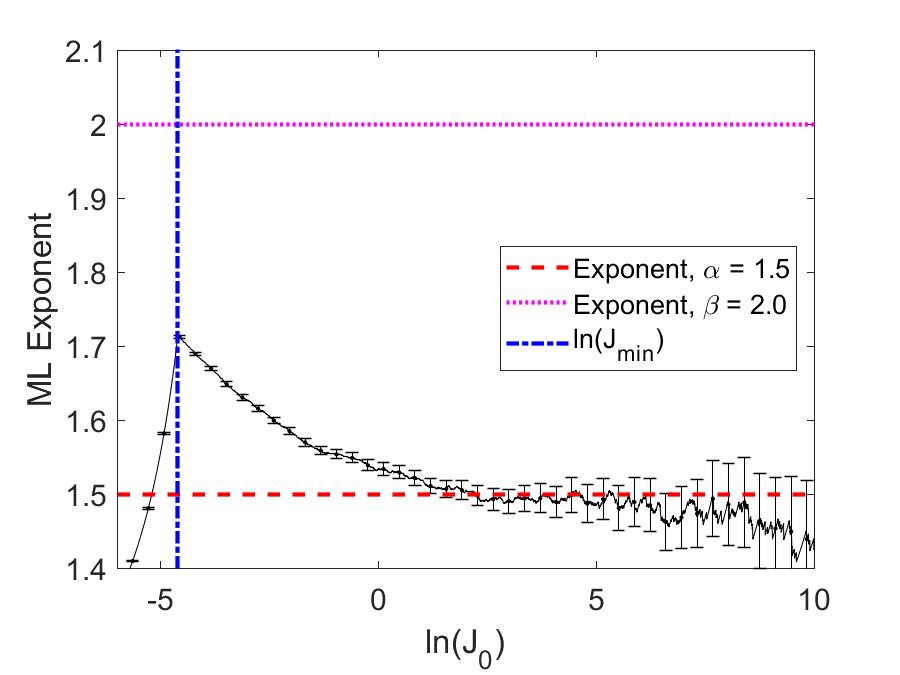}
		\subcaption{Corresponding maximum-likelihood (ML) fit of $40:60$ mixture. }
		\label{subfig:ML_4060}
	\end{subfigure}
	\begin{subfigure}[t]{0.48\textwidth}
		\includegraphics[width=1.0\textwidth]{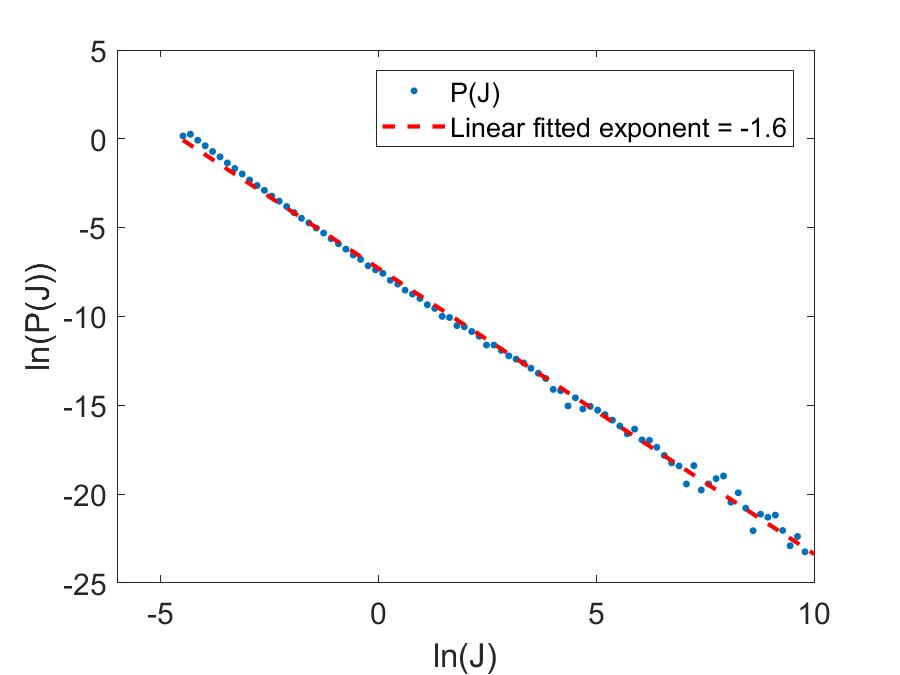}
		\subcaption{Log-log fit: $30:70$ mixture of exponents $\alpha=1.5$ and $\beta=2.0$.}
		\label{subfig:pjj_3070}	
	\end{subfigure}
	\hfill
	\begin{subfigure}[t]{0.48\textwidth}
		\includegraphics[width=1.0\textwidth]{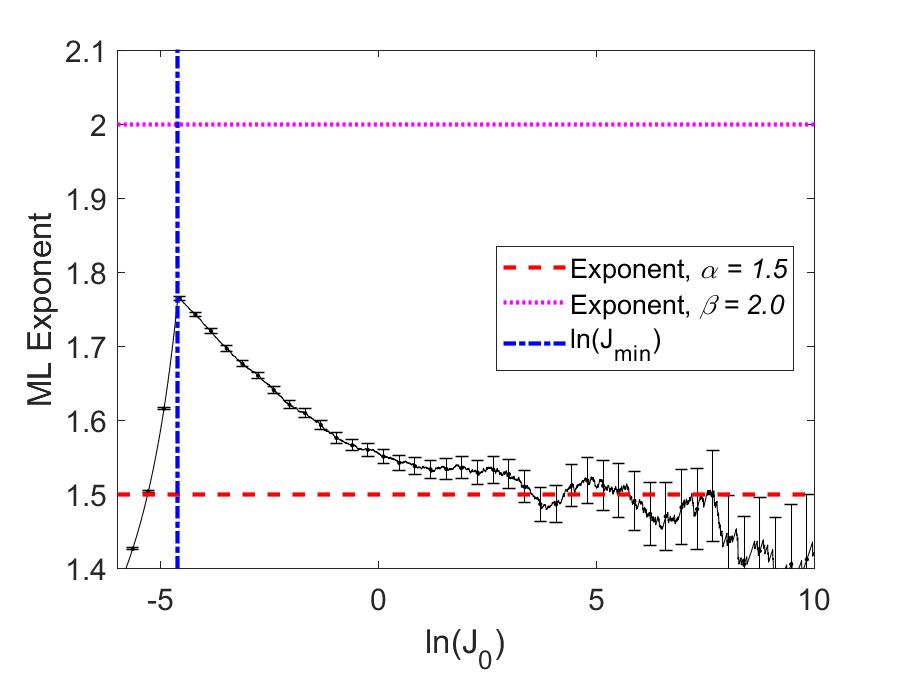}
		\subcaption{Corresponding maximum-likelihood (ML) fit of $30:70$ mixture. }
		\label{subfig:ML_3070}
	\end{subfigure}
	
	\centering
	\caption{(a) is the log-log distribution $P(J)$ vs $J$ plot of a randomly generated jerk spectrum that consists of two power-law exponents $\alpha = 1.5$ and $\beta = 2.0$ in a $40:60$ mixture. Linear fitting the distribution will give an exponent that lies between the two known exponents\cite{Salje2017}. (b) is the corresponding ML analysis which reveals the mixing nature by showing a kink at $J_0=J_\textrm{min}$ (blue dot-dashed line) that underestimates the higher exponent (pink dotted line), and a convergence to the lower exponent (red dashed line). (c) is a log-log distribution of exponents $\alpha = 1.5$ and $\beta = 2.0$ in a $30:70$ mixture instead with (d) as its respective ML plot. The kink is closer yet still underestimates the higher exponent and due to lack of statistically relevant data, the curve is unable to converge to the known exponent $\alpha = 1.5$.}
	\label{fig:twoexpo_4060linfit}
\end{figure}
A distribution with a mixture of two exponents is not immediately obvious to the eye, and an attempt to extract an exponent via linear-fit might be a pitfall, yielding an averaged exponent of the two mixtures\cite{Salje2017}.\\\par
A jerk spectrum with a total of $N=10^5$ jerks was randomly generated with two \textbf{known exponents} $\alpha=1.5$ and $\beta=2.0$ with a 40:60 ratio. The jerks generation method is briefly described in \textbf{\Cref{apn:MCgenexpomix}}. The distribution was plotted and linearly fitted in \Cref{subfig:pjj_4060}. Not knowing \textit{a priori} that the jerk spectrum consisted of two exponents, the fit in \Cref{subfig:pjj_4060} seemed good (with $R^2$ value $ = 0.9976$ given by Matlab\textsuperscript{\textregistered}) and a slope of $-1.6$ was fitted. Yet, we know this is not true; $-1.6$ is the result of an averaging effect of both exponents $\alpha$ and $\beta$. One way to think of it is to imagine an initial straight-line with gradient $\beta = 2.0$; the higher count at larger $J$ is mainly contributed by exponent $\alpha = 1.5$, reducing the overall steepness of the line. \\\par

Performing an ML analysis is an easy way of avoiding this pitfall. Shown in \Cref{subfig:ML_4060} is the ML curve of the 40:60 mixing jerk spectrum. As usual, a kink can be observed as $J_0 = J_\textrm{min}$ (blue dot-dashed line), but the kink gives an estimated exponent around $1.7$, which underestimates the value for exponent $\beta$ (pink dotted line), especially when the ratio of mixing is close\cite{Salje2017}. But as $J_0$ increases, jerks from exponent $\beta$ contribute less while jerk peaks from exponent $\alpha$ dominate, hence the ML curve converges to exponent $\alpha = 1.5$ (red dashed line). And as $J_0$ increases further, there are less statistically relevant data, the error increases and the ML curve fluctuates. \\\par
Finally, another jerk spectrum with the same exponents but with a mixture of 30:70 was generated. The respective ML curve is shown as \Cref{subfig:ML_3070}. Comparing between \Cref{subfig:ML_3070,subfig:ML_4060}, the kink for $30:70$ occurs at a higher exponent around $1.76$ due to fewer contributions from the now $30\%$ weighted exponent $\alpha = 1.5$. In addition, due to the fewer statistically relevant data points from exponent $\alpha$, the ML curve fails to converge to the known exponent as $J_0$ increases and deviates away.\\\par
So, a distribution that resembles a straight-line may be hiding a mixture of two power-laws. ML analysis must be performed in order to learn more about an avalanche system. Yet, even if a system is shown to demonstrate exponent mixing, the proper exponents are hard to determine. The higher exponent will be underestimated while the lower exponent will be overestimated if the jerk spectrum lacks relevant data \cite{Salje2017}.

\newpage


\subsection{Randomly Generated Jerk Spectra}
\subsubsection{Inversion Sampling Method}\label{apn:MCgensingpowlaw}

\begin{figure}[h]
	\centering
	\includegraphics[width=0.7\textwidth]{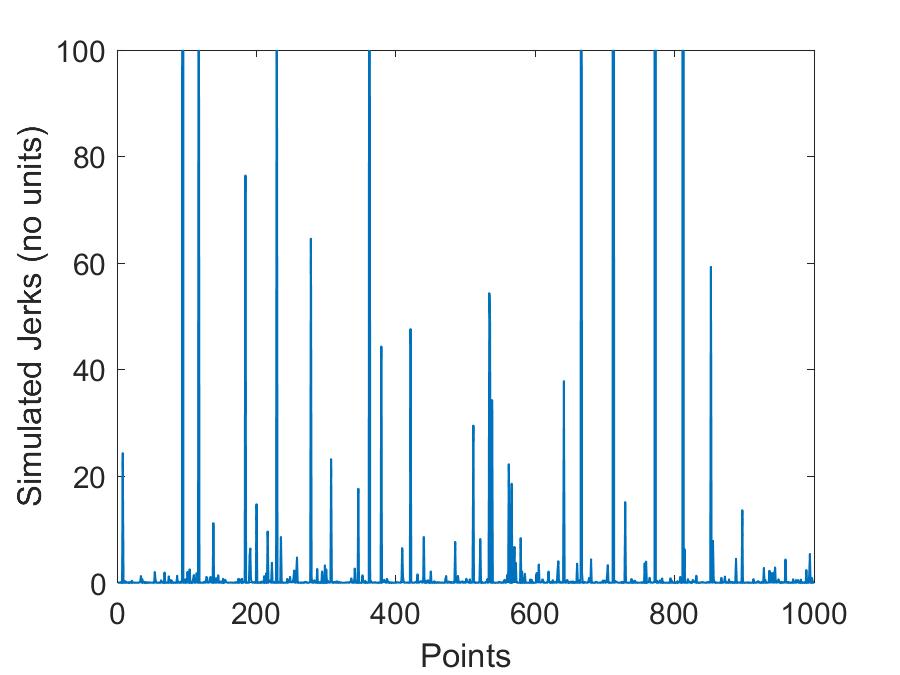}

	\caption{A randomly generated jerk spectrum of 1000 points. This jerk spectrum is analysed by ML and log-log fitting in \Cref{fig:perfpowlaw_comb}.}	
		\label{fig:simjerks_sing}
\end{figure}

Given a known probability distribution, in our case a power-law distribution, one can randomly generate a jerk spectrum by using a random number generator (RNG) and performing a probability distribution sampling. This sampling technique is the basis of many Monte Carlo simulations \cite{Vassiliev2017}.
To sample from the power-law probability distribution: 
\begin{equation}
P(J)=\frac{\alpha-1}{J_\textrm{min}}\left(\frac{J}{J_\textrm{min}}\right)^{-\alpha}\tag{\ref{eq:jerkpowerlaw}}
\end{equation}
one must invert the cumulative distribution function (CDF), $F(J)$\cite{Vassiliev2017} where
\begin{equation}
F(J)=\int_{-\infty}^{J'}P(J)dJ=\int_{-\infty}^{J'}\frac{\alpha-1}{J_\textrm{min}}\left(\frac{J}{J_\textrm{min}}\right)^{-\alpha}dJ
\end{equation}
where $J'$ is a dummy variable. 
Let $\xi$ be the call to a random number generator, and by replacing the lower-limit of the integral from $-\infty$ to $J_\textrm{min}$ (since $P(J) = 0$ for $J<J_\textrm{min}$), the recipe for the sampling algorithm will be
\begin{equation}
\begin{split}
\xi=\frac{\int_{J_\textrm{min}}^{J'}\frac{\alpha-1}{J_\textrm{min}}\left(\frac{J}{J_\textrm{min}}\right)^{-\alpha}dJ}{\int_{J_\textrm{min}}^{\infty}\frac{\alpha-1}{J_\textrm{min}}\left(\frac{J}{J_\textrm{min}}\right)^{-\alpha}dJ}\notag
\end{split}
\end{equation}
where the denominator is the normalisation condition, generating the jerk data in the range $J_\textrm{min}\leq J<\infty$. One can easily decrease the range by changing the upper limit of the integral. The equation reduces to:
\begin{equation}
\begin{split}
\xi&=\frac{\int_{J_\textrm{min}}^{J'}J^{-\alpha}dJ}{\int_{J_\textrm{min}}^{\infty}J^{-\alpha}dJ}=\frac{\left[J^{-\alpha+1}\right]^{J'}_{J_\textrm{min}}}
{\left[J^{-\alpha+1}\right]^{\infty}_{J_\textrm{min}}}\\
&=\frac{J'^{1-\alpha}-J_\textrm{min}^{1-\alpha}}{0-J_\textrm{min}^{1-\alpha}}\\
&=1-\left(\frac{J'}{J_\textrm{min}}\right)^{1-\alpha}
\end{split}
\end{equation}
From here, the CDF can be inverted and becomes:
\begin{equation}
J=J_\textrm{min}\left(1-\xi\right)^{\frac{1}{1-\alpha}}\label{eq:MCpowlawgen}
\end{equation}
where the dummy variable $J'$ is replaced by $J$. Therefore, for each random number $\xi$ called, a jerk data point will be generated that follows the power-law distribution of an exponent $\alpha$.
\subsubsection{Generating Jerk Spectra with Exponent Mixing}\label{apn:MCgenexpomix}
To generate a jerk spectra with two different exponents, the recipe above can be used with an additional call to the RNG to determine which distribution $J$ should be sampled from.\\\par
An example to generate a $40:60$ mixture of exponents $\alpha = 1.5 $ and $\beta = 2.0$ respectively is summarised below:
\begin{enumerate}
\item At the start of the algorithm, call a random number, $\xi_1$ and perform a check with the mixing ratio. 
\item If $\xi_1<0.4$, run \cref{eq:MCpowlawgen} for $\alpha$
\begin{equation}
J=J_\textrm{min}\left(1-\xi_2\right)^{\frac{1}{1-a}}\notag
\end{equation}
$\xi_1$ and perform a check with the mixing ratio. 
\item If $\xi_1>0.4$, run \cref{eq:MCpowlawgen} for $\beta$
\begin{equation}
J=J_\textrm{min}\left(1-\xi_3\right)^{\frac{1}{1-b}}\notag
\end{equation}
\end{enumerate}
Do note: $\xi_1,\xi_2,\xi_3$ are all \textbf{different} calls to the RNG.
\newpage
\section{Noise Analysis}\label{apn:noisean}
\begin{figure}[h]
	\centering
	\begin{subfigure}{0.48\textwidth}
		\includegraphics[width=1.0\textwidth]{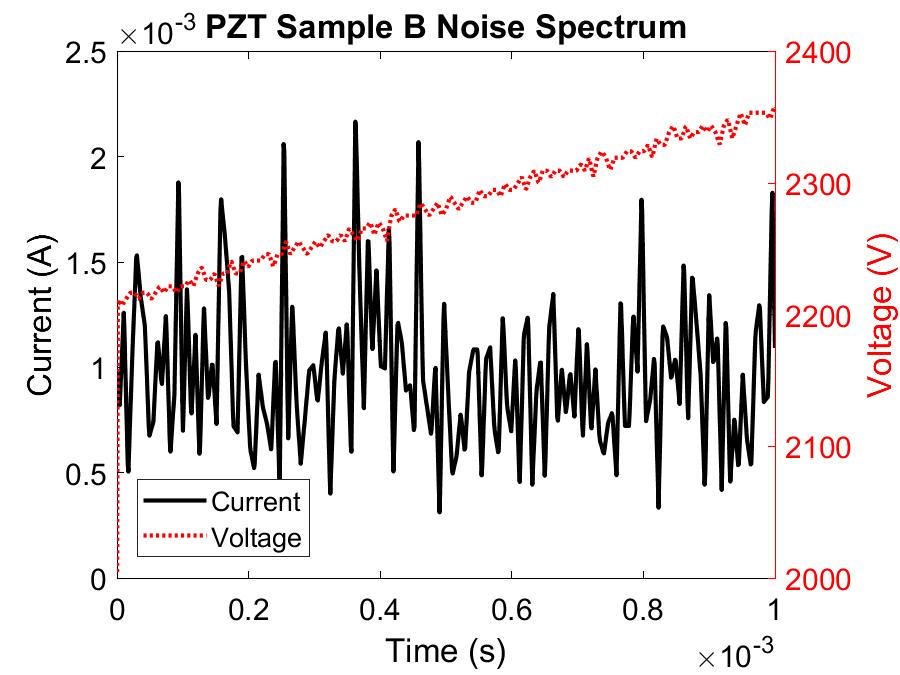}
		
		\subcaption{Current (black) and voltage (red) against time of sample B.}	
		\label{subfig:ew_pztb_noise}
	\end{subfigure}
	\hfill
	\begin{subfigure}{0.48\textwidth}
		\includegraphics[width=1.0\textwidth]{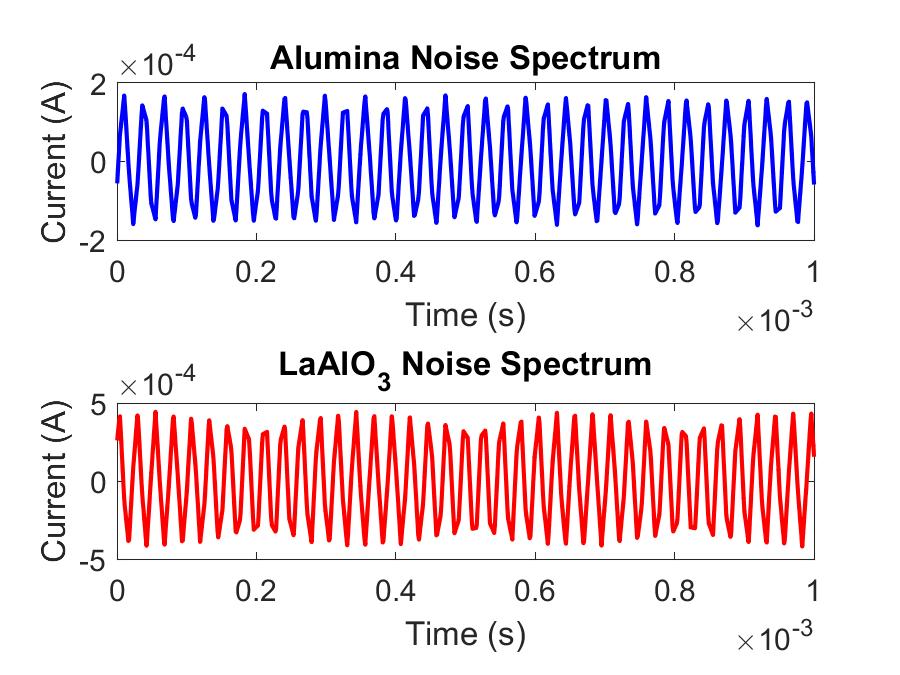}
		
		\subcaption{Alumina (top blue) and LaAlO\textsubscript{3} (bottom red) current respones.}
		\label{subfig:ew_perov_noise}
	\end{subfigure}
	
	\caption{(a): The current and voltage response of PZT sample B. The applied UpSwP has a rate of $200$ kVs$^{-1}$ and a maximum voltage of $2.4$ kV. Only the data for the last millisecond of UpSwP were recorded, where the Barkhausen noise were presumed to be abundant. Note that the final voltage data point did not really achieve $2.4$ kV, which was due to the limitations of the instrument. (b): The top blue current spectrum belongs to the alumina sample while the bottom red spectrum belongs to lanthanum aluminate, LaAlO\textsubscript{3}. The noise generated from PZT in (a) seemed random while the noises from both Al$_2$O$_3$ and LaAlO\textsubscript{3} in (b) were oscillatory and resembled ringing noise.}
	\label{fig:ew_noise_raw}
	\begin{multicols}{2}
		\includegraphics[width=0.4\textwidth]{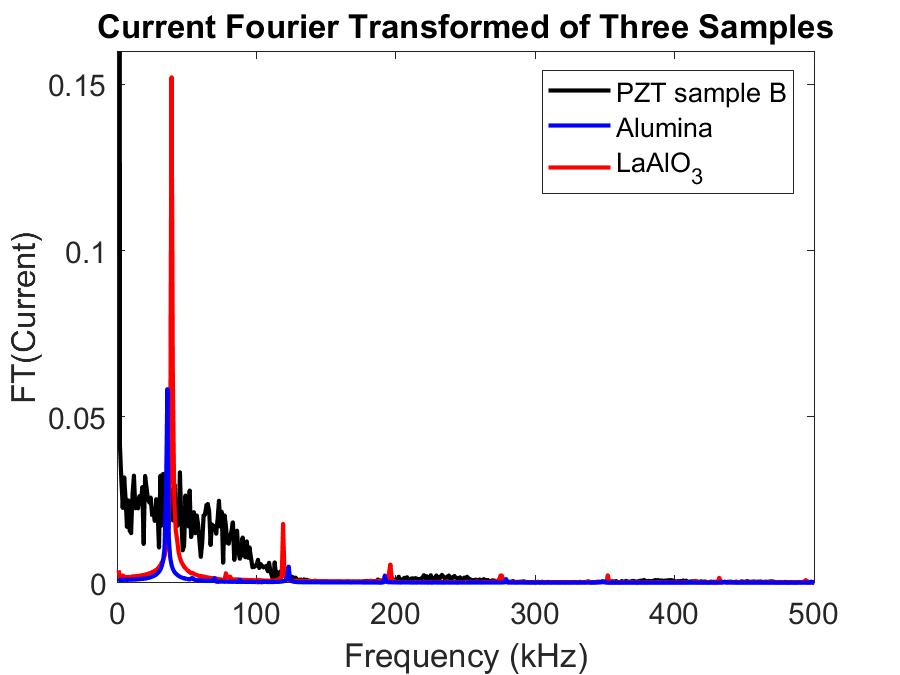}
		\caption{The current spectra (from \Cref{fig:ew_noise_raw} above) of PZT (black), Al$_2$O$_3$ (blue) and LaAlO$_3$ (red) Fourier transformed. The Fourier spectra of both Al$_2$O$_3$ and LaAlO$_3$ showed multiple peaks that represent periodicities that were caused by the oscillatory current responses. We suspect these oscillations were sample dependent and may be voltage ``echoes". These ``echo" effects bear a resemblance to crystal resonators\cite{Johannsmann2015}.}
		\label{fig:ew_three_fft}
	\end{multicols}
\end{figure}

To ensure those spikes were not merely electrical artefacts, the two pulse experiments were conducted on two other non-ferroelectric samples, alumina, Al\textsubscript{2}O\textsubscript{3} and lanthanum aluminate\footnote{Coincidentally, lanthanum aluminate is also perovskite-structured\cite{Kawabe2000}.}, LaAlO\textsubscript{3}, shown as \Cref{fig:ew_noise_raw}. \par
A pair of antisymmetrical voltage pulses with maximum magnitude, $\left|V_\textrm{max}\right|=2.4$ kV with rate $=200$ kVs\textsuperscript{$-1$} were applied to PZT sample B, alumina and LaAlO$_3$; the final one millisecond of UpSwP (where the presumed Barkhausen noise of sample B was the most evident) was recorded for all three samples. From \Cref{subfig:ew_pztb_noise}, the noise generated from PZT sample B seemed jerky and random. Both Al$_2$O$_3$ (top blue) and LaAlO$_3$ however produced oscillatory ringing noise. The current spectra from \Cref{fig:ew_noise_raw} were then Fourier transformed (\Cref{fig:ew_three_fft}, where Al$_2$O$_3$ and LaAlO$_3$ exhibited strong periodicities. These ringing ``echoes" may be sample dependent due to the peaks occurring at various frequencies. We suspect these ``echo" effects may also be similar to crystal resonators\cite{Johannsmann2015}.\\\par

\section{Removing Smooth Baselines}\label{apn:RemBase_pztf}
\begin{figure}[h]
	\centering
	\begin{subfigure}[t]{0.48\textwidth}
		\includegraphics[width=1\textwidth]{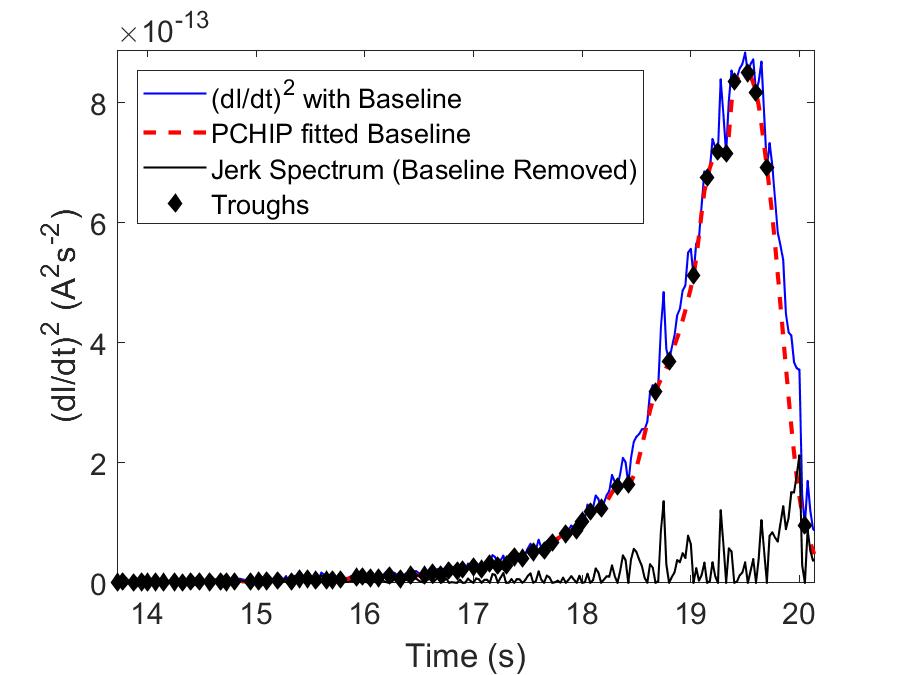}
		\caption{Removing baseline of jerk spectrum in PZT B Run 1}
		\label{subfig:pztb_blrem}
	\end{subfigure}\hfill
	\begin{subfigure}[t]{0.48\textwidth}
		\includegraphics[width=1\textwidth]{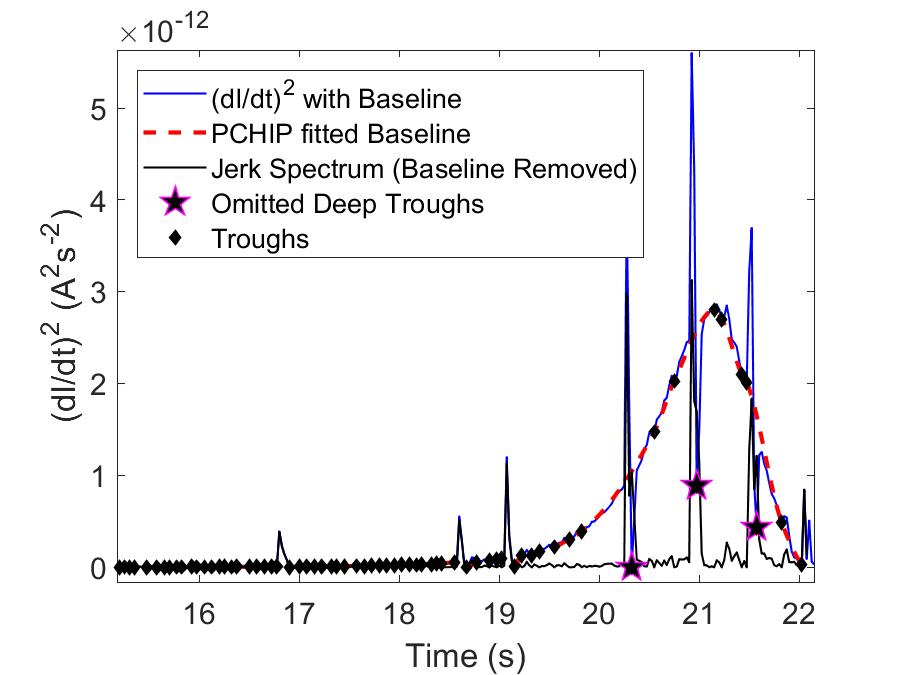}
		\caption{Removing baseline of jerk spectrum in PZT F Run 1}
		\label{subfig:pztf_blrem}
	\end{subfigure}
	
	\caption{(a): Procedure of removing the baseline from the jerk spectrum of PZT sample B run 1. The baseline was formed by interpolating the troughs of the $(dI/dt)^2$ curve using the Piecewise Cubic Hermite Interpolating Polynomial (PCHIP) method in Matlab\textsuperscript{\textregistered} to preserve the shape\cite{Fritsch1980}. The baseline was then subtracted, leaving the jerk peaks. (b): This was not as straight-forward for PZT sample F where there were huge spanning avalanche peaks\cite{He2016}. The deep troughs (pentagram shaped points) were ignored in order to allow the PCHIP interpolation to form a smooth baseline. A trough is defined as a point where both the entries immediately before and after it are greater.}
	\label{fig:blrem}
\end{figure}

For sample B's jerk spectra in \textbf{\Cref{subsec:srun_pztb,subsec:mrun_pztb}}, the troughs\footnote{A trough is defined as a point $i$, let the magnitude of the point be $x_i$, where $x_{i-1}>x_i$ and $x_{i+1}>x_i$.} in the jerk spectra were interpolated by the Piecewise Cubic Hermite Interpolating Polynomial (PCHIP) function in  Matlab\textsuperscript{\textregistered}, forming shape-preserving\cite{Fritsch1980} baselines (\Cref{subfig:pztb_blrem}). These makeshift baselines were then subtracted from the $(dI/dt)^2$ curve, leaving the jerk peaks for analysis. The baseline removal procedure is essential when there is a superposition of jerks on a smooth background\cite{Salje2014,Gallardo2010}.\\\par
Sample F however had deep troughs in the $(dI/dt)^2$ curve due to those current spikes. The Matlab\textsuperscript{\textregistered} routine that was previously used for interpolating was then modified to ignore the deep troughs (pentagram labelled in \Cref{subfig:pztf_blrem}). This allowed the \textit{interpolator} to omit those points and form a smooth baseline.

\newpage
\section{Other Jerk Analyses Results}\label{apn:othjerkres}

\subsection{PZT Sample B: Runs 2 to 5}\label{apn:pztb_2_5}
\subsubsection{Runs 2 and 3}
\begin{figure}[h]
	\centering
	\begin{subfigure}[t]{0.45\textwidth}
		\includegraphics[width=1.0\textwidth]{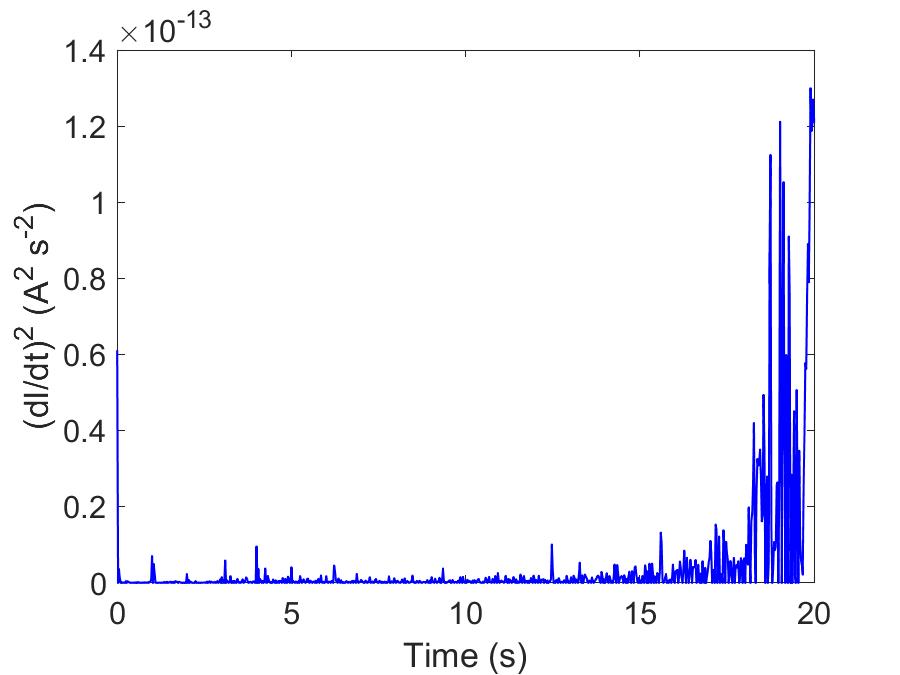}
		\caption{Run 2}
		\label{subfig:pzt_jspec_run_2}
	\end{subfigure}
	\hfill
	\begin{subfigure}[t]{0.45\textwidth}
		\includegraphics[width=1.0\textwidth]{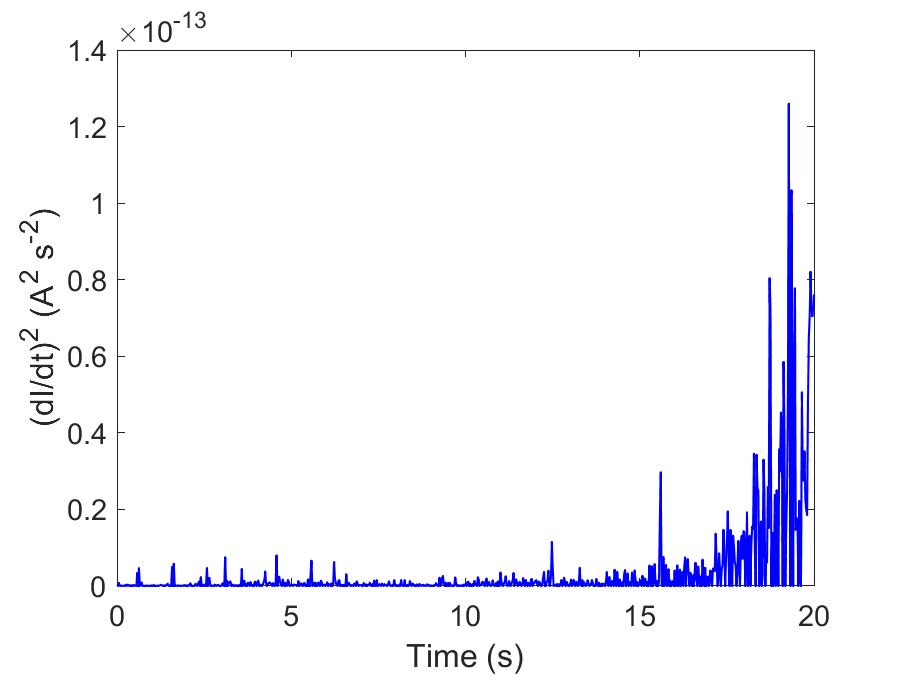}
		\caption{Run 3}
		\label{subfig:pzt_jspec_run_3}
	\end{subfigure}
	
	\centering
		\begin{subfigure}[t]{0.45\textwidth}
		\includegraphics[width=1\textwidth]{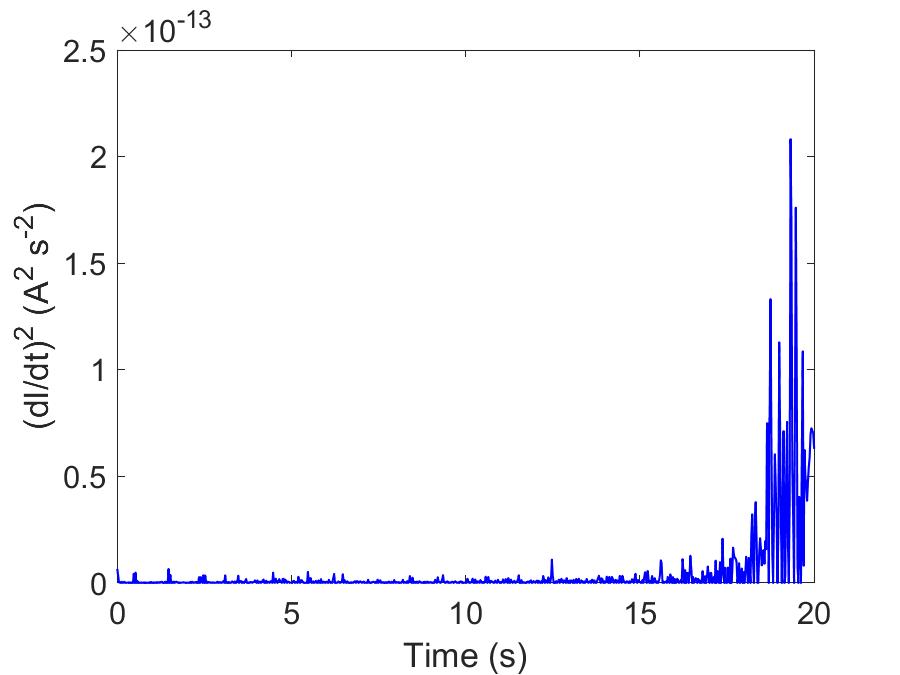}
		\caption{Run 4}
	\end{subfigure}
	\hfill
	\begin{subfigure}[t]{0.45\textwidth}
		\includegraphics[width=1\textwidth]{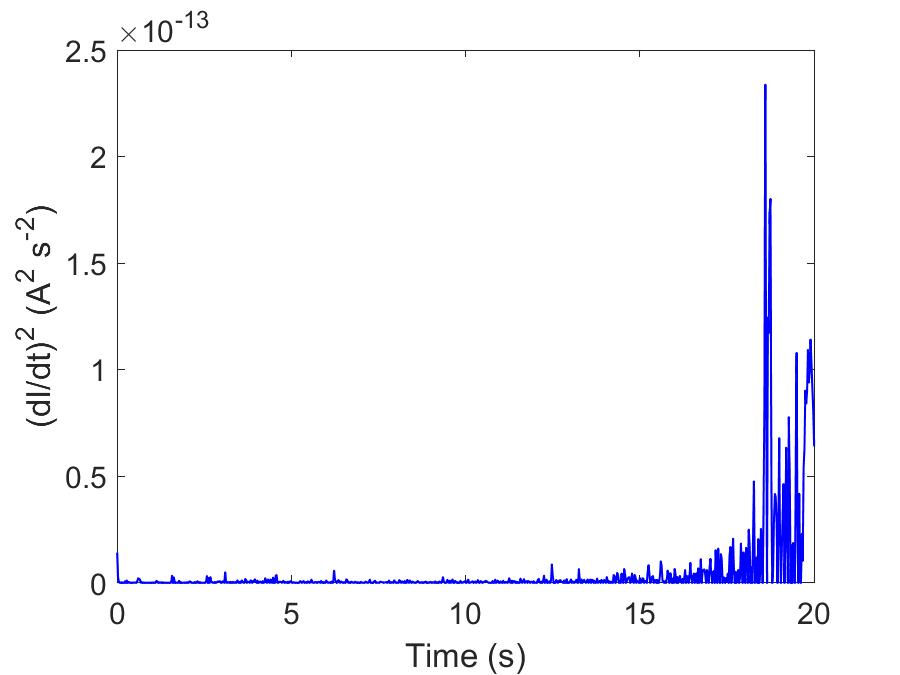}
		\caption{Run 5}
	\end{subfigure}
	
	\caption{(a-d) are the jerk spectra of runs number 2 to 5 respectively from PZT sample B. The peaks are used for separate ML analyses (\Cref{fig:pztb_MLs_2_5}) and are later normalised and combined for a \textit{grand} analysis.}
	\label{fig:pztb_runs_2_5}
\end{figure}

\begin{figure}[h]
	\centering
	\begin{subfigure}[t]{0.48\textwidth}
		\includegraphics[width=1.0\textwidth]{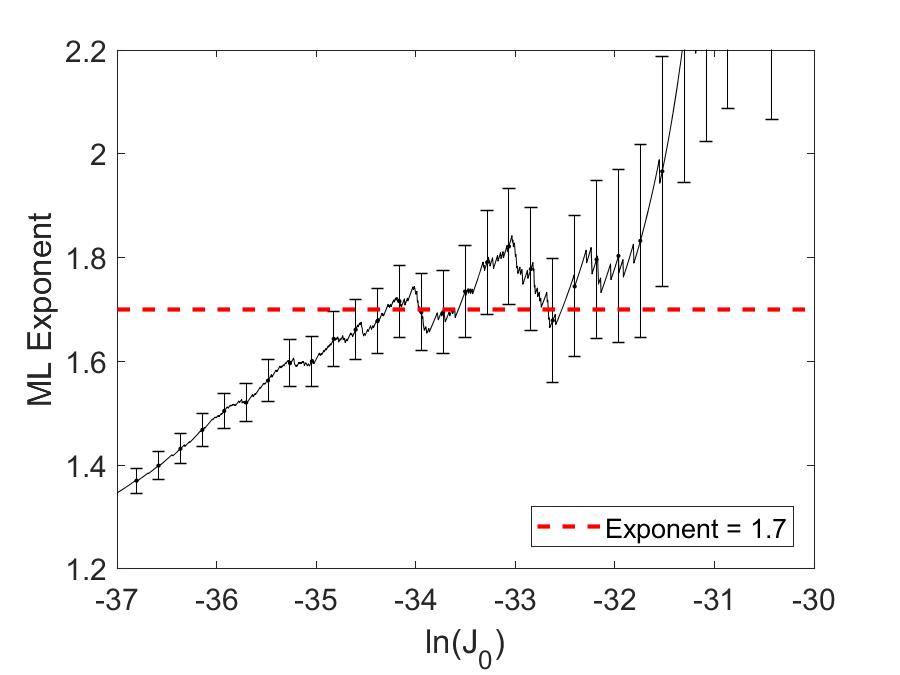}
		\caption{Run 2: ML Analysis}
		\label{subfig:pzt_ML_run_2}
	\end{subfigure}
	\hfill
	\begin{subfigure}[t]{0.48\textwidth}
		\includegraphics[width=1.0\textwidth]{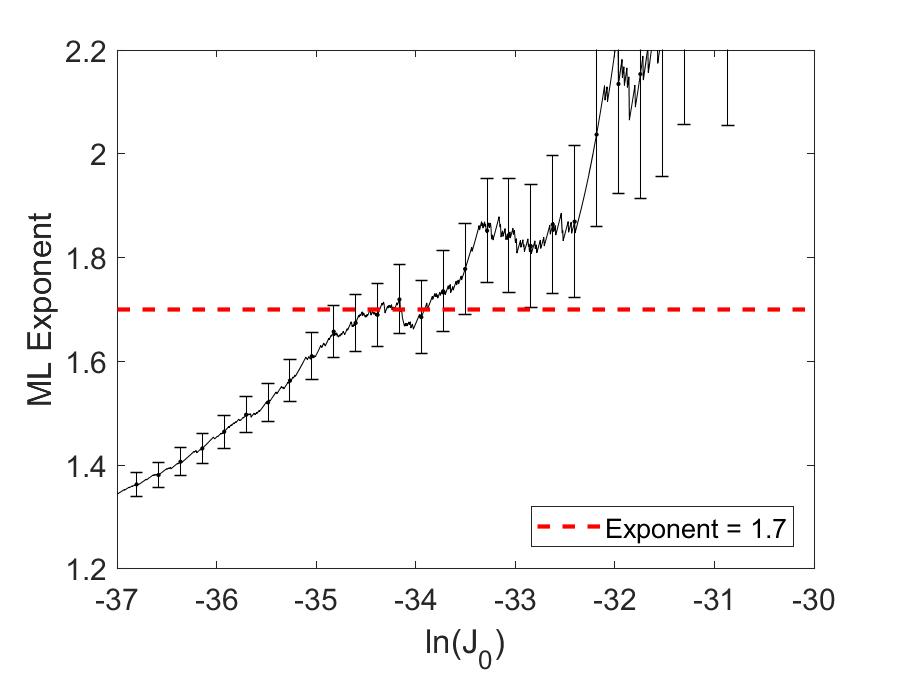}
		\caption{Run 3: ML Analysis}
		\label{subfig:pzt_ML_run_3}
	\end{subfigure}
\\
	\begin{subfigure}[t]{0.48\textwidth}
		\includegraphics[width=1.0\textwidth]{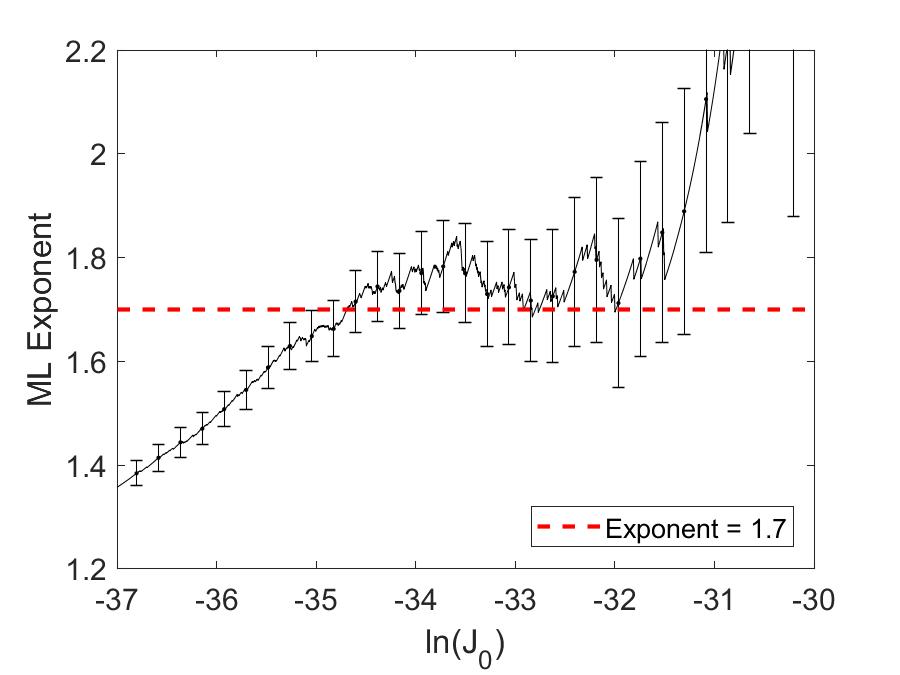}
		\caption{Run 4: ML Analysis}
	\end{subfigure}
	\hfill
	\begin{subfigure}[t]{0.48\textwidth}
		\includegraphics[width=1.0\textwidth]{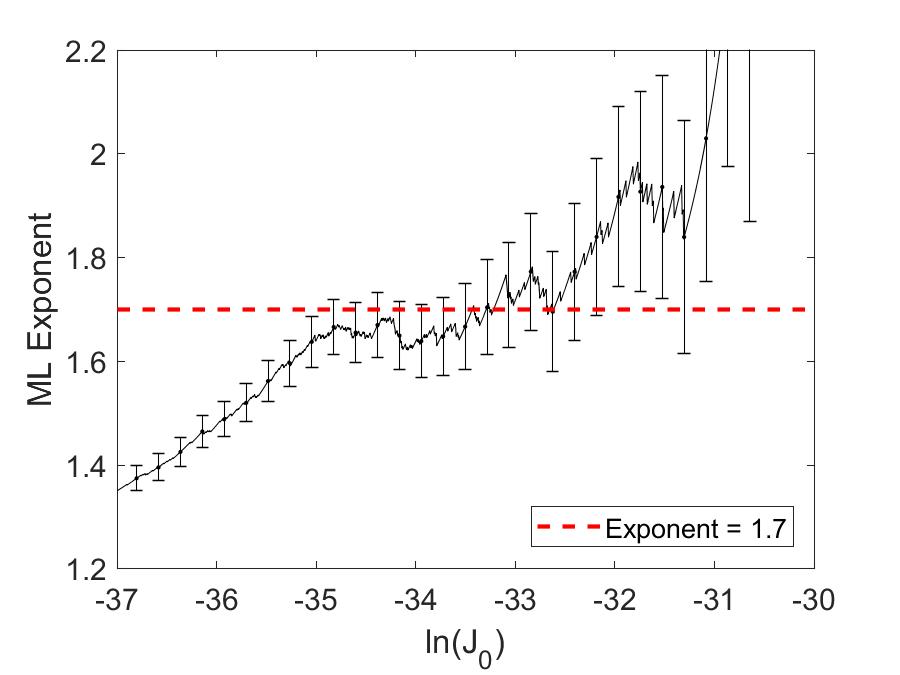}
		\caption{Run 5: ML Analysis}
	\end{subfigure}\\
	\caption{(a-d): ML analyses for runs 2 to 5 respectively. Four out of five ML analyses showed plateaus that spanned a few decades. The red dashed line denotes exponent $1.7$ that was used as a reference point. }
	\label{fig:pztb_MLs_2_5}
\end{figure}

A total of $5$ runs were recorded and analysed for PZT sample B, the jerk spectra are displayed in in \Cref{fig:pztb_runs_2_5} while the ML analyses are shown in \Cref{fig:pztb_MLs_2_5}. These jerk spectra were analysed independently from the ML method to show that they were power-law distributed.\\\par

\newpage
For Run $2$, a plateau that spans for 3 decades around $\ln(J_0)\approx-35$ to $-32$ with an exponent around $1.7$ can be seen. The kink however starts a little lower at $\ln(J_0)$ around $-35$ with an exponent of $1.6$. The ML result for Run $3$ however is not as simple as the previous two runs. Although a kink can be seen again at $\ln(J_0)$ around $-35$ with an exponent of $1.7$ in the ML curve, the plateau is not as strong as the others. Both runs 4 and 5 shows an exponent plateau around $1.7$ that spans for at least two decades. The kinks were observed similar to Run $2$, at around $\ln(J_0)=-35$. This solidifies our idea that these jerks do obey a power-law!\\\par

\newpage
\phantomsection 
\addcontentsline{toc}{section}{References} 
\bibliographystyle{ieeetr}
\bibliography{FinalYearReportWriting}
\end{document}